\documentclass[%
singlecolumn,
amsmath,amssymb,
aps,
notitlepage
]{revtex4-1}

\pdfoutput=1
\usepackage{subfig}
\usepackage{graphicx}
\usepackage{dcolumn}
\usepackage{bm}
\usepackage{hyperref}

\DeclareMathAlphabet{\mathsfbi}{OT1}{\sfdefault}{bx}{sl}
\DeclareMathVersion{sfletters}
\SetSymbolFont{letters}{sfletters}{OML}{ntxsfmi}{b}{it}

\makeatletter
\newcommand{\mathbfsbilow}[1]{%
  \text{\mathversion{sfletters}$\m@th#1$}%
}

\DeclareRobustCommand{\tensor}[1]{%
  \begingroup
  \ifcat\noexpand #1\relax
    \edef\greek@test{\detokenize{#1}}%
    \edef\greek@test{\expandafter\@cdr\greek@test\@nil}%
    \edef\greek@test{\expandafter\@car\greek@test\@nil}%
    \edef\x{\the\lccode\expandafter`\greek@test}%
    \edef\y{\number\expandafter`\greek@test}%
    \ifnum\x=\y\relax
      \mathbfsbilow{#1}%
    \else
      \mathsfbi{#1}%
    \fi
  \else
    \mathsfbi{#1}%
  \fi
  \endgroup
}
\makeatother

\begin{document}

\preprint{APS/123-QED}

\title{Algebraic disturbances and their consequences in rotating channel flow transition}

\author{Sharath Jose}
\email{josesk@tifrh.res.in}
\affiliation{TIFR Centre for Interdisciplinary Sciences, Tata Institute of Fundamental Research, 21 Brundavan Colony, Narsingi, Hyderabad 500075 India.}
\author{Vishnu Kuzhimparampil}
\affiliation{Shell Technology Center Bangalore (STCB), RMZ Centennial Campus B, Whitefield, Bangalore 560048, India}
\author{Beno\^it Pier}
\affiliation{Laboratoire de m\'ecanique des fluides et d'acoustique, CNRS -- \'Ecole centrale de Lyon -- Universit\'e Claude-Bernard Lyon 1 -- INSA, 36 avenue Guy-de-Collongue, F-69134 \'Ecully, France}
\author{Rama Govindarajan}
\affiliation{TIFR Centre for Interdisciplinary Sciences, Tata Institute of Fundamental Research, 21 Brundavan Colony, Narsingi, Hyderabad 500075 India.}

\date{}
\begin{abstract}
  It is now established that subcritical mechanisms play a crucial role in the transition to turbulence of non-rotating plane shear flows. The role of these mechanisms in rotating channel flow is examined here in the linear and nonlinear stages. Distinct patterns of behaviour are found: the transient growth leading to nonlinearity at low rotation rates $Ro$, a highly chaotic intermediate $Ro$ regime, a localised weak chaos at higher $Ro$, and complete stabilization of transient disturbances at very high $Ro$. At very low $Ro$, the transient growth amplitudes are close to those for non-rotating flow, but Coriolis forces already assert themselves by producing distinct asymmetry about the channel centreline. Nonlinear processes are then triggered, in a streak-breakdown mode of transition. The high $Ro$ regimes do not show these signatures, here the leading eigenmode emerges as dominant in the early stages. Elongated structures plastered close to one wall are seen at higher rotation rates. Rotation is shown to 
reduce non-normality in the linear operator, in an indirect manifestation of Taylor--Proudman effects. Although the critical Reynolds for exponential growth of instabilities is known to vary a lot with rotation rate, we show that the energy critical Reynolds number is insensitive to rotation rate. It is hoped that these findings will motivate experimental verification, and examination of other rotating flows in this light.
\end{abstract}

\pacs{Valid PACS appear here}
\maketitle

\section{\label{sec:Introduction}Introduction}
Rotation of the system, in a number of flow situations, plays an important role in the stability and turbulence characteristics. Rotational effects are seen to influence the evolution of several flow phenomena of practical interest ranging from engineering to geophysics. Atmospheric and oceanic flows offer a myriad of not entirely understood phenomena, which are affected by Earth's rotation in addition to other physics \cite{Gill_82book,Vallis_06book}. In industrial situations, the modelling of rotational effects of flows is a crucial aspect in the design procedure of several technologies: pumps and turbines for example \cite{Childs_10book}. 

The effect of rotation on shear flow instabilities is not immediately obvious, and it depends largely on the strength of the rotation. Rotation introduces a body force which is a function of space and time, and bears some analogy to density stratification. At high $Ro$, the flow is expected to obey Taylor--Proudman behaviour \cite{Taylor_1917PRSLSA,Proudman_1916PRSLSA}, by which variations parallel to the rotation axis are strongly suppressed. In the manner of swirling flows, an inviscid criterion for instability in parallel flows of the form $\boldsymbol{U} = (U(y),0,0)$, with the rotation vector $\boldsymbol{\Omega} = (0,0,\Omega)$ perpendicular to the plane of the flow, can be formulated as follows \cite{Bradshaw_69JFM,Pedley_69JFM} 
\begin{align}
 2\Omega\left(-\frac{\partial U}{\partial y} + 2\Omega\right) < 0. \label{eq:rayleigh_dim}
\end{align}
This states that an instability can occcur if at any point in the flow the absolute vorticity of the base flow and the rotation vector are anti-parallel. Subsequent studies have shown that this simple analogue of the Rayleigh criterion provides good predictions in many parallel flows even in the presence of viscous effects \cite{Hart_71JFM,Johnston_etal_72JFM,Lezius_Johnston_76JFM,Alfredsson_Persson_89JFM,Yanase_etal_93PFA}. 

One of the most commonly studied systems is the pressure driven flow between two stationary, parallel plates that is rotated about the spanwise coordinate, which is also the geometry of our interest. This is an appealing system for investigation as it is a simple rotating shear flow which offers regions that are both stable and unstable as per the inviscid criterion given above. Henceforth this system will be referred to simply as rotating channel flow. On neglecting the effect of end walls, one sees that the base flow is described by the familiar parabolic velocity profile \cite{Lezius_Johnston_76JFM}. Close to these walls, a secondary flow in the form of a double vortex is set up \cite{Speziale_Thangam_83JFM}. 

This flow is characterised by two parameters, the Reynolds number $Re = U_{0}d/\nu$, and the rotation number $Ro =\Omega d/U_{0}$, where $U_0$ is the centreline velocity in the channel, $d$ its half-width, $\Omega$ the rotation rate, and $\nu$ the kinematic viscosity of the fluid. It was found experimentally that the critical Reynolds number $\Re_{cr}$, below which no exponential instabilities exist, may be up to two orders magnitude lower than that of a non-rotating channel \cite{Hart_71JFM,Alfredsson_Persson_89JFM}. This critical Reynolds number shows a non-monotonic variation with the strength of rotation, and is very sensitive to it. Just past $Re_{cr}$, the first unstable mode corresponds to a stationary streamwise-invariant disturbance. As we move further into the unstable part of the parameter space, we may find oblique modes that have growth rates comparable to the streamwise-invariant mode \cite{Wall_Nagata_06JFM}. At high $Ro$, Taylor--Proudman behaviour sets in and these streamwise-invariant 
rotation modes are suppressed. The two-dimensional spanwise-invariant Tollmien--Schlichting (TS) mode can still be triggered for values of $Re$ above its critical value 5772  \cite{Drazin_Reid_04book,Wallin_etal_13JFM}. But in the regime where both the TS mode and the rotation mode are present, the rotation mode is expected to win over due to a much larger growth rate \cite{Wall_Nagata_06JFM}. 

Secondary instabilities of the travelling wave type with short and long wavelengths which eventually broke down to turbulence had been observed in experiments \cite{Alfredsson_Persson_89JFM}. Merging and splitting of vortex pairs through a nonlinear wavelength selection process was also seen. These types of motions were further confirmed by numerical studies \cite{Finlay_90JFM,Finlay_92JFM,Yang_Kim_91PFA}. Matsubara \& Alfredsson attribute the secondary instability to the spanwise inflectional profile resulting from the saturation of the primary disturbance \cite{Matsubara_Alfredsson_98JFM}. The existence of secondary and tertiary saturated solutions of rotating shear flows have also been investigated \cite{Wall_Nagata_06JFM,Wall_Nagata_13JFM,Daly_etal_14JFM}. The turbulent rotating channel flow has been studied extensively for a wide range of Reynolds numbers through experiments and simulations \cite{Johnston_etal_72JFM,Kristoffersen_Andersson_93JFM,Lamballais_etal_98TCFD,Grundestam_etal_08JFM,Xia_etal_16JFM}. 

Sub-critical linear processes have been shown to be very relevant in determining the conditions of transition to turbulence in a variety of shear flows \cite{Reddy_etal_93JFM,Schmid_Henningson_01book,Schmid_07ARFM}. Very little attention has been given thus far to the effect of sub-critical mechanisms for pressure-driven shear flows with system rotation. One instance of such work is that of Yeckp \& Rossi on the asymptotic suction boundary layer \cite{Yecko_Rossi_04PF}. In the astophysical community, the effect of sub-critical mechanisms in non-magnetised accretion disks modelled as plane shear flows with rotation have been studied \cite{Yecko_04AA,Chagelishvili_etal_03AA,Mukhopadhyay_etal_06ASR}. The focus of this article is on the role of sub-critical mechanisms on transitions from the laminar state at different rotation regimes. 

At a given $Re$, there are two stable regions to the left and the right of the neutral stability boundary, at low and high $Ro$ respectively. We show that these two regions display markedly different behaviour. At low $Ro$, we find the optimal disturbances to be similar to those obtained in the non-rotating case - streamwise independent disturbances that develop linearly through lift-up effect \cite{Landahl_80JFM}. The difference is that the Coriolis force effects a break in symmetry of the dominant structures about the channel centreline. At high $Ro$, in a demonstration of the Taylor--Proudman effect, the optimal disturbances are those that vary very slowly with the spanwise coordinate and the dynamics is dominated by the Orr mechanism \cite{Orr_1907PRIAA}. 

Further, we investigate the nonlinear evolution resulting from linear transiently growing disturbances. We fix the Reynolds number at 1500, and traverse the $Ro$ line. For a fair comparison, a fixed initial condition is chosen, corresponding to the one which produces optimal transient growth at low $Ro$. For comparison we also introduce the least stable eigenmode as initial condition. At low $Ro$, we show that transition does take place, leading to sustained unsteady flows which are asymmetric in the mean. As we increase the rotation rate, into the linearly unstable regime, the nonlinear behaviour is highly chaotic, up to $Ro \sim 0.2$. Beyond this $Ro$, the secondary flow is restricted to the high pressure side of the channel, leaving the other side laminar. These secondary flows become weaker as $Ro$ is increased. Past the neutral boundary on the right, the flow returns rapidly to the laminar state. 

\section{System}
\subsection{Base flow}
\begin{figure}
 \centering
 \includegraphics[width=5.75in]{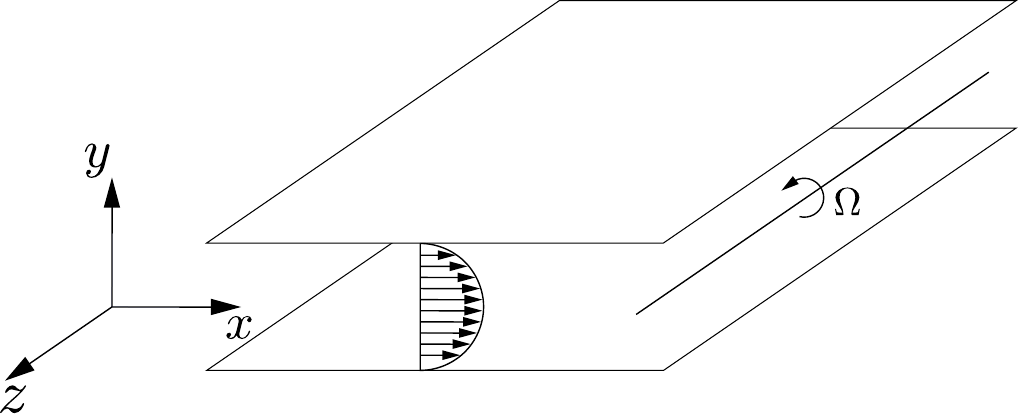}
 \caption{The rotating channel, with a parabolic streamwise velocity ($U = 1- y^2$) and a rotation rate $\mathbf{\Omega}$ about the spanwise coordinate.}
 \label{fig_base_conf}
\end{figure}
Our system, consisting of a pressure driven flow between parallel fixed walls, is subjected to rotation about the spanwise direction with a constant angular velocity $\boldsymbol{\Omega} = (0,0,\Omega)$ (figure \ref{fig_base_conf}). For the purposes of the analysis to follow, we consider the two parallel walls to extend infinitely, i.e. there are no end walls. It is also convenient to work in a frame of reference that is rotating along with the channel. Then the governing equations for the velocity $\boldsymbol{u^*}=(u^*,v^*,w^*)$ and pressure $p^*$ are the incompressible Navier-Stokes equations in the rotating frame given by
\begin{align}
\partial_{t^*}\boldsymbol{u^*} + \boldsymbol{u^*}\cdot\nabla^{*}\boldsymbol{u^*} &= -\frac{1}{\rho}\nabla^{*}p^* + \nu\Delta^{*}\boldsymbol{u^*} - 2\boldsymbol{\Omega}\times\boldsymbol{u^*}, \label{eq:ns_dim}\\
 \nabla^{*}\cdot\boldsymbol{u^*} &= 0. \label{eq:cont_dim} 
\end{align}
Here $\rho$ is constant density of the fluid. The centrifugal force has been absorbed into the pressure term. With $U_0$ and $d$ as the velocity and length scales respectively, and the Reynolds and Rotation numbers as defined in the previous section, and $\hat{\boldsymbol{z}}$ being the unit vector along the spanwise coordinate, the governing equations in the nondimensional form are as follows: 
\begin{align}
 \partial_{t}\boldsymbol{u} + \boldsymbol{u}\cdot\nabla\boldsymbol{u} &= -\nabla p+ \frac{1}{Re}\Delta\boldsymbol{u} - 2Ro\hat{\boldsymbol{z}}\times\boldsymbol{u}, \label{eq:ns_ndim}\\
 \nabla\cdot\boldsymbol{u} &= 0. \label{eq:cont_ndim}
\end{align}

\begin{figure}
 \centering
 \includegraphics[width=5.75in]{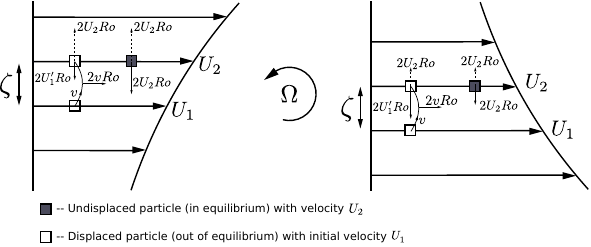}
 \caption{On the left we have a situation where the displaced particle is further driven away from its initial position due to the vertical pressure gradient force (dashed line with arrow pointing upwards) being higher than the Coriolis force (vertical solid line with arrow pointing downwards). On the right the displaced particle will overcome the pressure gradient and return to its initial state. The figure here is motivated by the one used by \cite{Tritton_Davies_85BookChap} for explaining the same mechanism.}
 \label{fig_Tritton_mech}
\end{figure}
When the effects of the end walls are neglected, the base flow adopts a parabolic streamwise velocity profile $U = (1-y^2)$  \cite{Lezius_Johnston_76JFM}. Note that we use upper case to denote base flow. The transverse and wall-normal velocity components are zero. As is standard for rotating flows, a mean pressure gradient is sustained in the wall-normal direction $y$, balancing Coriolis forces, and may be obtained from equation (\ref{eq:ns_ndim}) as
\begin{align}
  &\partial_{y}P = -2 U Ro. \label{eq:base_y_comp}
\end{align}

Using figure \ref{fig_Tritton_mech}, we briefly recapitulate the manner in which instabilities are driven by this wall-normal pressure gradient as put forth elegantly by Tritton \& Davies \cite{Tritton_Davies_85BookChap}. Consider a fluid parcel that, by a velocity perturbation $v$, is displaced in the wall-normal direction by a distance $\zeta$ from a position where it had a velocity $U_{1}$. Owing to a streamwise Coriolis force $2vRo$, this parcel will experience a change in its streamwise velocity, attaining a value $U_{1}' \neq U_{1}$. Surrounding fluid at the new level has a velocity $U_{2}$. Now, the Coriolis force $2U_{1}'Ro$ on the displaced parcel is no longer balanced by the pressure gradient $-2U_{2}Ro$ (see equation (\ref{eq:base_y_comp})). If $U_{2} > U_{1}'$, the parcel is driven continuously out of equilibrium, as seen in the left part of figure \ref{fig_Tritton_mech}. When $U_{2} < U_{1}'$, the effect is stabilising, as depicted on the right of the figure. Comparing the difference in final 
and initial velocities of the displaced particle and the difference in velocities of undisturbed fluid particles at the two levels, we can obtain the inviscid instability criterion introduced earlier in the following (non-dimensionalised) form, 
\begin{align}
\phi(y) = 2Ro\left(-\frac{\partial U}{\partial y} + 2Ro\right) < 0. \label{eq:rayleigh_ndim}
\end{align}

Examining our system in this context, we see that for a given sense of rotation, one side of the channel is inviscidly stable and the other unstable. Since we have a base flow which is symmetric about the centreline, if the sense of rotation were to be reversed, we would merely have a switch in which side is unstable, and all results would merely be mirror images. We therefore fix our rotation to be anticyclonic, i.e., $Ro > 0$. However we caution that, for asymmetric shear flows, positive and negative $Ro$ would need to be studied separately \cite{Tritton_Davies_85BookChap,Sipp_etal_99PF,Yecko_Rossi_04PF}. 

\subsection{Linear perturbation} \label{ssec_sys_linp}
On introducing perturbations that are nominally small compared to the base state quantities, and by linearising the governing equations, we can study linear stability characteristics of the base flow. In the current setting where the channel extends infinitely in the streamwise and the spanwise directions, we can consider the disturbances to be periodic in these directions with a specific wavenumber $\boldsymbol{k}=(\alpha,\beta)$, with $\alpha$ the streamwise wavenumber and $\beta$ the spanwise wavenumber, so disturbances take on the form $f = \hat{f}(y,t)e^{\textrm{i}(\alpha x + \beta z)}$. In terms of the wall-normal velocity disturbance ($v = \hat{v}(y,t)e^{\textrm{i}(\alpha x + \beta z)}$) and the wall-normal vorticity disturbance($\eta = \hat{\eta}(y,t)e^{\textrm{i}(\alpha x + \beta z)}$), the resulting equation is
\begin{align}
\label{eq_pert_eqs}
 &\frac{\partial \hat{\boldsymbol{q}}}{\partial t} = \mathsfbi{L} \hat{\boldsymbol{q}}, \;\; \hat{\boldsymbol{q}}(t=0)=\hat{\boldsymbol{q}}_{_0}, \\
 \textrm{where  } \hat{\boldsymbol{q}} = \left[\begin{array}{c} \hat{v} \\ \hat{\eta} \end{array}\right], \; \textrm{and  } &\mathsfbi{L} = \begin{bmatrix} D^2 - k^2 & 0 \\ 0 & 1 \end{bmatrix} ^{-1} \begin{bmatrix} L_{os} & -2\textrm{i}Ro\beta \\ -\textrm{i}\beta U'+ 2\textrm{i}Ro\beta & L_{sq} \end{bmatrix}. \nonumber
\end{align}
Here $D(.) = \partial(.)/\partial y$, a prime denotes $d(.)/dy$, and $k^2 = \alpha^2 + \beta^2$. $L_{os}$ and $L_{sq}$ are the Orr--Sommerfeld and Squire operators given by
\begin{align}
 &L_{OS} = \textrm{i}\alpha U'' - \textrm{i}\alpha U(D^2 - k^2) + \frac{1}{Re}(D^2 - k^2)^2 \\
 &L_{SQ} = -\textrm{i}\alpha U + \frac{1}{Re}(D^2 - k^2).
\end{align}
The perturbation pressure can then be obtained as the solution of a Poisson equation. The boundary conditions for the above system of equations are
\begin{align}
 & \hat{v}(\pm1,t) = D\hat{v}(\pm1,t) = \hat{\eta}(\pm1,t) = 0. \label{eq:b_val}
\end{align}
For the rest of the article, we simply refer to wall-normal components of velocity and vorticity as normal components unless suggested otherwise. 

\subsection{Transient growth calculations} \label{ssec_sys_trans}
The linearised problem is addressed as an initial value problem with a view of finding the initial condition that maximizes an objective functional, i.e., in this case the disturbance kinetic energy \cite{Schmid_Henningson_01book}. The maximum possible gain at a given time $G(t)$ and its global maximum $G_{max}$ can be defined for a set of fixed values of parameters $Re$ and $Ro$ as follows,
\begin{align}
 &G(t;\alpha,\beta) = \sup_{{\hat{\boldsymbol{q}}_{_0}}} \frac{||\hat{\boldsymbol{q}}(t)||^{^2}_{_E}}{||\hat{\boldsymbol{q}}_{_0}||^{^2}_{_E}}, \; \; \textrm{and }\; G_{max} = \sup_{{t\geq0, \alpha, \beta}}\;G(t;\alpha,\beta). \; 
\end{align}
The disturbance kinetic energy norm that is to be maximzed is defined in terms of the normal velocity $v$ and normal vorticity $\eta$ as
\begin{align}
\label{eq:above}
||\hat{\boldsymbol{q}}(t)||^{^2}_{_E} = \frac{1}{2k^2} \int_{-1}^1 \hat{\boldsymbol{q}}^H(t) \begin{bmatrix} k^2 - D^2 & 0 \\ 0 & 1 \end{bmatrix}\hat{\boldsymbol{q}}(t)\;\mathrm{d}y.
\end{align}

For obtaining the optimal initial condition, the eigenvectors of the linearised operators can be used as a basis as they are complete in the present bounded geometry \cite{Prima_Habetler_69ARMA}. Upon inspecting the pseudospectra of the linearised operator, we find that a resolution of $N = 65$ Chebyshev collocation points in the normal direction is sufficient to form a complete basis of eigenvectors. Calculations performed with $N = 81$ or $121$ produce the same result up to at least $9$ decimal places. Defining $\boldsymbol{\Lambda}$ as the diagonal matrix consisting of eigenvalues of the operator $\mathsfbi{L}$, and $\mathsfbi{\tilde{q}}$ as the corresponding set of eigenvectors of $\mathsfbi{L}$, we may express solutions of  equation (\ref{eq_pert_eqs}) in variables separable form as 
\begin{align}
 \hat{\boldsymbol{q}}(y,t) = \mathsfbi{\tilde{q}}(y)\boldsymbol{k}(t) .
\end{align}

This allows us to deal with a computationally simpler problem as we now have $\boldsymbol{k}(t) = e^{\boldsymbol{\Lambda} t}\boldsymbol{k}(0)$. Eigenfunctions which decay extremely rapidly have no consequence to the evolution of the transient disturbance, and may be ignored. We choose a decay rate of $-3$ as the cut-off, and find disturbances that are linear combinations of eigenfunctions with slower decay rate. The computations were performed in MATLAB. Our code uses a differentiation suite for the Chebyshev grid developed by Weideman \& Reddy \cite{Weideman_Reddy_00ACM}. The objective functional was maximised by using the MATLAB generic nonlinear constrained optimisation package fmincon. The code has been validated by confirming the eigenspectra for the non-rotating channel flow at $Re=2000$ up to 8 significant digits (Appendix A.7 in \cite{Schmid_Henningson_01book}).

\section{Non-modal stability characteristics}
If the spectrum of the linearised operator $\mathsfbi{L}$ in equation (\ref{eq_pert_eqs}) contains an eigenvalue with a positive real part, then there is an exponentially growing mode that causes the base flow to transition to another state. If there is no such growing eigenmode, we may conclude that the flow is asymptotically stable. A flow which is asymptotically stable need not be stable in the sense of energy of the disturbance \cite{Joseph_76book}. If the energy of any small disturbance decays monotonically for all time, then the flow is considered to be stable from an energy point of view. A flow which is asymptotically stable but not energy-stable may display a transiently growing feature, which, but for nonlinear effects, will eventually decay. This growth of disturbances can sometimes be sufficient to trigger nonlinearities, in which case the flow need not return to the initial base state. We denote the critical Reynolds number, below which no disturbance mode grows exponentially, as $Re_{crM}$ (
where $M$ stands for modal), and the energy critical Reynolds number as $Re_{crE}$, above which $G_{max}$ first exceeds $1$. It is clear that $Re_{crE} \le Re_{crM}$. In shear flows typically there is a stark difference in the critical Reynolds numbers by the two measures. This is due to the fact the linearised operator $\mathsfbi{L}$ is non-normal. 

\begin{figure}
 \centering
 \includegraphics[width=5.75in]{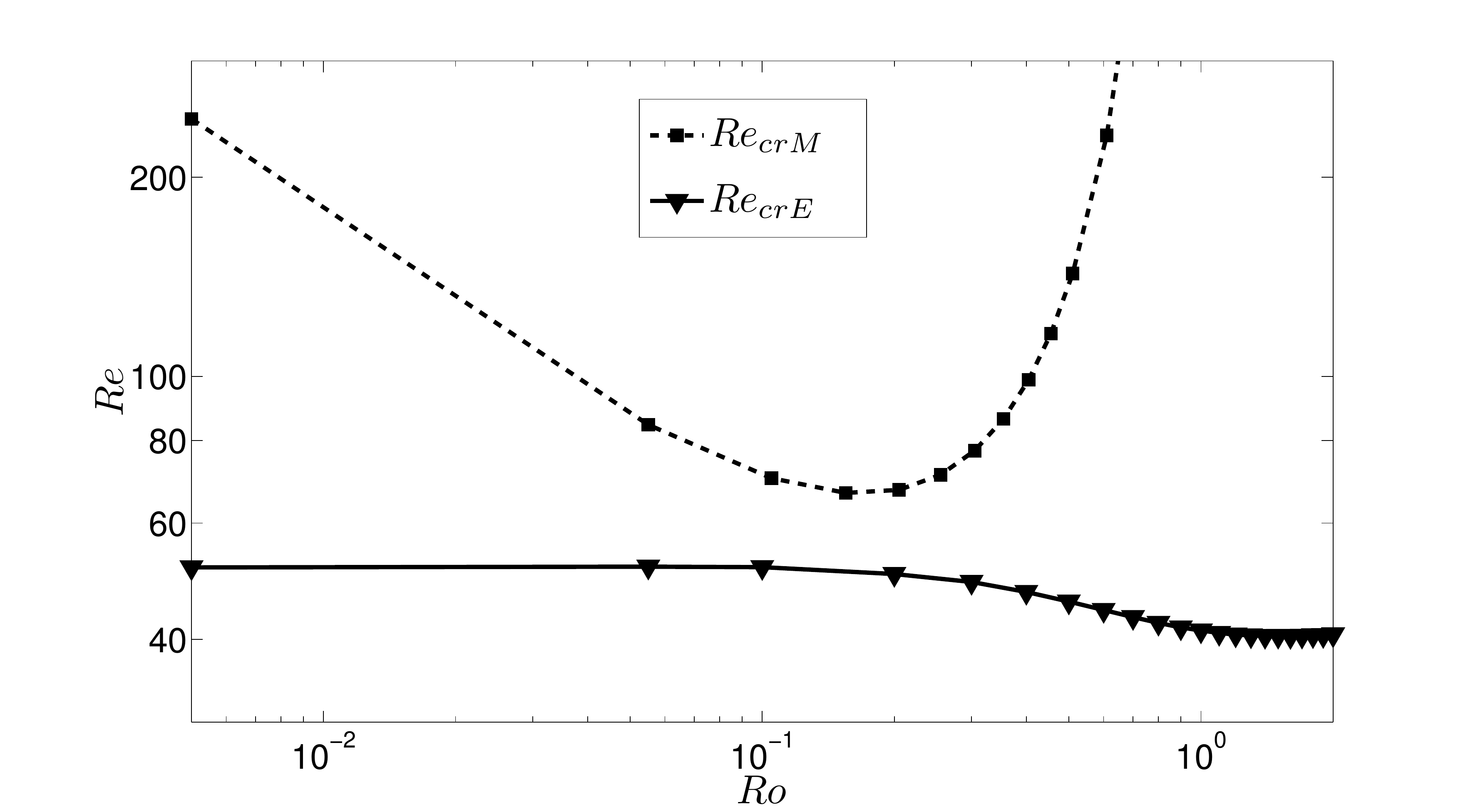}
 \caption{The stability boundaries as per the energy and the eigenvalue methods. It is seen that the critical Reynolds number $Re_{crM}$ obtained by modal analysis is highly sensitive on the rotation rate. By energy considerations, we see that $Re_{crE}$ is far less sensitive to changes in the rotation rate. It changes from $51.43$ to $41.16$ over 3 decades of magnitude of $Ro$.}
 \label{fig_ener_asym}
\end{figure}

In the rotating channel, we have discussed that $Re_{crM}$ is sensitive to the rotation rate. What about $Re_{crE}$? We begin the presentation of our results by plotting this quantity in Figure \ref{fig_ener_asym} for different rotation rates. The neutral stability boundary, defined by $Re_{crM}$, is shown for comparison. In configurations between the two curves, disturbances may grow in energy for some time by a linear mechanism. It is seen that, using the energy method, the critical Reynolds numbers obtained for low rotation rates are close to 49.60. This is in agreement with the value obtained by Joseph \& Carmi \cite{Joseph_Carmi_69QAM} for the nonrotating channel flow (also see \cite{Joseph_76book}). 

\begin{figure}
\centering
  \subfloat[$Ro = 0.001$]{
    \includegraphics[width=3.125in]{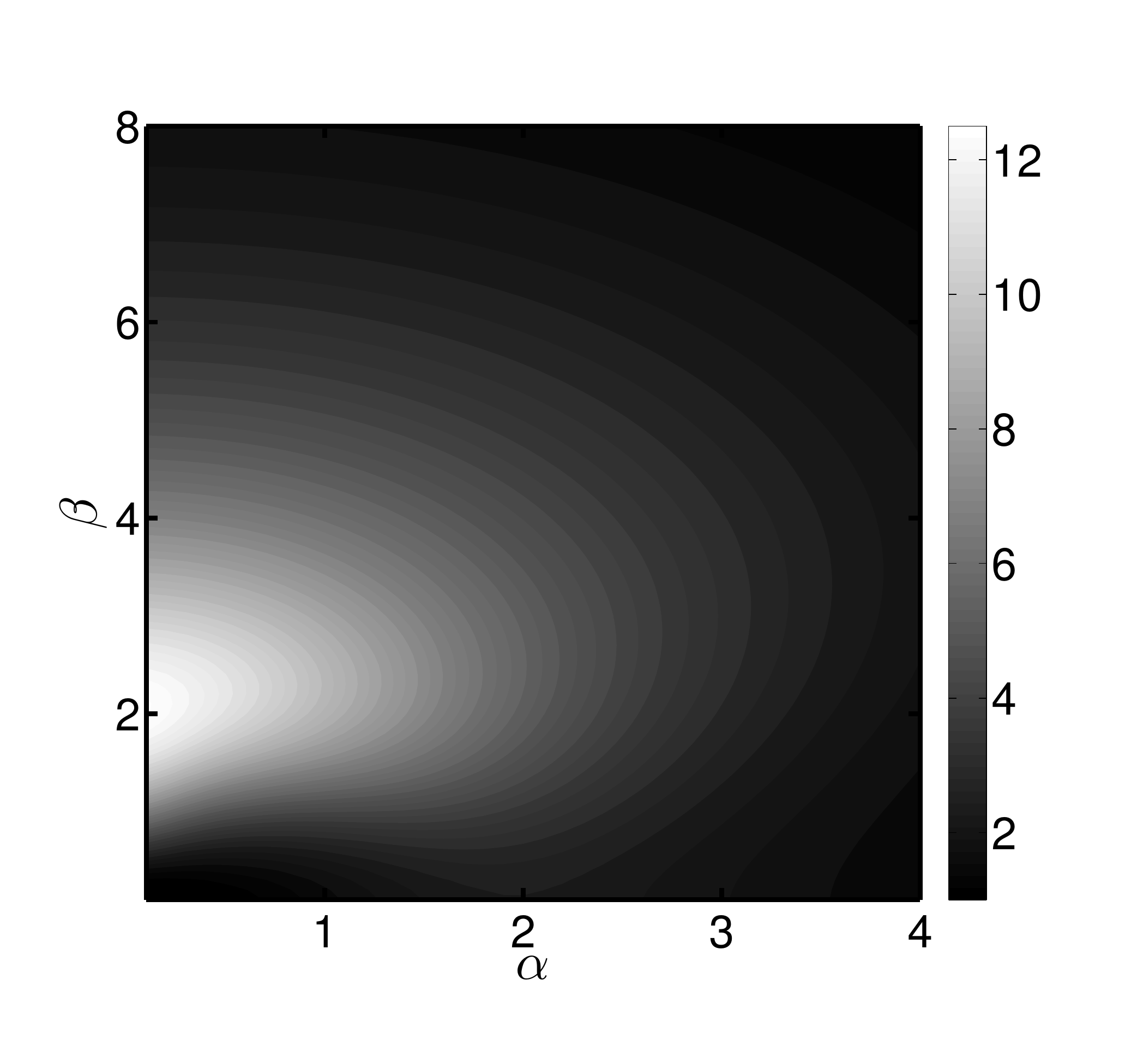}}
    \hspace{0.05in}
  \subfloat[$Ro = 0.8$]{
    \includegraphics[width=3.125in]{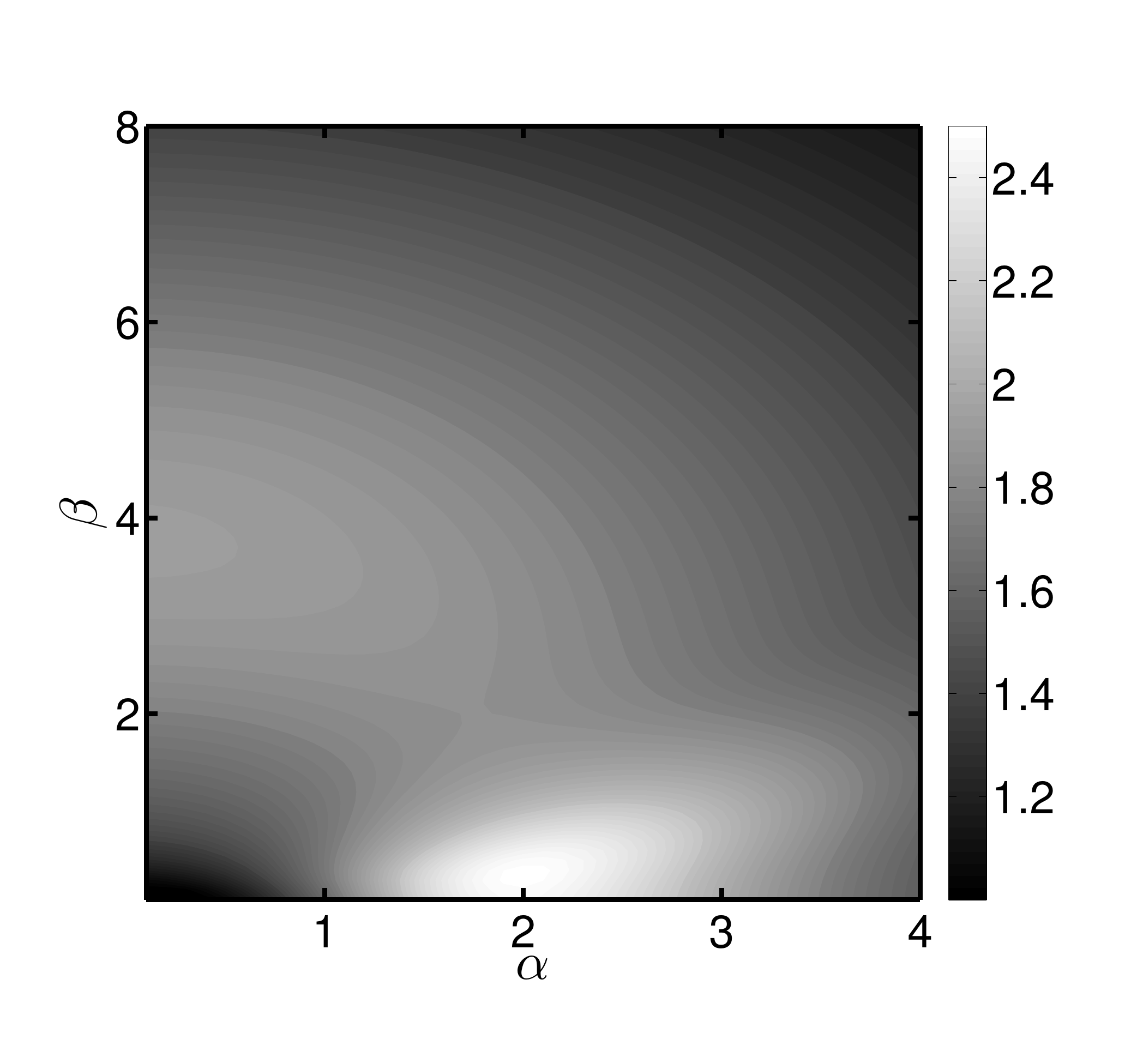}}
  \caption{The $G_{max}$ contours for $Re = 250$ for different $Ro$. The optimal parameters are (a) $T_{opt} = 18.72$, $\alpha_{opt} = 0.0$, $\beta_{opt} = 2.05$, $G_{max} = 12.64$; (b) $T_{opt} = 4.85$, $\alpha_{opt} = 2.00$, $\beta_{opt} = 0.36$, $G_{max} = 2.52$}
\label{fig_cont_g}
\end{figure}
We next study the transient growth characteristics in different regions of the $Re$-$Ro$ parameter space where exponential instabilities are absent. In figure \ref{fig_cont_g} we demonstrate by a typical example that transient growth is qualitatively different to the left and the right of the neutral stability boundary. For a fixed $Re$ ($=250$), the figure shows contours of $G_{max}$ for representative low and high rotation rates. The $Ro = 0.001$ case is not markedly different from the corresponding results for a stationary channel at this Reynolds number. There too, the disturbances that yield the largest growths are streamwise-invariant \cite{Reddy_etal_93JFM}. They form rolls that evolve into streaks as a consequence of the vortex tilting lift-up mechanism \cite{Landahl_80JFM}. On the other hand, at $Ro = 0.8$, the optimal disturbances are almost aligned along streamwise direction with a very small dependence on the spanwise coordinate. This suggests that the Orr mechanism is likely to be the more 
dominant energy amplification mechanism \cite{Orr_1907PRIAA}. It may be noticed that the maximum achievable algebraic growth is lower when compared to the situation at low $Ro$. 

\begin{figure}
 \centering
 \includegraphics[width=5.75in]{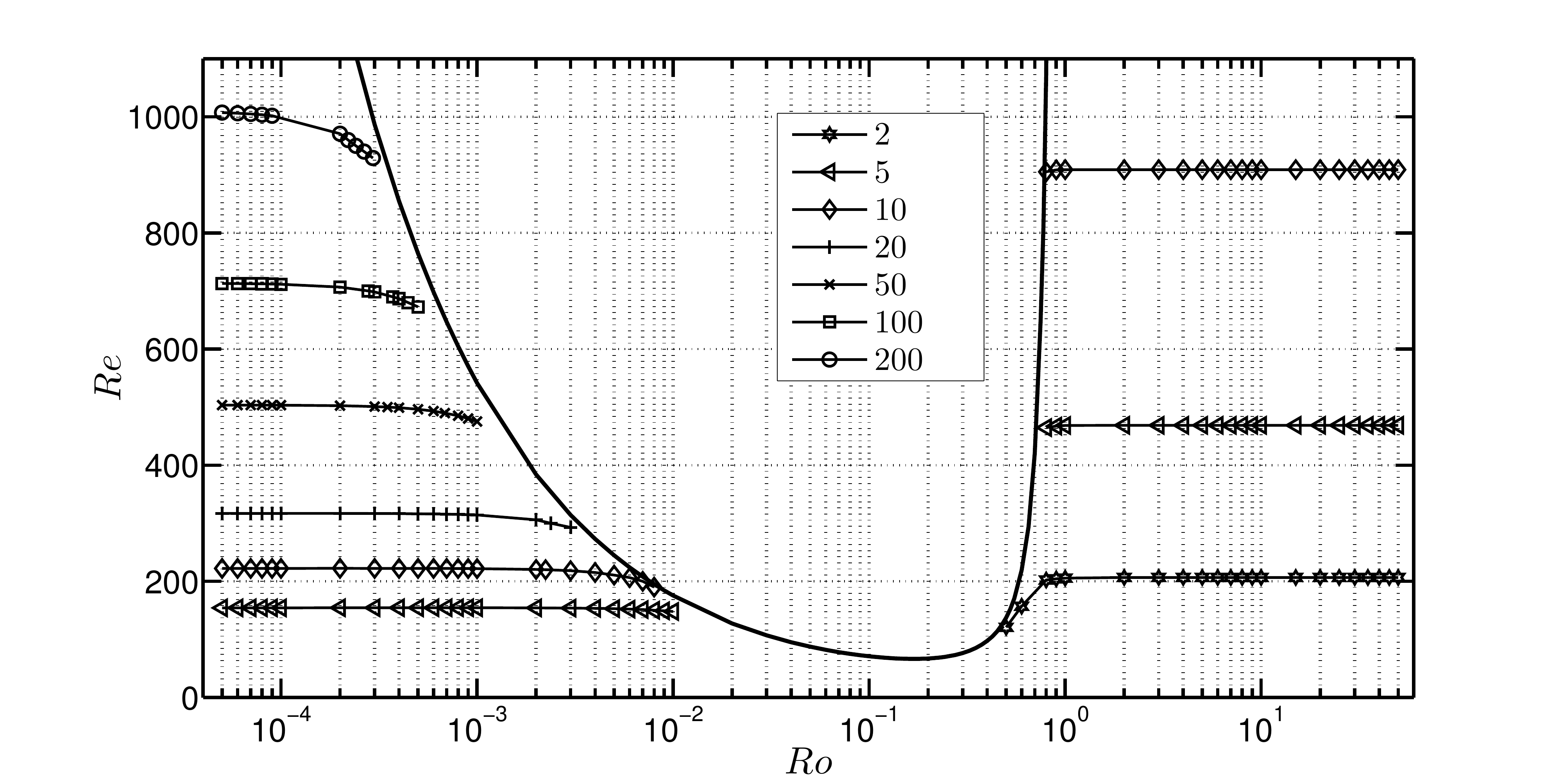}
 \caption{Level curves of $G_{max}$ are given. For low $Ro$, the behaviour resembles that of the optimal disturbances of the non-rotating channel. A large drop in the amplification levels is also seen at high $Ro$.}
 \label{fig_main_res}
\end{figure}
In figure \ref{fig_main_res} we show level curves of $G_{max}$ in the $Re$-$Ro$ plane (outside the linearly unstable region). Two findings are immediately apparent. As in the example above, transient growth levels are much smaller everywhere on the right of the neutral boundary as compared to a corresponding Reynolds number on the left. Secondly, on a given side of the neutral boundary, $G_{max}$ depends primarily on $Re$ and is rather insensitive to changes in $Ro$. We can gain insight into these observations by examining the different sources of non-normality in the linearised equations (equation (\ref{eq_pert_eqs})). The strength of the rotation then determines which of the sources of non-normality will emerge stronger.

The major source of non-normality is the forcing due to the normal velocity $\hat{v}$ in the normal vorticity equation, and it is this term that is responsible for the lift-up mechanism. An examination of the structure of the stability operator makes it evident that, at a given $\beta$, the departure from normality due to these operators decreases as the rotation rate increases. As in the non-rotating case, the largest amplifications are seen for disturbances that are streamwise-independent at low rotation rates. For low rotation rates, the terms involving $Ro$ serve to act as small corrections to the linearised operator. Hence the growth in disturbance energies is similar to the growth seen in the non-rotating case. This translates to the lack of the dependence on $Ro$ of $G_{max}$ in the low rotation regime in figure \ref{fig_main_res}.

The other source of non-normality in the linearised equations is that the Orr--Sommerfeld operator $L_{OS}$ itself is not self-adjoint. This gives rise to a much weaker transient growth in two-dimensions, and is completely independent of the rotation rate. As the rotation rate is increased, consistent with Taylor--Proudman arguments, the motion of the fluid is restricted to the plane perpendicular to the rotation axis. Thus, while disturbances favouring the lift-up mechanism are suppressed strongly, the disturbances amplified by Orr mechanism can still be excited at higher rotation rates. Disturbances initially having spanwise variation rapidly evolve to become two-dimensional with no flow along the axis of rotation. Thus the optimal disturbances in this regime evolve transiently only due to the Orr mechanism; i.e. $\beta_{opt} = 0$. Evidence of the Orr mechanism leading to the largest amplifications can be seen in figure \ref{fig_main_res} at high $Ro$, where the level curves become horizontal and thus 
displaying insensitivity to the rotation rate.

\begin{figure}
 \centering
 \centerline{\includegraphics[width=3.125in]{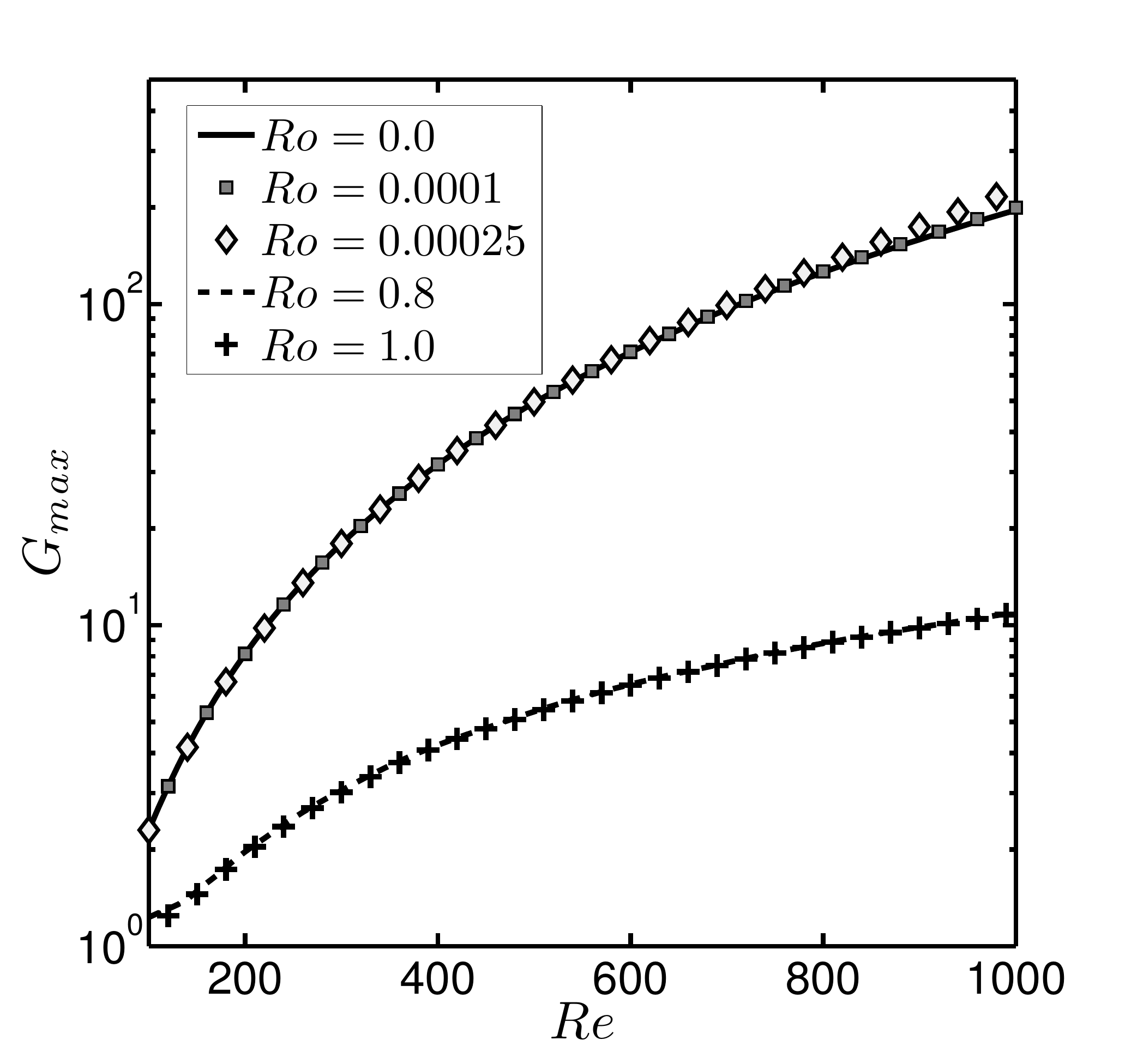}}
 \caption{A contrast between maximum transient energy growth at high and low rotation numbers. On a given side of the neutral boundary of figure \ref{fig_main_res}, there is virtually no difference in the $G_{max}$ due to $Ro$, whereas there is an order of magnitude of difference in the values of $G_{max}$ between low and high $Ro$.}
 \label{fig_Re_var}
\end{figure}

In figure \ref{fig_Re_var},  we have plotted the $G_{max}$ for specific values of $Ro$ in different rotation regimes as a function of $Re$. The values obtained at higher $Ro$ are shown to be at times an order of magnitude lower than for a small $Ro$ for a given $Re$. It is seen here  as well that $G_{max}$ does not vary too much as the $Ro$ is varied in different rotation regimes for large ranges of $Re$. At low rotation rates, as long as $Re$ is not sufficiently close to the critical value at given $Ro$, the energy amplification obtained is found to obey the scaling laws due to Gustavsson \cite{Gustavsson_91JFM} as the different curves coincide. It can be seen that for $Ro = 2.5\times10^{-4}$, as we increase $Re$, deviations from the $Ro = 0$ curve start to appear. This is a result of the values of $Re$ approaching the neutral boundary. Thus it would be of interest to examine the regions close to the stability boundary in more detail.

In the high rotation rate modally stable regime, very close to the stability boundary, it can be seen in figure \ref{fig_main_res} that the level curves rise very slightly with $Ro$. This implies that the value of $Re$ which yields a given energy growth increases with the rotation rate. It is important to note that if $\beta$ were to be identically zero, there would be no effect of the rotation on transient growth, since $Ro$ would completely drop out of equation (\ref{eq_pert_eqs}) ($Ro$ appears in the linearised equation only in the form $\beta Ro$). The independence of the Orr mechanism on rotation was also demonstrated for the rotating Couette flow in the thesis of Daly \cite{Daly_04thesis}. Thus the level curves not being perfectly horizontal near the stability boundary implies an oblique optimal structure; i.e. $\beta_{opt} \neq 0$. As we move to a region in $Re$-$Ro$ parameter space further away from the stability boundary, the optimal disturbance corresponds to the disturbance best amplified by the 
Orr mechanism at a given $Re$ as the flow exhibits behaviour consistent with Taylor--Proudman theory. 

\begin{figure}
\centering
  \subfloat[$G_{max}$ vs $Ro$]{
    \includegraphics[width=3.125in]{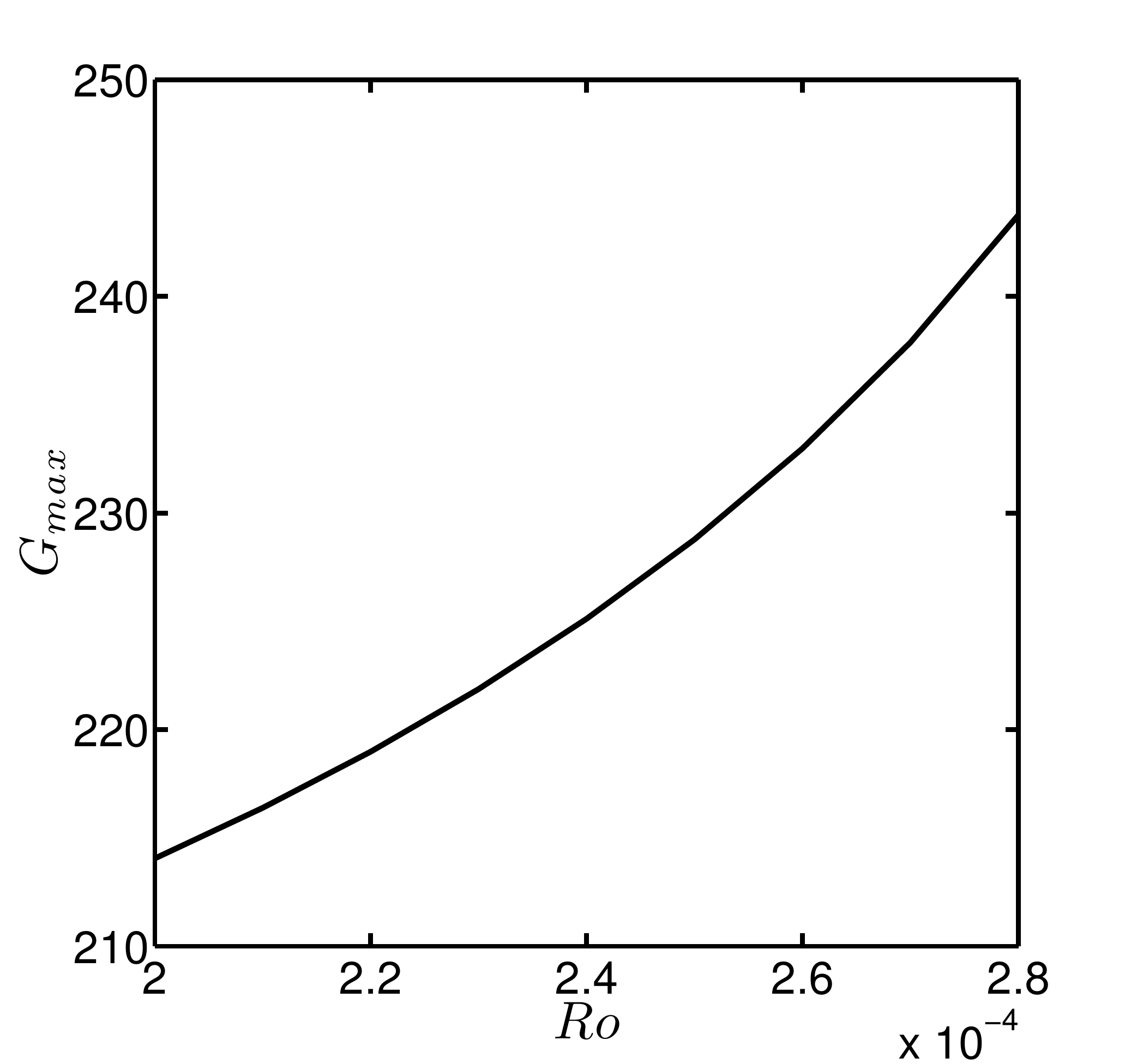}}
    \hspace{0.025in}
  \subfloat[$T_{opt}$ vs $Ro$]{
    \includegraphics[width=3.125in]{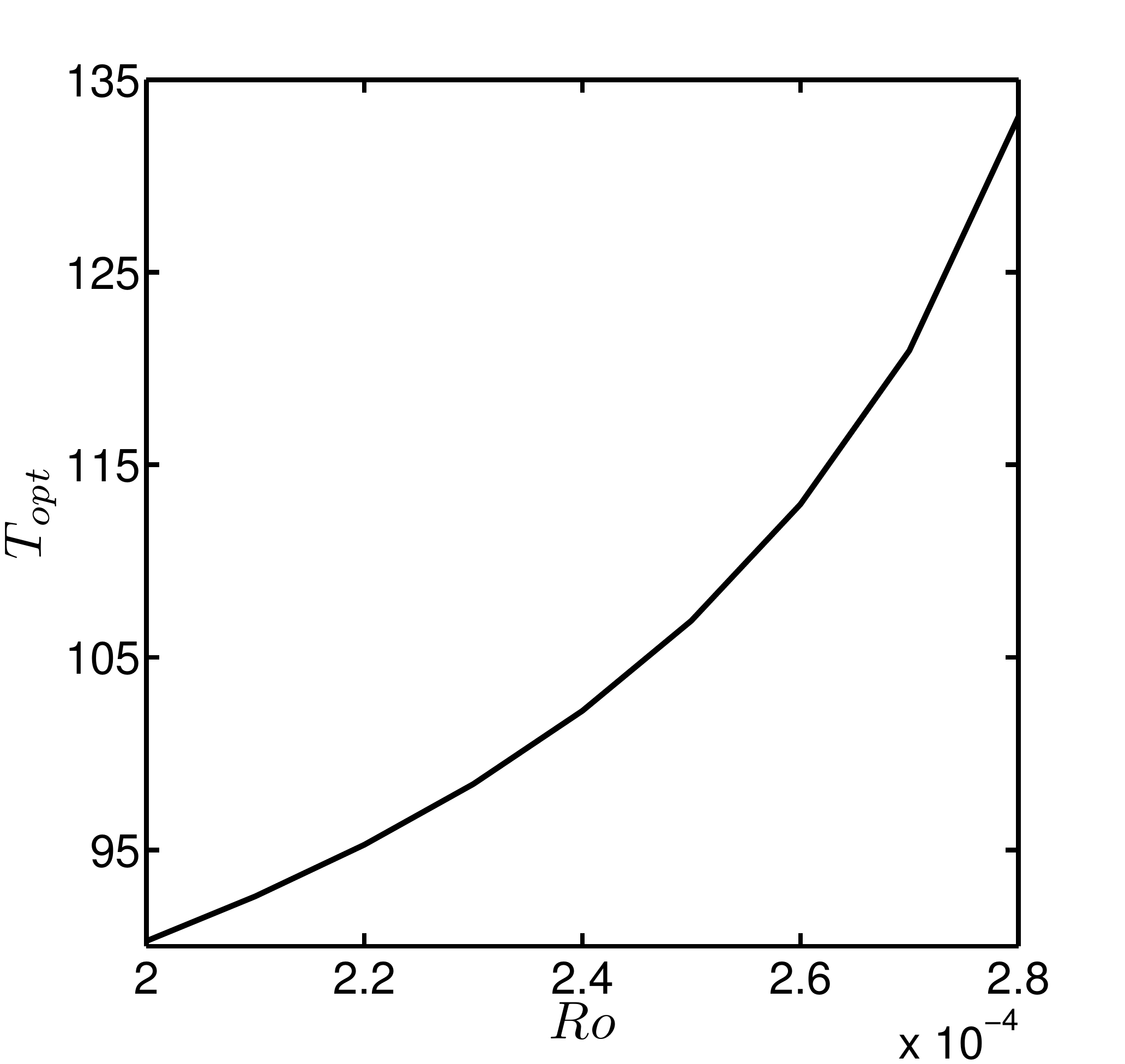}}
  \caption{Optimal growth behaviour as we approach the neutral stability boundary on the low $Ro$ side. On the left the variation of $G_{max}$ is shown as a function of $Ro$ at $Re = 1000$. On the right is shown the corresponding time at which this $G_{max}$ was attained. The disturbance grows transiently to slightly higher levels and grows for a longer duration as we approach the exponentially unstable regime. This can be attributed to the slower decay of the modes comprising the disturbance.}
\label{fig_Re1000_Roran}
\end{figure}

As we approach the neutral stability curve from the low $Ro$ side, we see in figure \ref{fig_main_res} that the level curves of $G_{max}$ noticeably dip towards a lower $Re$. This means that for a fixed $Re$, we have an increase in $G_{max}$ as we approach the stability boundary. The typical behaviour of optimal growth with $Ro$ (for $Re = 1000$) is shown in figure \ref{fig_Re1000_Roran}. The corresponding time at which this optimal growth is attained is also shown. The first modes that go linearly unstable are streamwise independent. The least stable modes have smaller decay rates as we approach the neutral boundary from the left, and hence the time before the modes individually decay is slightly longer. This allows for the lift up effect to persist for a slightly longer time. 

\begin{figure}
\centering
  \subfloat[$Ro = 0.0002$]{
    \includegraphics[width=3.4in]{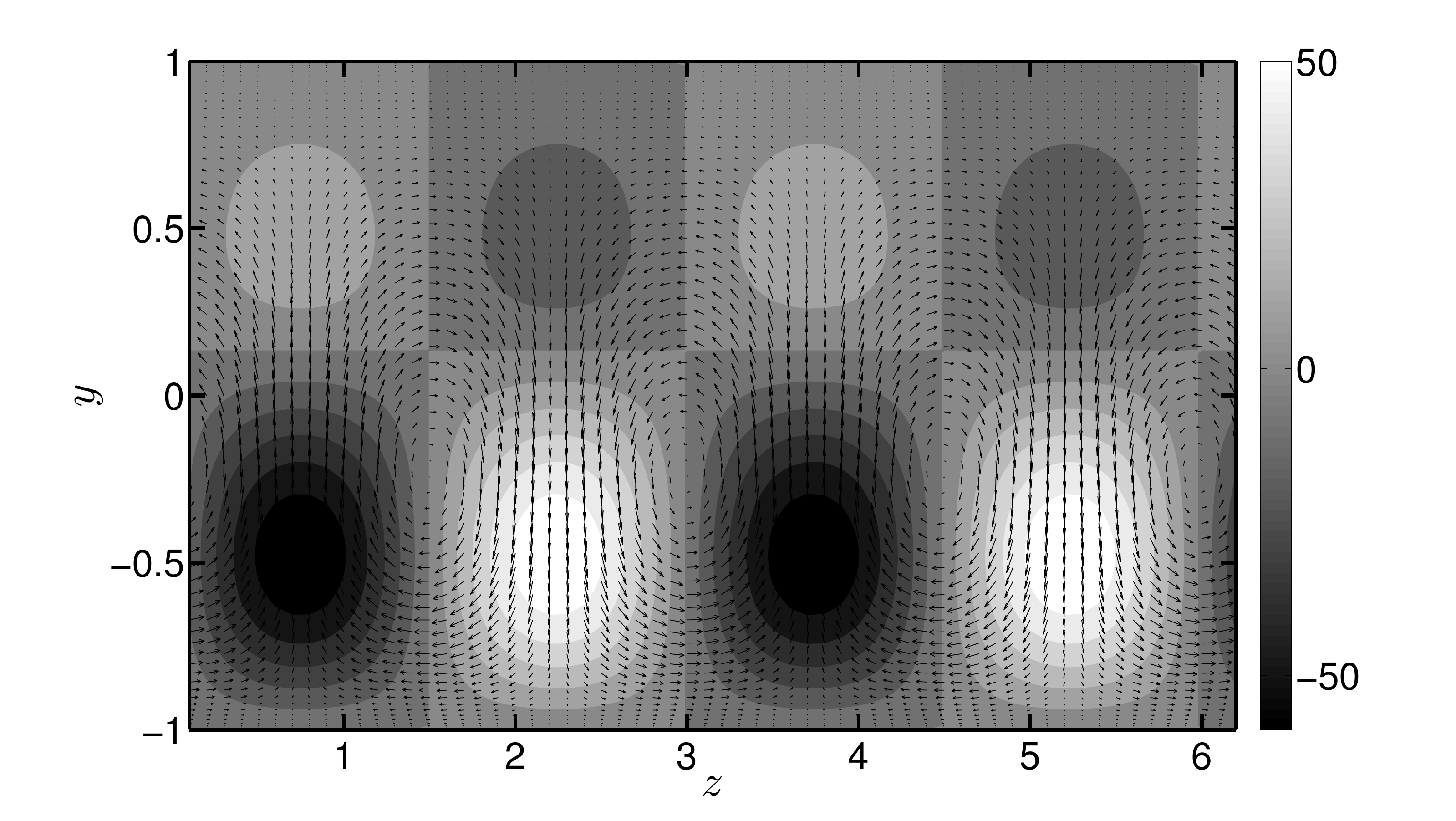}}
    \hspace{0.005in}
  \subfloat[$Ro = 0$]{
    \includegraphics[width=3.4in]{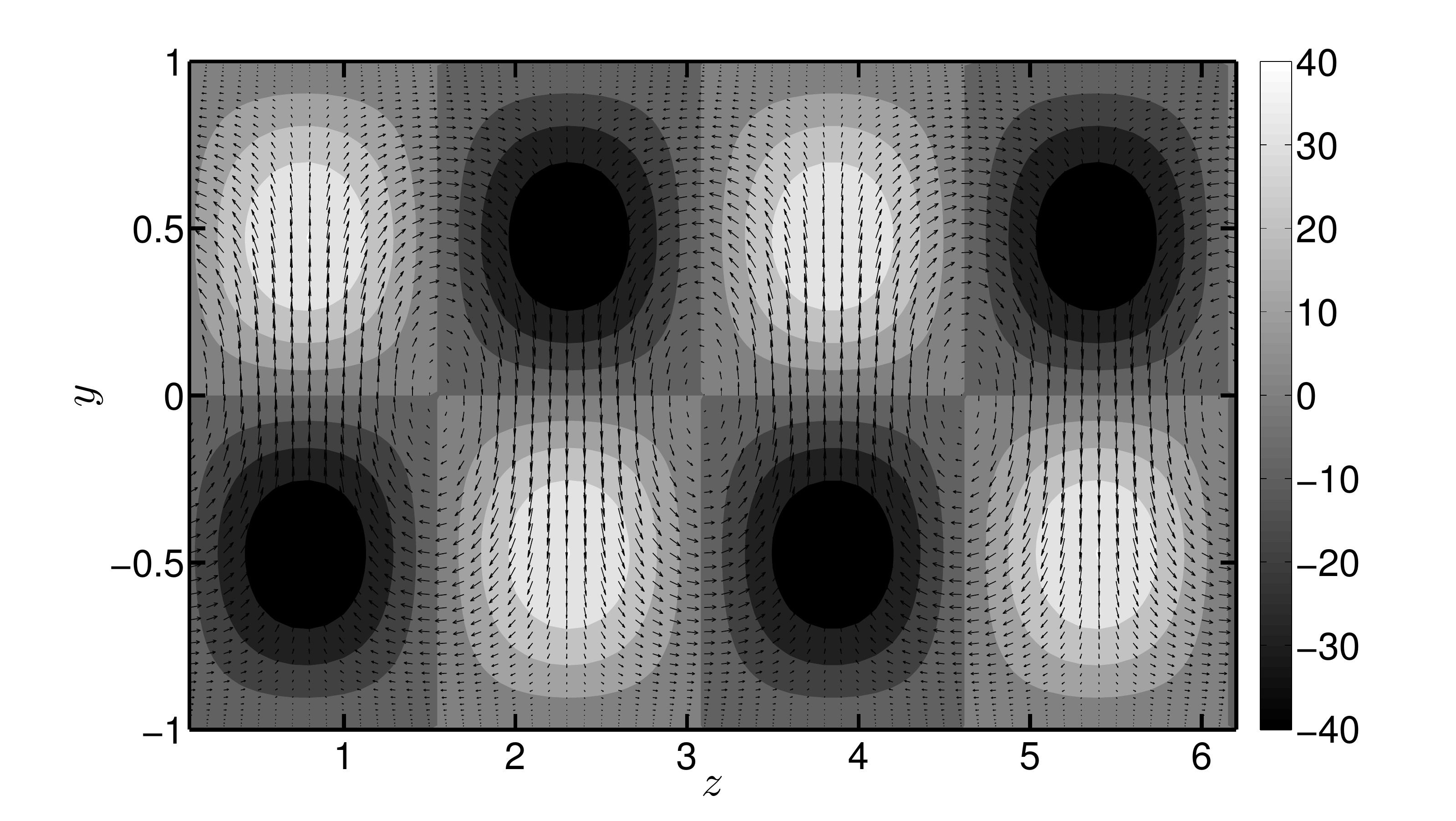}}
  \caption{The optimal structures for $Re = 1000$ for low $Ro$ (a), and for the non-rotating case (b), at the optimal times. The contours depict the magnitude of the streamwise velocity component. The velocity field $(v,w)$ are expressed through the vectors. The optimal parameters are (a) $T_{opt} = 90.27$, $\alpha_{opt} = 0$, $\beta_{opt} = 2.10$, $G_{max} = 214.05$; (b) $T_{opt} = 75.68$, $\alpha_{opt} = 0$, $\beta_{opt} = 2.04$, $G_{max} = 196.17$.}
\label{fig_opt_struct}
\end{figure}

As mentioned earlier, the optimal structures obtained at low rotation rates are streamwise rolls that develop into streaks. This is similar to the non-rotating case in the sense that streaks are formed. However due to the additional Coriolis force, the streaks are not symmetric about the centreline, with one side of the channel displaying a stronger streak than the other. This feature is more pronounced close to the stability boundary. Figure \ref{fig_opt_struct} shows the optimal structure for a typical low rotation ($Ro = 0.0002$) and compares this to the non-rotating case. The velocities in the two cases are comparable. 

\begin{figure}
 \centering
 \includegraphics[width=4.5in]{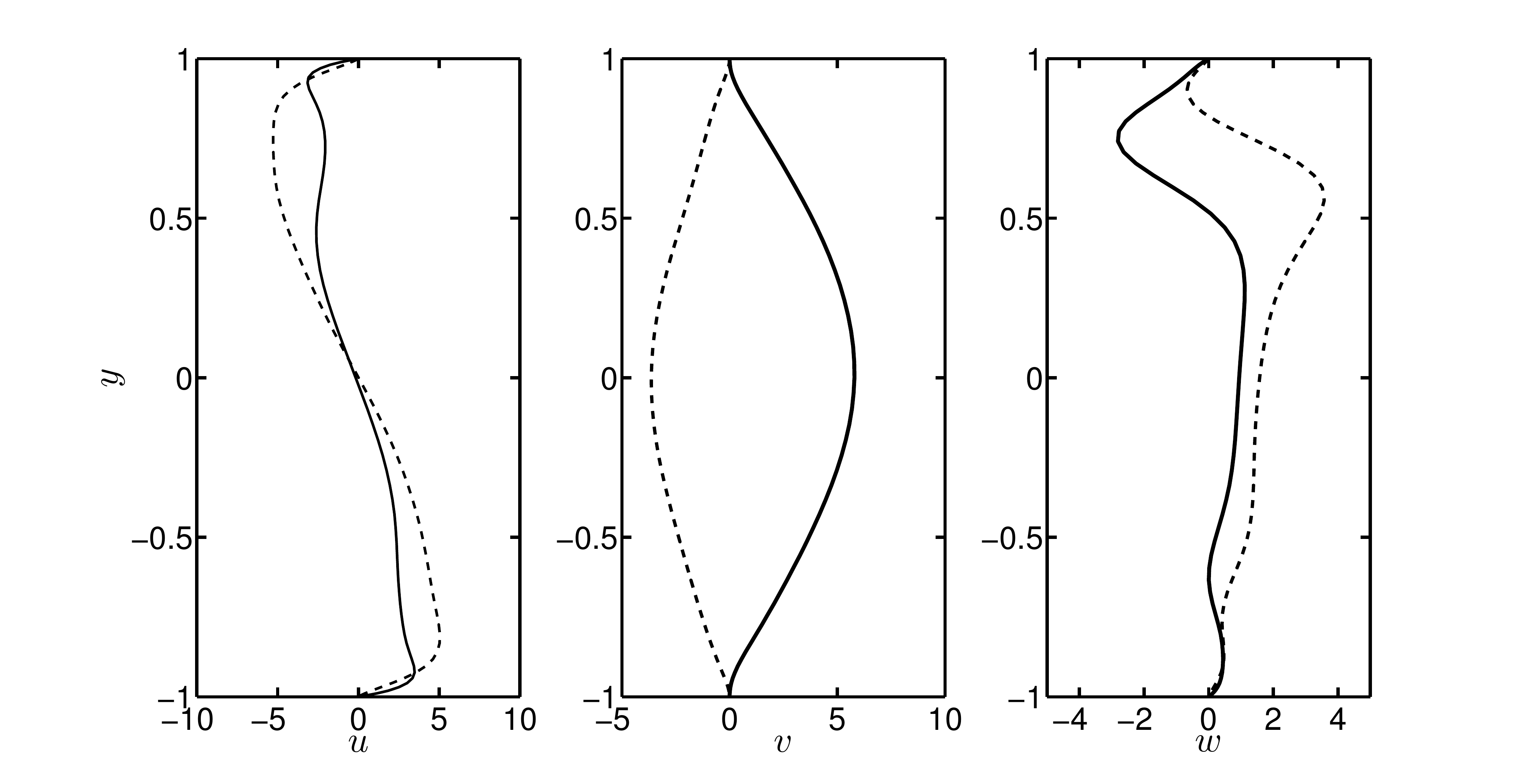}
 \caption{The velocity components of the optimal disturbance at the optimal time when $Ro = 0.8$ and $Re = 1000$. The real and the imaginary parts are given by the solid line and the dashed line respectively. The optimal parameters are: $T_{opt} = 8.52$, $\alpha_{opt} = 1.75$, $\beta_{opt} = 0.169$, $G_{max} = 11.03$.}
 \label{fig_opt_struct_highRo_norm}
\end{figure}

The optimal disturbance is given in figure \ref{fig_opt_struct_highRo_norm} for $Re = 1000$ and $Ro = 0.8$. We can see that the rotation does not bias the occurrence of a secondary disturbance velocity towards any particular wall despite the high rotation rates. The strong rotation does little to alter the features of a flow that is largely confined to the plane normal to the rotational axis i.e., the $x$-$y$ plane. Hence disturbances may evolve by the Orr mechanism unhindered by the rotation. In the following section we examine how these observations relate to nonlinear evolution of perturbations.

\section{Nonlinear Simulations}
In cases where a modal perturbation grows exponentially, or where transient growth is large, a nonlinear study is imperative to understand the next stage of evolution. We carry this out by direct numerical simulations of the three-dimensional Navier--Stokes equations in this flow, to characterise the transition to a new (steady or unsteady) state of the channel flow at different rotation rates. As discussed above, the rotational channel flow is a well studied problem from both numerical and experimental points of view. In these studies, typically transitions away from the parabolic profile are achieved by the introduction of noise at a sufficiently high level such that instabilities are triggered, and the flow is allowed to evolve nonlinearly \cite{Yang_Kim_91PFA,Yanase_Kaga_04JPSJ}. These studies were interested in the linearly unstable regime. Our primary focus on the other hand will be regimes in $Re$-$Ro$ space which show stability in terms of modal growth, and we shall contrast this dynamics with 
behaviour within the linearly unstable regime. To make the discussion simpler we present results at a Reynolds number of $1500$, deeming them to be typical of the simulations we have carried out at other Reynolds numbers. 

\subsection{Methodology}
The simulations were performed using the SIMSON code developed in KTH Mechanics, Stockholm \cite{Chevalier_etal_07Tech}. A pseudospectral method is employed with Fourier expansions in the streamwise and spanwise directions, and a Chebyshev discretization is employed in the normal direction. For the results to follow, the horizontal directions are discretized using $64$ Fourier modes each, and $81$ Chebyshev polynomials are used for discretizing the normal coordinate. A second-order Crank--Nicolson scheme was used to discretize the linear terms, and the nonlinear terms were discretized by a four-stage Runge--Kutta (RK3) scheme. Periodic boundary conditions are used in the streamwise and spanwise directions and at the walls, no slip and no penetration are imposed. 

\begin{figure}
 \centering
 \includegraphics[width=3.125in]{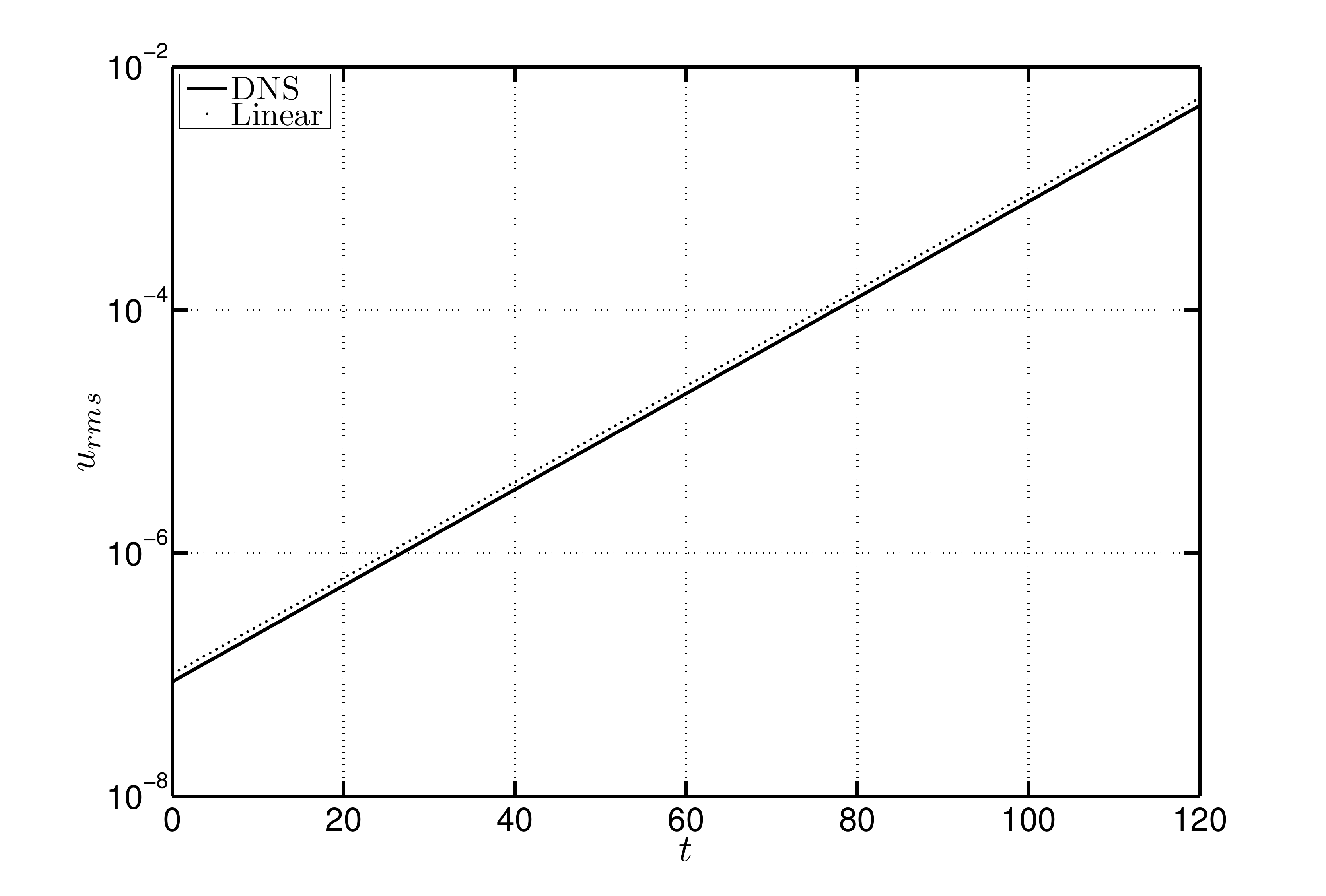}
 \caption{Validation of the nonlinear code was done by introducing eigenmodes of the linearised operator as a perturbation at an amplitude of $10^{-6}$ times the centreline base velocity. The perturbation is shown to grow with the predicted linear growth rate. Here $Re = 1000$, $Ro = 0.02$, and the perturbation wavenumber chosen is $\mathbf{k} = (\alpha,\beta) = (0,2)$.}
 \label{fig_val}
\end{figure}

We use two types of initial conditions. In one, we impose the least stable eigenmode as a perturbation, with a velocity amplitude of $10^{-6}$ times the centreline base velocity. The code was validated by imposing the least stable eigenmode as the initial condition, and comparisons were made with linear stability predictions. A sample is shown in figure \ref{fig_val}. As the other initial condition, we impose a relevant optimal perturbation. The initial kinetic energy has an amplitude of $25 \times 10^{-6}$ per box of size one wavelength of the disturbance in both streamwise and spanwise directions. In case of a zero wavenumber in either direction, an arbitrary length is fixed in that direction to define the box. The initial amplitude chosen is the lowest that leads to transition in the non-rotating case, and this is in agreement with the threshold values obtained in previous studies \cite{Reddy_etal_98JFM}. This second initial condition is particularly important in the regime of transient growth. At 
different Reynolds numbers, transient growth at $Ro=10^{-7}$ was checked to produce identical structures, and the same energy growth rates, as results available for a non-rotating channel.

Particular care must be taken when the initial perturbation is streamwise independent, since these perturbations, through the nonlinear term in the governing equation, would act to excite only higher harmonics of the initial spanwise wavenumber while remaining independent of the streamwise coordinate. In order to excite other streamwise spatial frequencies we also introduce noise at a very low level at the start of the simulation \cite{Reddy_etal_98JFM,Schmid_Henningson_01book}. The noise is introduced in the form of Stokes modes for a few non-zero streamwise wavenumbers. The total energy content of the noise is prescribed to be half a percent of that of the optimal perturbation, which ensures that noise is not the dominant factor in the dynamics and serves to trigger secondary instabilities. We have verified that the flow does not undergo transition to a new state when noise alone is introduced. The noise is added primarily to excite secondary instabilities. 

\begin{figure}
 \centering
 \includegraphics[width=4.5in]{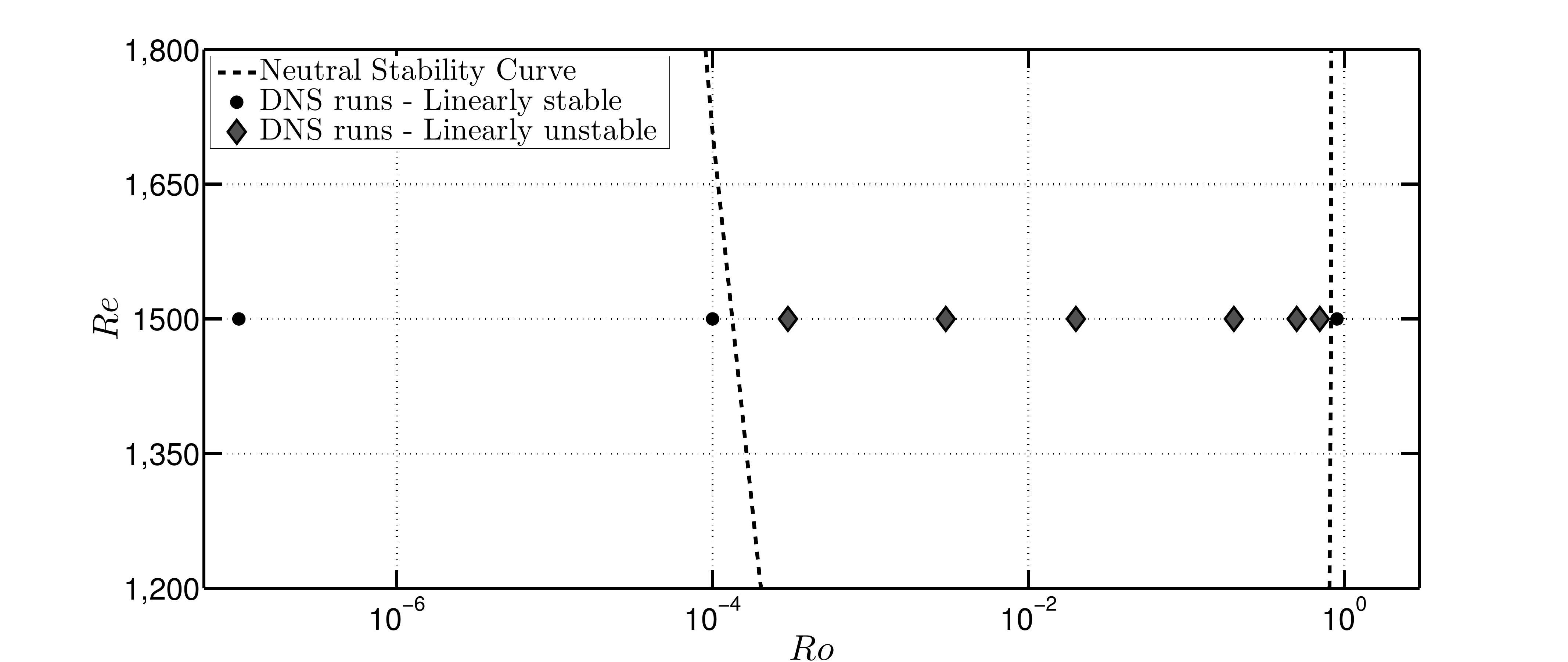}
 \caption{The nonlinear simulations presented here are for the parameter values denoted by the symbols. The rotation numbers are varied from the linear stable region on the left, through the region within the neutral boundary where the flow is linearly unstable, to the stable region on the right, while $Re$ is fixed at 1500.}
 \label{fig_DNS_cases}
\end{figure}

For all the results here we fix $Re = 1500$, a value where subcritical transition was previously observed in the non-rotating case \cite{Reddy_etal_98JFM}. To study transient growth, the initial condition chosen is the optimal disturbance with wavenumber vector $\boldsymbol{k}=(\alpha,\beta) = (0,2)$ for $Ro = 10^{-4}$. We thus have the same computational box for all simulations. The computational region measures two wavelengths of the perturbation in the spanwise coordinate. The optimal spanwise wavenumber and optimal structure across $Ro$ on the left of the neutral boundary are very close to those at $Ro=10^{-4}$. Within the regime of linear instability, we continue to use these initial conditions. In the next subsection, we describe simulations with the least stable eigenmode as the initial condition. On the right of the stability boundary, all initial perturbations, whether the optimal (Orr-like) structures or lift-up structures, decay within a short time for this Reynolds number, and the parabolic 
velocity profile is recovered.

The rotation rate is varied across some decades of $Ro$ to clearly elucidate what happens in different regimes. In figure \ref{fig_DNS_cases} we show the different cases considered and where they lie in the $Re$--$Ro$ space with respect to the neutral stability boundary. The lowest $Ro$ we choose is $1\times 10^{-7}$, and this practically corresponds to the non-rotating case. On the low rotation side we also study $Ro = 1\times10^{-4}$, which is just outside the stability boundary. Within the region where exponential instabilities may occur, we choose several points. The highest $Ro$ we study is $0.9$ which does not yield exponentially growing disturbances at $Re = 1500$, i.e., it lies to the right of the stability boundary. For ease of discussion we shall refer to cases with $Ro \leq 0.003$ as low rotation, and to higher $Ro$ as high rotation cases. 

To get a measure of whether the flow is chaotic, we define an entropy $Q$ as follows.
\begin{align}
 Q = \frac{1}{2x_{l}z_{l}}\int_{box} \left[\omega_{x}(x,y,z,\tau + \Delta t) - \omega_{x}(x,y,z,\tau)\right]^2 \;\mathrm{d}\boldsymbol{x}, \label{eq_entropy2}
\end{align}
where $x_{l}$ and $z_{l}$ are the streamwise and spanwise lengths of the box, and $2x_{l}z_{l}$ is the volume of the periodic box. Streaks that have evolved from streamwise vortices have been seen as precursors to the transition process in several shear flows \cite{Elofsson_etal_99PF,Klingmann_92JFM}. Hence we define $Q$ based on the streamwise vorticity as this gives us a picture of the fluid motion in the $y$-$z$. A similar approach had been employed by previously to quantify chaotic motion due to a flow past an inline oscillating cylinder \cite{Srikanth_etal_11PF}. Other measures will give qualitatively the same results. We choose the reference time $\tau$ to be later than the time at which the initial transient behaviour dies down. At $\Delta t=0$ we have $Q=0$. In a strictly periodic flow with period $T$, $Q$ will return to zero when $\Delta t=nT$, where $n$ is any positive integer. 

As we are dealing with cases within the linearly unstable regime, we have to examine if the unstable mode does play a role in the dynamics. Is it that the algebraically growing mode only serves as a noisy environment from which the unstable mode gets excited? Now for a given rotation rate, we shall denote the unstable eigenmode as $\hat{\boldsymbol{q}}_u$. We shall analyse the evolution of the initial perturbation for different rotation rates in the linearised setting governed by equation (\ref{eq_pert_eqs}). $\hat{\boldsymbol{q}}(t)$ denotes the disturbance state vector at different times during the linear evolution of the perturbation. At this juncture, for every time, we define a vector $\hat{\boldsymbol{p}}(t)$ that is obtained by normalising $\hat{\boldsymbol{q}}(t)$ to have unit kinetic energy as per equation (\ref{eq:above}). Thus, for $\hat{\boldsymbol{p}}(t)$, we simply have
\begin{align}
 &\hat{\boldsymbol{p}}(t) = \frac{\hat{\boldsymbol{q}}(t)}{||\hat{\boldsymbol{q}}(t)||_{_E}} .
\end{align}

To see if $\hat{\boldsymbol{q}}(t)$ is indeed coincident with the unstable mode $\hat{\boldsymbol{q}}_u$, we now take advantage of equation (\ref{eq:above}) and define a new quantity $M$ as 
\begin{align}
 &M(t) = \frac{1}{2k^2} \int_{-1}^1 \hat{\boldsymbol{p}}^H(t) \begin{bmatrix} k^2 - D^2 & 0 \\ 0 & 1 \end{bmatrix}\hat{\boldsymbol{q}}_u\;\mathrm{d}y. \label{eq:proj}
\end{align}
We can interpret $M$ simply as a measure of the projection of the disturbance onto the unstable eigenmode. When $M = 1$, the disturbance has evolved such that it exactly coincides with the unstable eigenmode. It follows that $M = 1$ for all subsequent times after this point while the system can still be considered linear.

In the table \ref{tab:growth_rates}, we have specified that the growth rates of the least stable eigenmodes with wavenumber vector $\boldsymbol{k}=(\alpha,\beta) = (0,2)$ when $Re = 1500$ to serve as a reference. 
\begin{table}
  \begin{center}
  \caption{The exponential growth rates of the least stable mode with $\boldsymbol{k}=(\alpha,\beta) = (0,2)$ when $Re = 1500$ at different rotation rates.}
  \begin{ruledtabular}
  \begin{tabular}{cdcdcd}
      $Ro$ & \multicolumn{1}{c}{\textrm{Growth rate}} & $Ro$ & \multicolumn{1}{c}{\textrm{Growth rate}} & $Ro$ & \multicolumn{1}{c}{\textrm{Growth rate}}\\
      \colrule
       $1 \times 10^{-7}$ & -0.00431158 & 0.003 & 0.03165619 & 0.5 & 0.19551277 \\
       $1 \times 10^{-4}$ & -0.00087845 & 0.02 & 0.09718189 & 0.7 & 0.05017800 \\
       $3 \times 10^{-4}$ & 0.00398466 & 0.2 & 0.25758712 & 0.9 & - 0.01717999 \\
  \end{tabular}  
  \label{tab:growth_rates}
  \end{ruledtabular}
  \end{center}
\end{table}

\subsection{Nonlinear results -- low rotation rates} \label{ssec_nonlin_lowRo}

To start off, we shall first consider the cases where the rotation rates are small. As a measure of transition, we examine the time evolution of the root mean square (rms) values of the different components of the velocity for all the cases. For all the cases there is an initial period where there is transient amplification of the disturbance. Several wavenumbers then start to gain energy (not shown) through the nonlinear terms aided by the initial noise. The transient amplification seen at early times is then inhibited by nonlinear effects. For both low and high rotation cases, we seek to describe the flow characteristics well after the initial transients have run their course, and the resulting flow is fully nonlinear. 

\begin{figure}
\centering
  \subfloat[]{
    \includegraphics[width=3.125in]{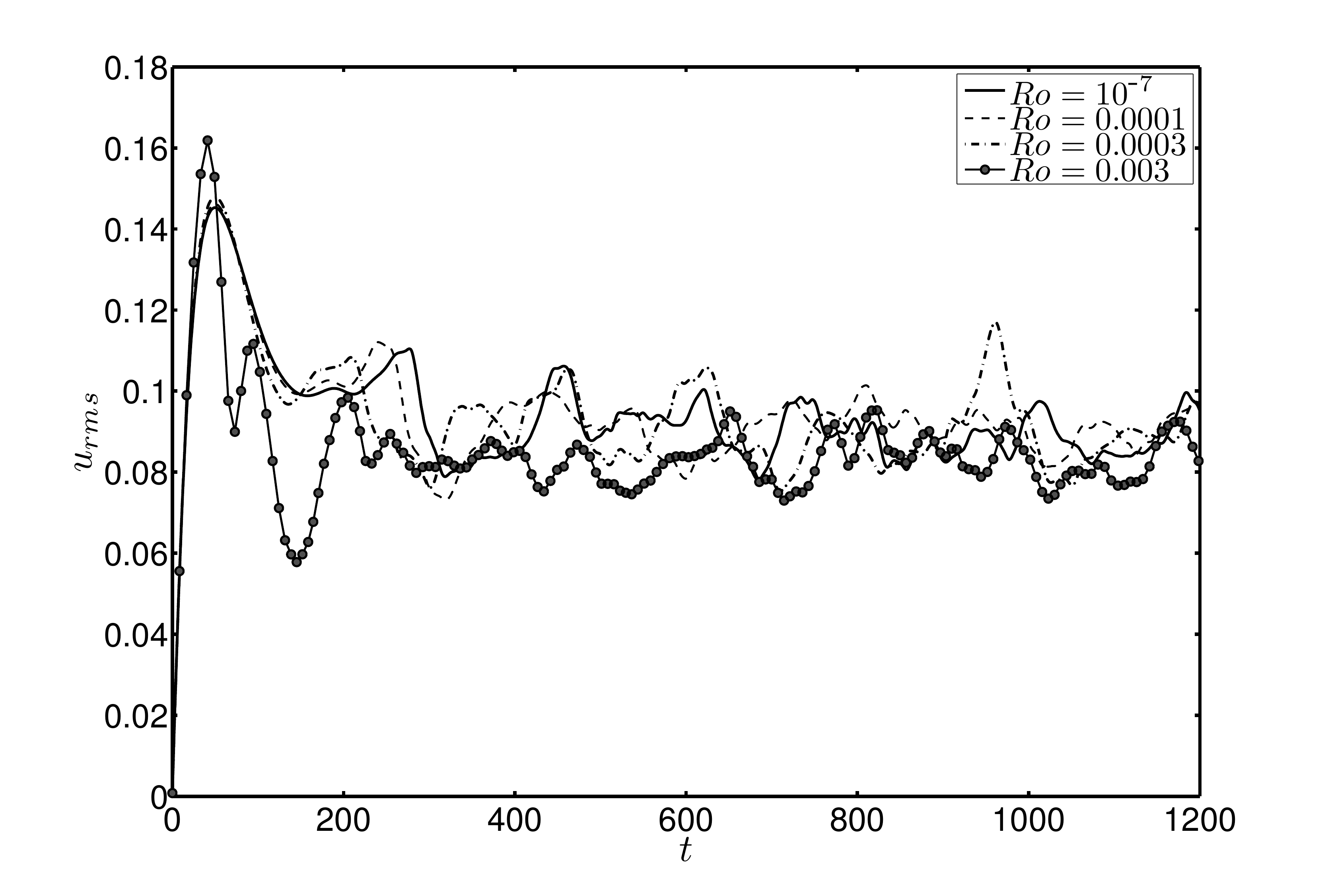}}
    \hspace{0.015in}
  \subfloat[]{
    \includegraphics[width=3.125in]{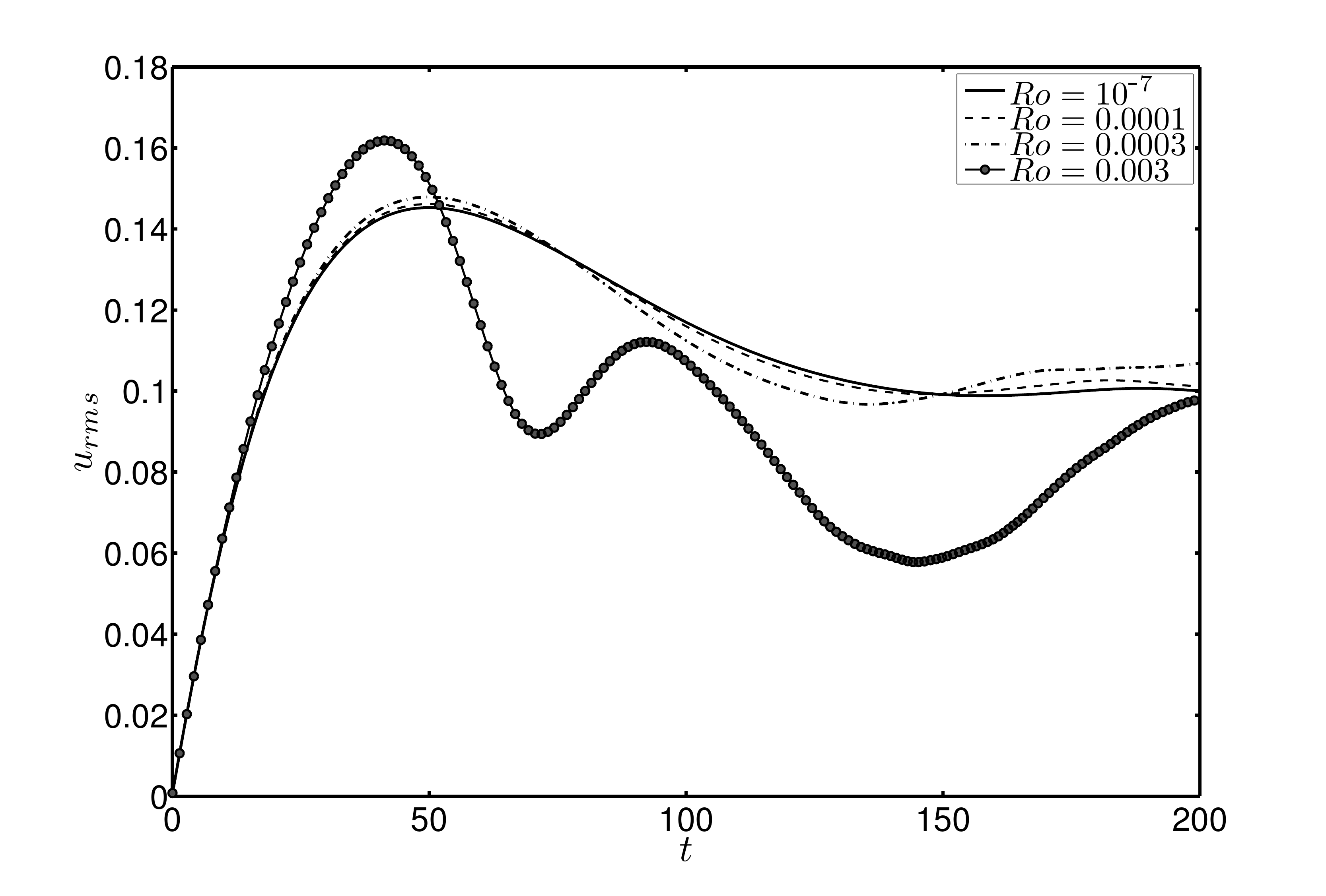}}
  \caption{Evolution of root mean square (rms) values of the streamwise velocity component. The secondary flow is dominant in the streamwise direction. (a) We see that the rms values, at later times, settle within a small range for different rotation rates. (b) This is a magnified portion of (a) where it is seen that the effect of rotation is not very different during the early stages of the evolution of the secondary flow. We start to see larger differences at early times as the rotation rate is increased, i.e., when $Ro = 0.003$.}
\label{fig_rms_lowRo}
\end{figure}

It can be seen in figure \ref{fig_rms_lowRo} that the rms values of the streamwise velocity $u$ settle within a range of amplitudes not very dependent on the rotation rate. A transient spurt in rms values of the various velocity components is followed by a settling down into a time-dependent state at a lower mean energy level than the maximum transient. In this state the rms values show an apparently chaotic signal for all the cases with similar time averaged behaviour. The observations when other components of the velocity are considered offer similar trends. We emphasise that this is the case regardless of the fact that some of the configurations here ($Ro = 3 \times 10^{-4}, 0.003$) can support exponential instabilities. We also show in figure \ref{fig_rms_lowRo} that the initial evolution of the secondary flow appears to be similar while the nonlinear terms have not fully come into play. Departures from this type of behaviour are seen when the rotation rate is increased. This prepares us for pronouncedly 
different behaviour at high rotation rates. 

\begin{figure}
\centering
  \subfloat[$Ro = 10^{-7}$ (Subcritical)]{
    \includegraphics[width=3.15in]{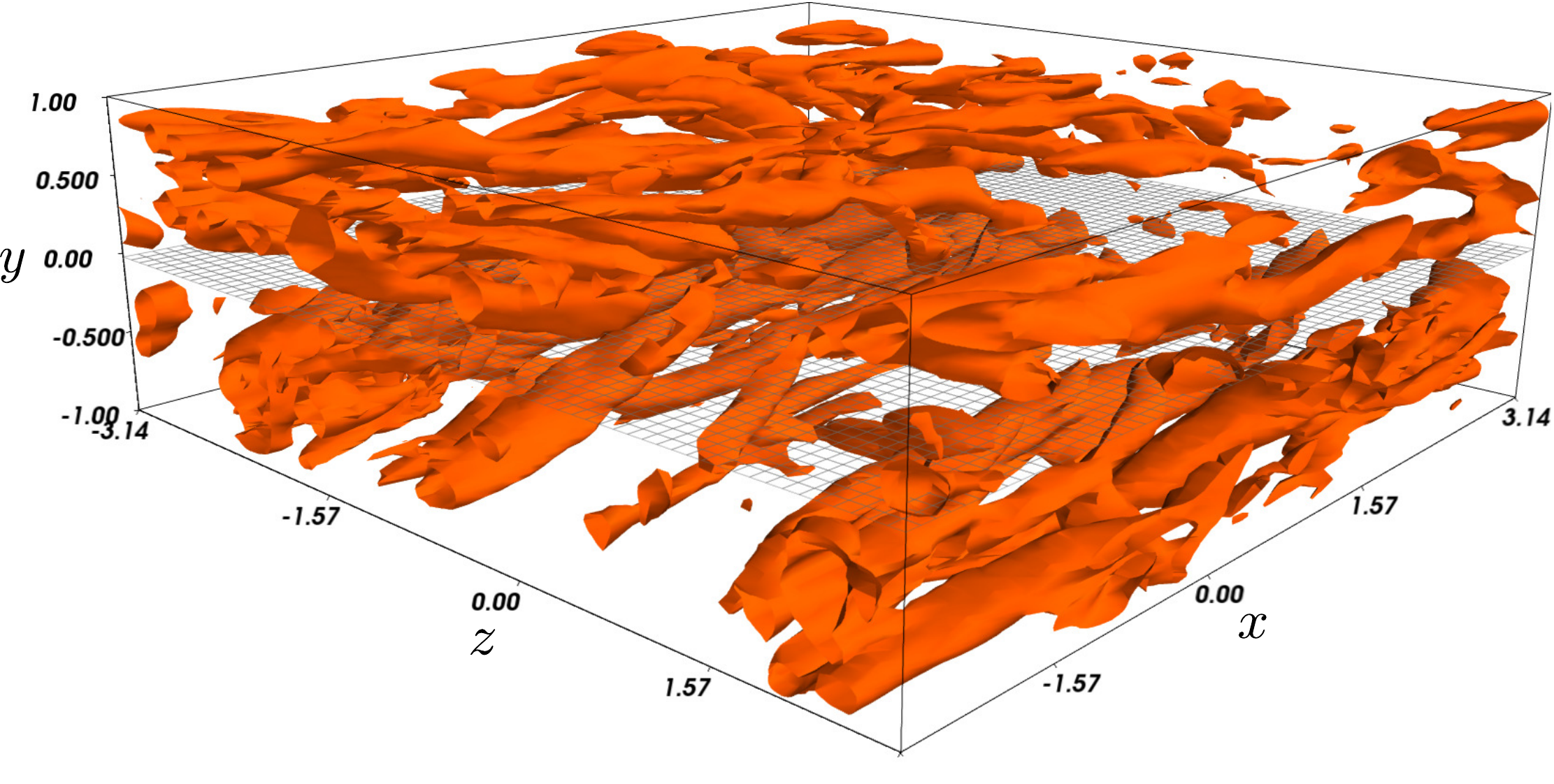}}
    \hspace{0.01in}
  \subfloat[$Ro = 10^{-4}$ (Subcritical)]{
    \includegraphics[width=3.15in]{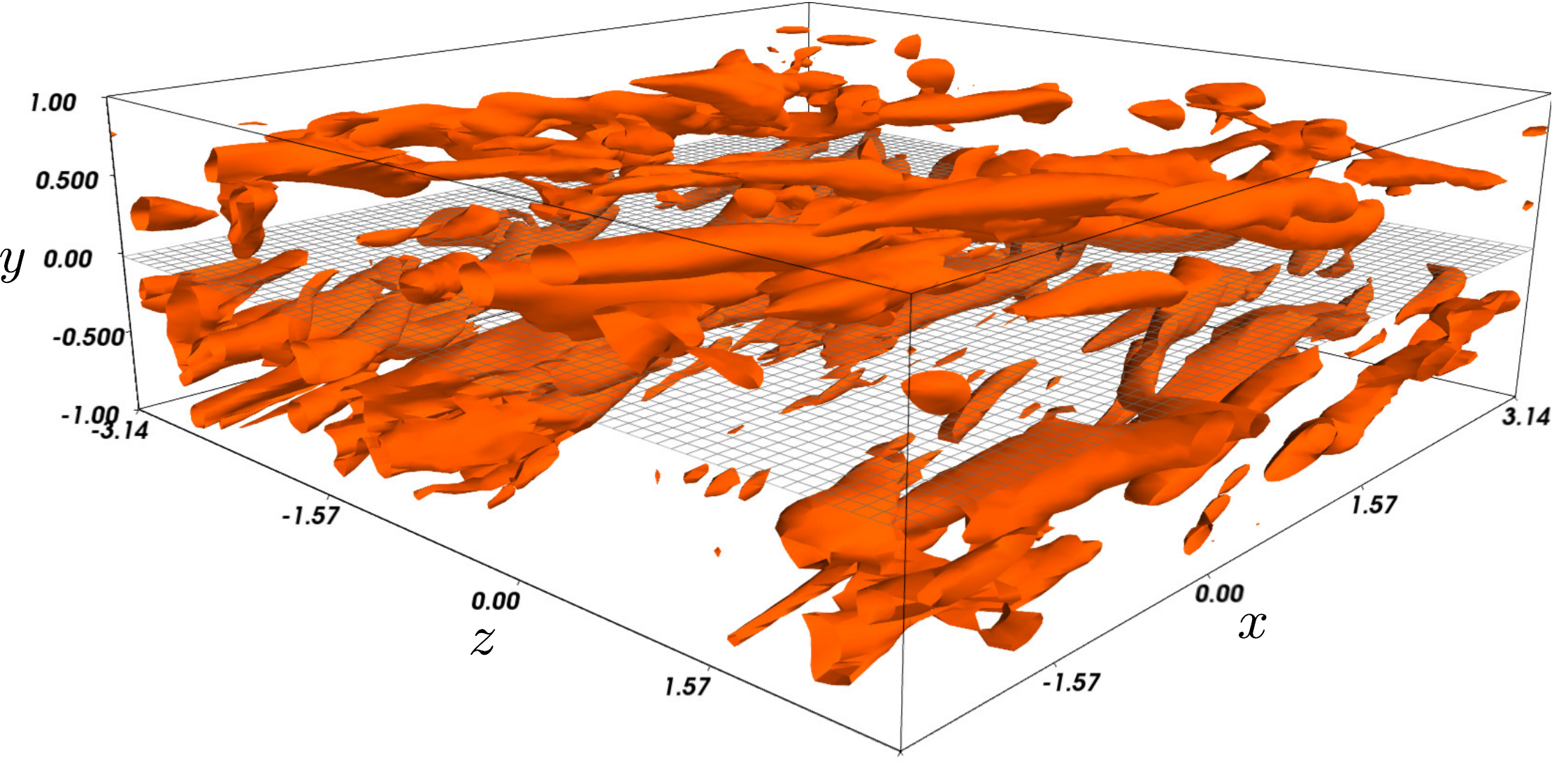}}
    \vspace{0.01in}
  \subfloat[$Ro = 3\times 10^{-4}$ (Supercritical)]{
    \includegraphics[width=3.15in]{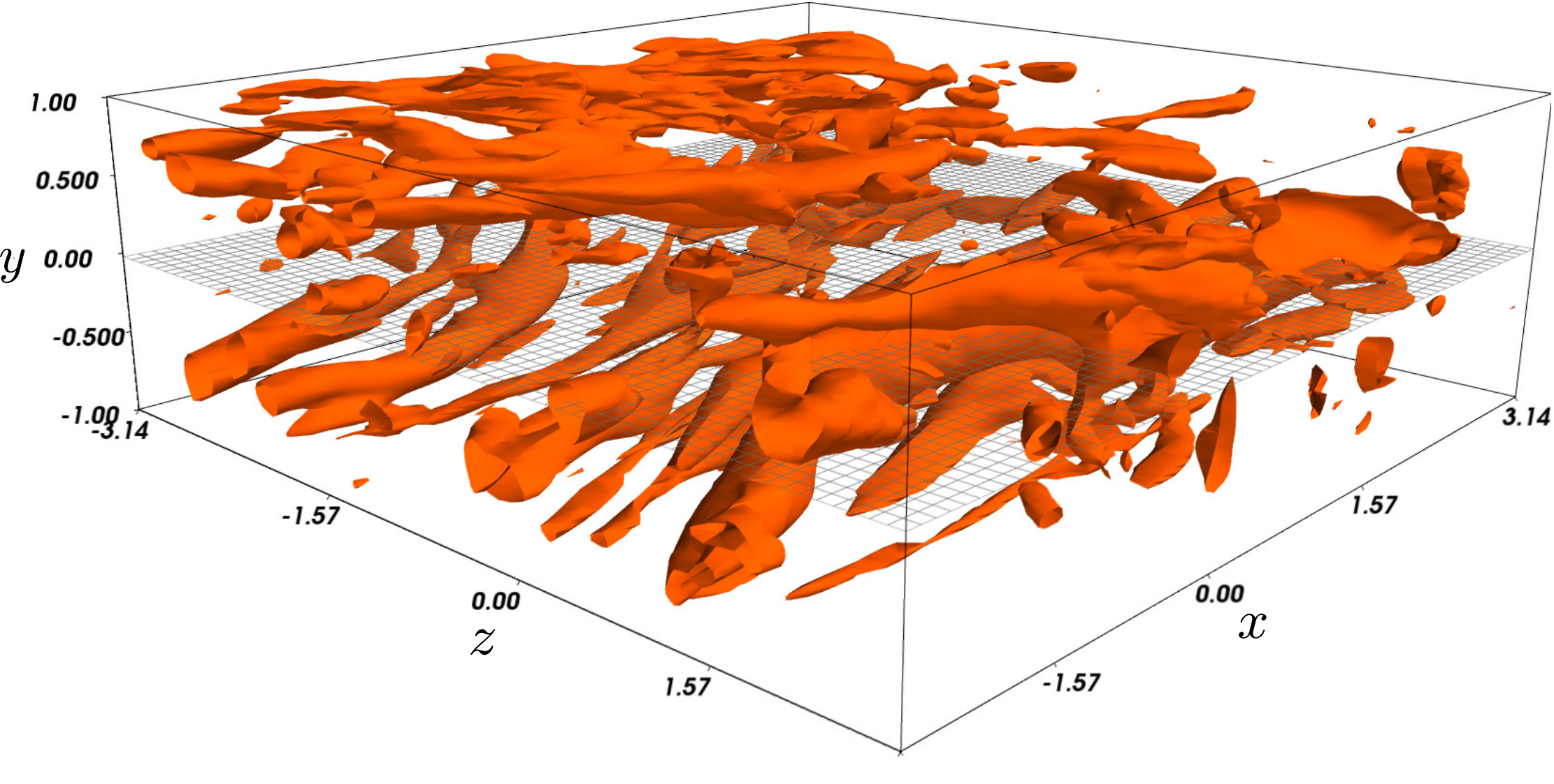}}
    \hspace{0.01in}
  \subfloat[$Ro = 3\times 10^{-3}$ (Supercritical)]{
    \includegraphics[width=3.15in]{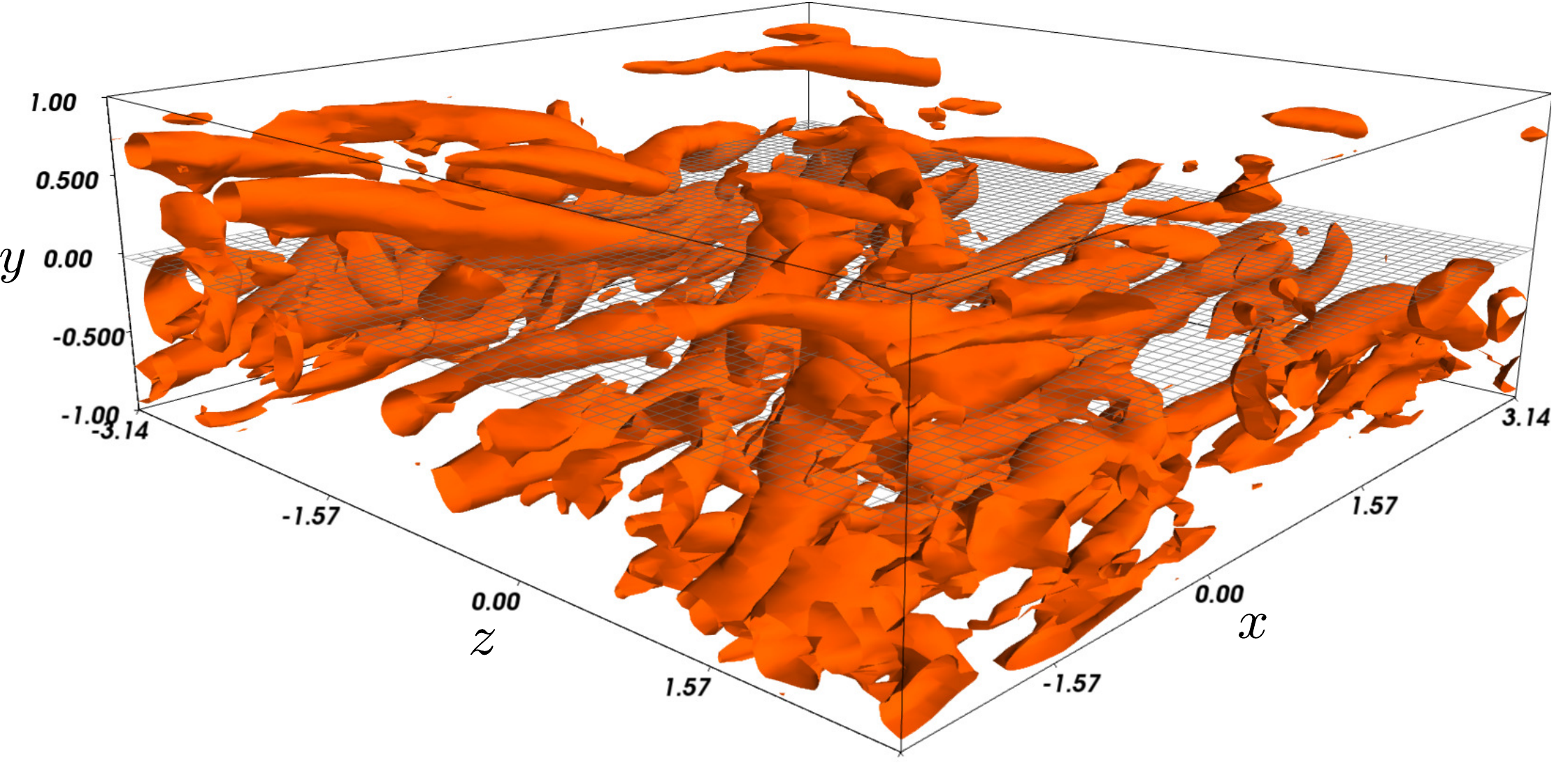}}
  \caption{The isocontours of constant $\lambda_{2} = -0.05$ at various rotation rates. The vortex cores are identified in each of the cases at $t = 500$, a characteristic time after the initial transient behaviour has died out. $Ro = 3\times 10^{-4},\; 0.003$ are cases where unstable modes may be excited. At low rotation rates, we see that there are no significant changes in the distribution of the identified vortex cores. }
  \label{fig_lambda2_lowRo}
\end{figure}

The rms values do not however tell us about where the secondary flow is set up and what the dominant structures are. The linear stability results lead us to expect that the Coriolis force biases the flow towards stronger secondary flow near the high pressure side of the channel. In figure \ref{fig_lambda2_lowRo}, we show vortex core regions identified by use of the $\lambda_2$ criterion \cite{Jeong_Hussain_95JFM}. $\lambda_{2}$ is the second eigenvalue of the tensor $\mathsfbi{S}^2 + \mathsfbi{R}^2$, where $\mathsfbi{S}$ is the strain rate tensor and $\mathsfbi{R}$ is the antisymmetric part of the velocity gradient tensor (the rotation tensor multiplied by 0.5). We choose a characteristic time $t = 500$ for comparing the different cases such that the initial transient behaviour has run its course and the flow is fully nonlinear. 

When the rotation rate is low, we see the vortex structures to be distributed across the channel and to be disordered. The behaviour is qualitatively similar in all cases, and similar to that in a stationary channel. At early times (not shown), the secondary flow initially gets set up as aligned in the streamwise direction, and then breaks down to give a seemingly chaotic flow \cite{Reddy_etal_98JFM}. Thus the flow at low $Ro$, even if within the linearly unstable regime, can undergo transition via mechanisms by which subcritical transition occurs in the non-rotating channel flow. At this point, no clear indication of the role of the unstable mode is seen. 

\begin{figure}
\centering
  \subfloat[$Ro = 10^{-7}$ (Subcritical)]{
    \includegraphics[width=3.125in]{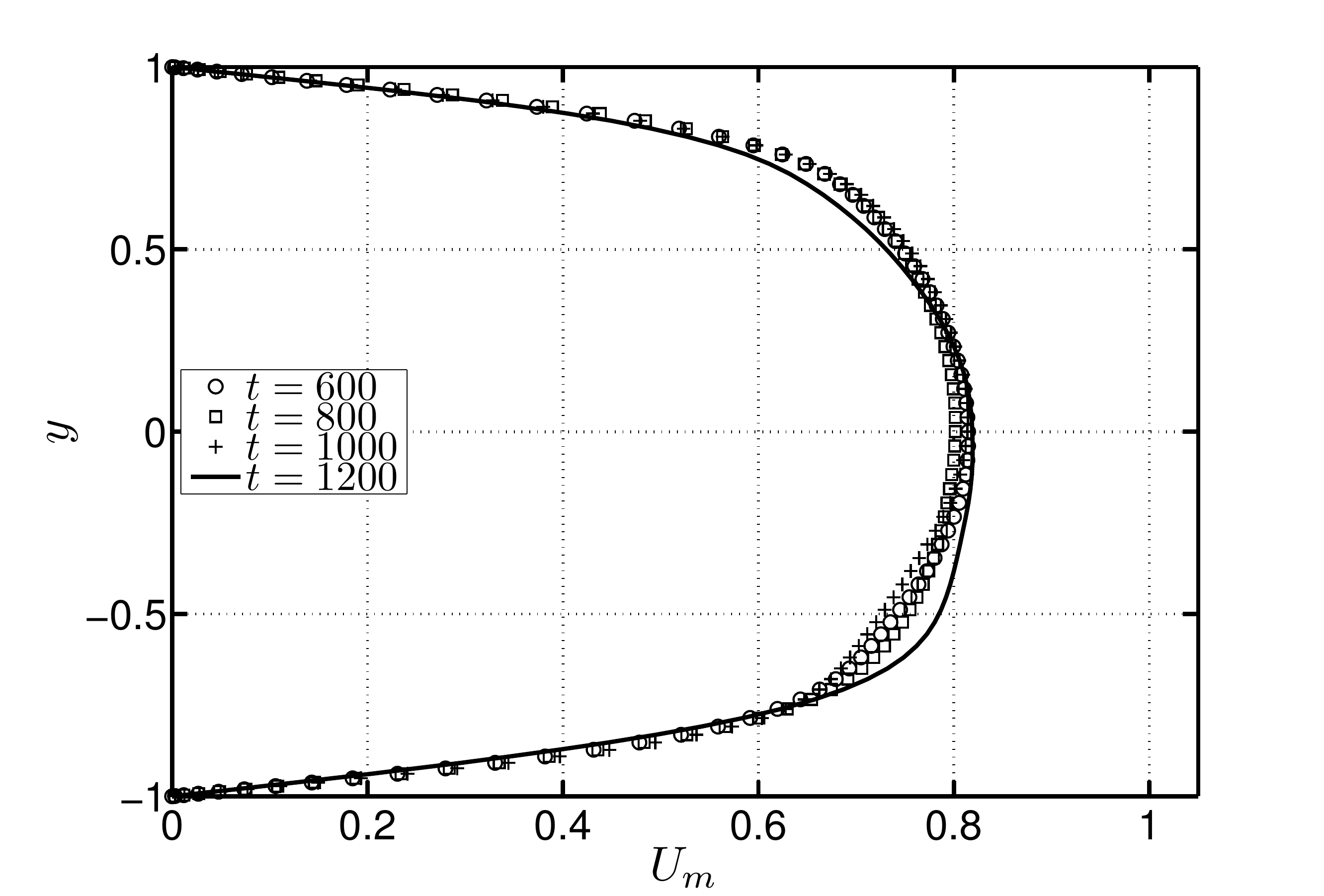}}
  \hspace{0.01in}
  \subfloat[$Ro = 10^{-4}$ (Subcritical)]{
    \includegraphics[width=3.125in]{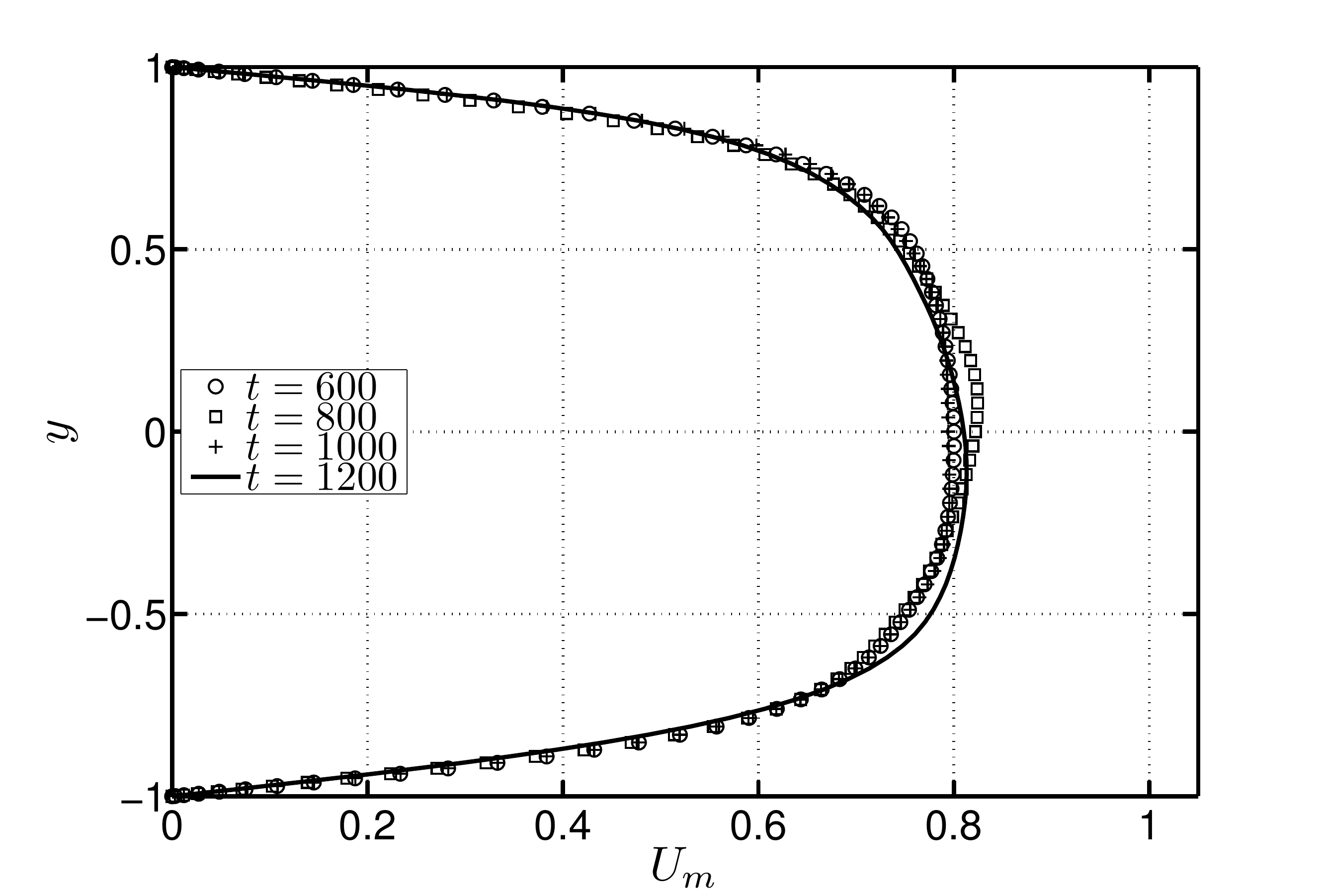}}
  \vspace{0.015in}
  \subfloat[$Ro = 3\times 10^{-4}$ (Supercritical)]{
    \includegraphics[width=3.125in]{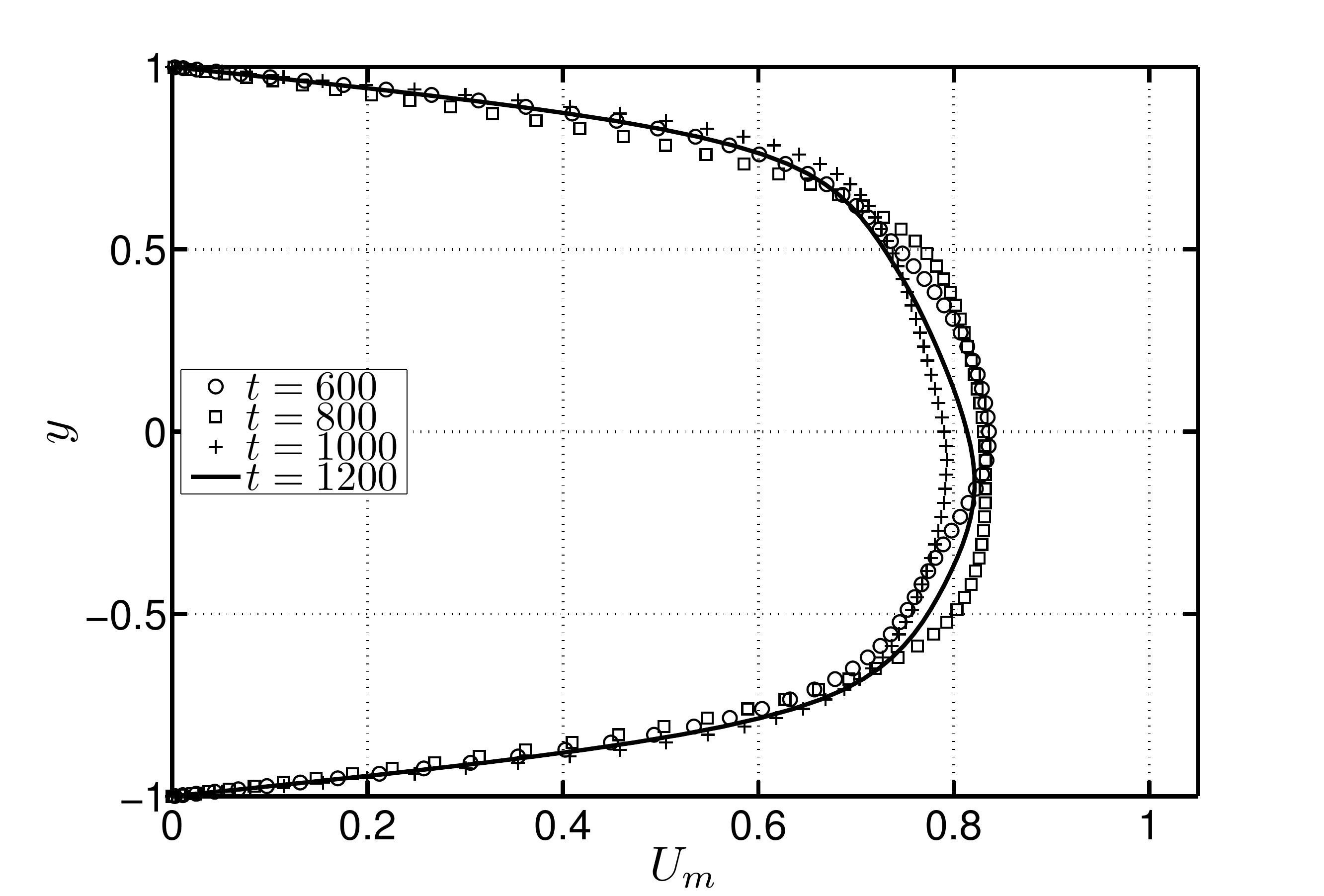}}
  \hspace{0.01in}
  \subfloat[$Ro = 3\times 10^{-3}$ (Supercritical)]{
    \includegraphics[width=3.125in]{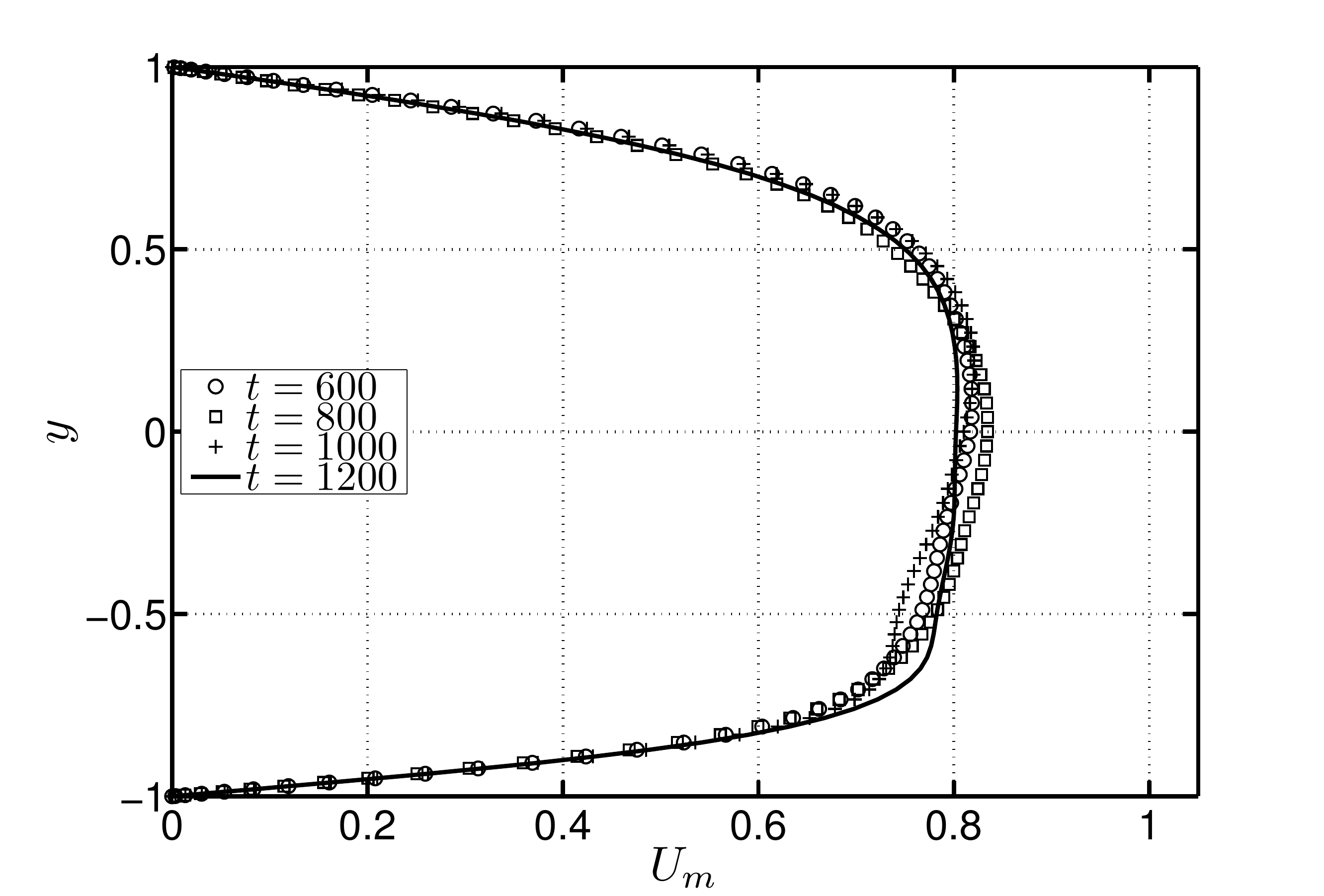}}
  \caption{Spatially averaged streamwise velocity profiles for different rotation rates, shown at different times. The familiar profile of a turbulent channel flow is obtained for low rotation rates.}
  \label{fig_U_mean_lowRo}
\end{figure}

In figure \ref{fig_U_mean_lowRo} are shown mean flow profiles at different times of the present simulation for different values of $Ro$ considered. The mean flow is derived by averaging in the streamwise and spanwise coordinates at a given time. Consistent with the observation that the rms velocity components display significant fluctuations in time, we notice variations in time of the space-averaged mean flow. The profiles obtained resemble that of non-rotating turbulent channel flow. It was seen in earlier work that the mean flow in the rotating channel is no longer symmetric about the centreline due to the Coriolis force when the flow becomes turbulent \cite{Kristoffersen_Andersson_93JFM,Lamballais_etal_98TCFD}. Despite being in the linearly unstable regime, here we do not see strong manifestations of the asymmetry in the mean flows as the Coriolis force is relatively weak. 

So far we have examined structures and the mean flow at different times (figures \ref{fig_lambda2_lowRo} and \ref{fig_U_mean_lowRo}). The rms values in figure \ref{fig_rms_lowRo} suggest a strongly fluctuating velocity field. To get a sense of how chaotic each of the resulting flows are, we now turn to the entropy measure $Q$ defined in equation \ref{eq_entropy2}. We choose a reference time $\tau = 500$. The plots make it evident that we have chaotic flow, since in no case do we have $Q$ returning to zero. In addition, the range of values of $Q$ seen for the different cases is not drastically different. 
    
\begin{figure}
\centering
  \subfloat[$Ro = 10^{-7}$ (Subcritical)]{
    \includegraphics[width=3.125in]{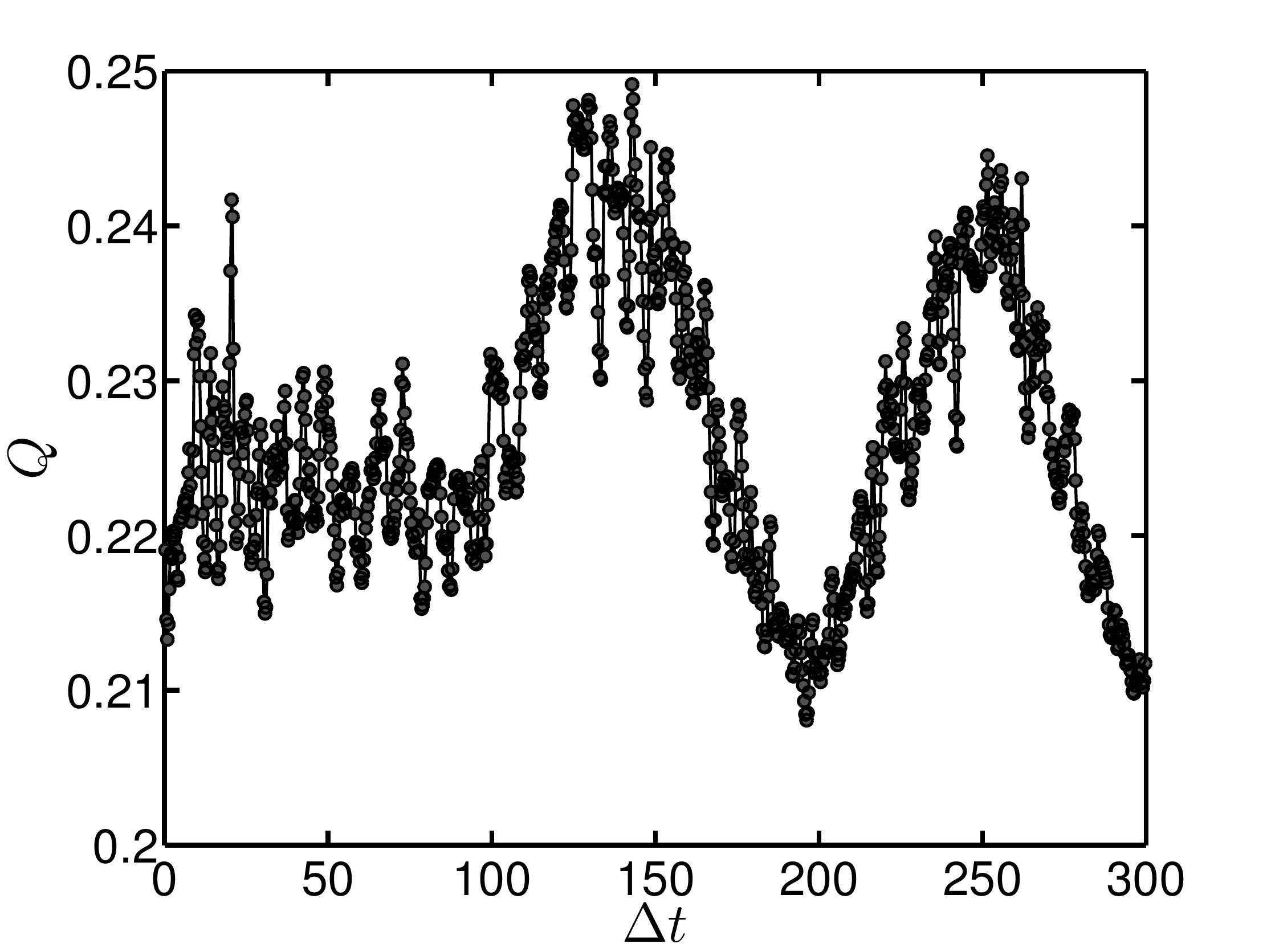}}
  \hspace{0.15in}
  \subfloat[$Ro = 10^{-4}$ (Subcritical)]{
    \includegraphics[width=3.125in]{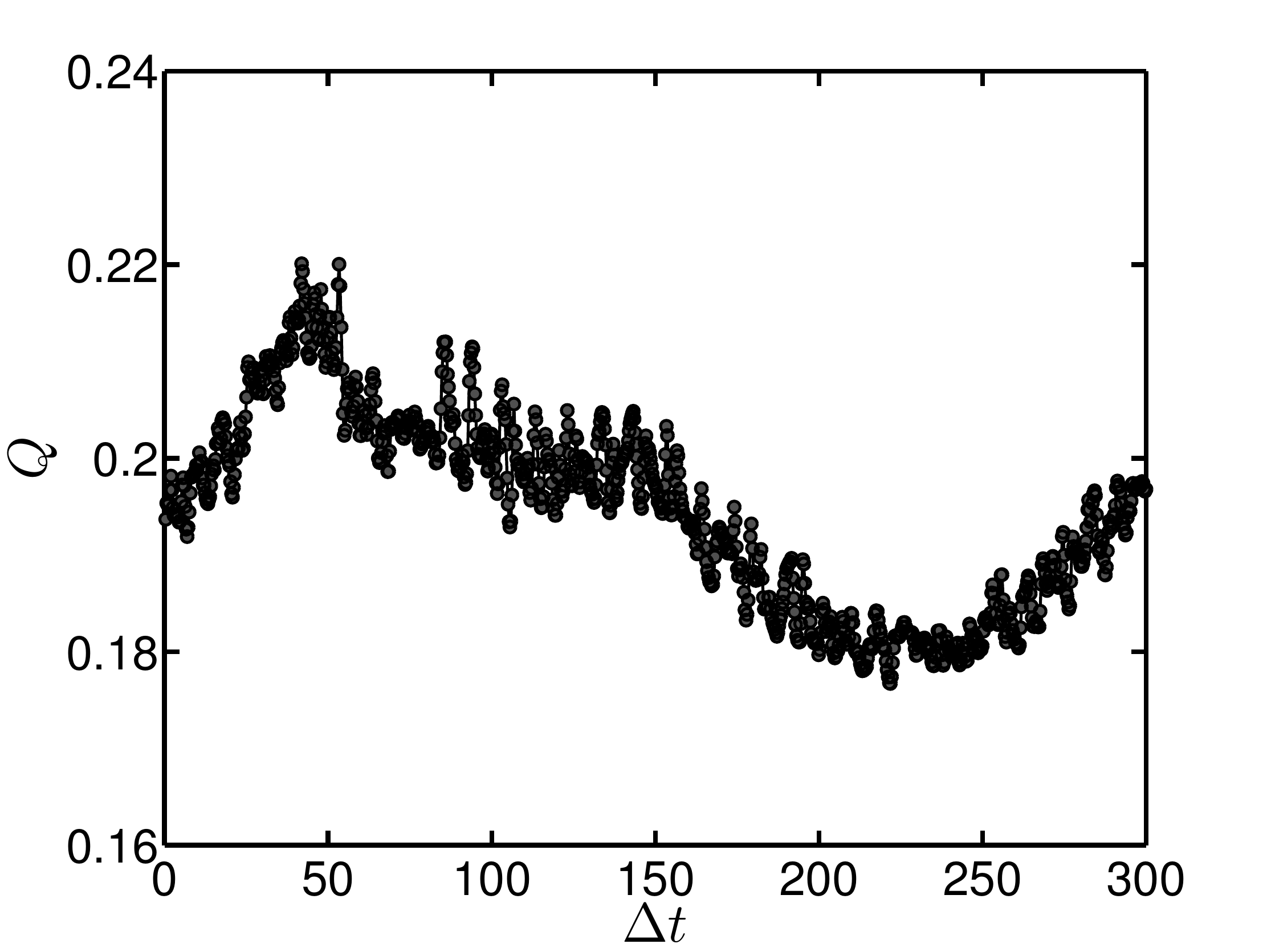}}
  \vspace{0.01in}
  \subfloat[$Ro = 3\times 10^{-4}$ (Supercritical)]{
    \includegraphics[width=3.125in]{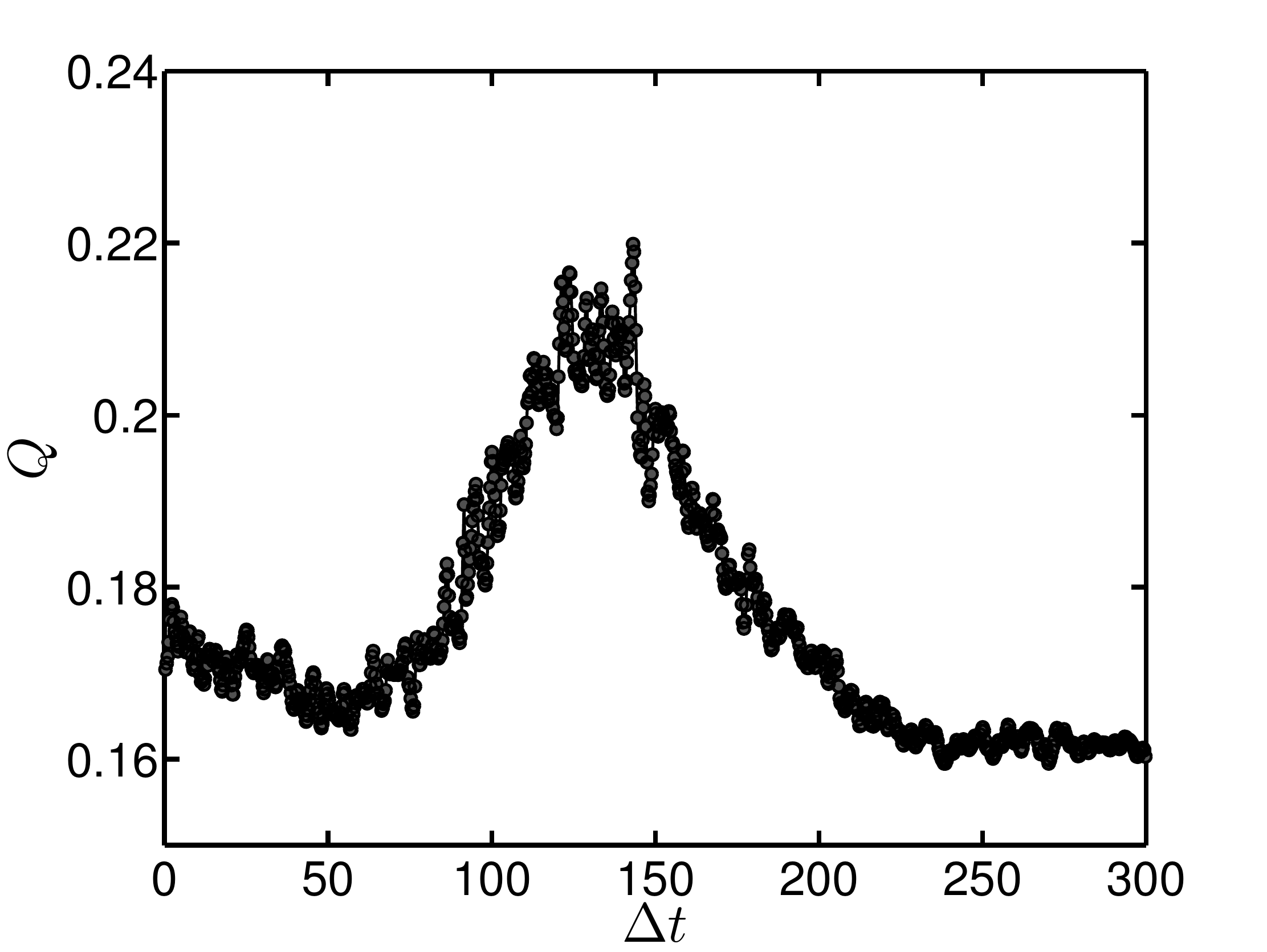}}
  \hspace{0.15in}
  \subfloat[$Ro = 3\times 10^{-3}$ (Supercritical)]{
    \includegraphics[width=3.125in]{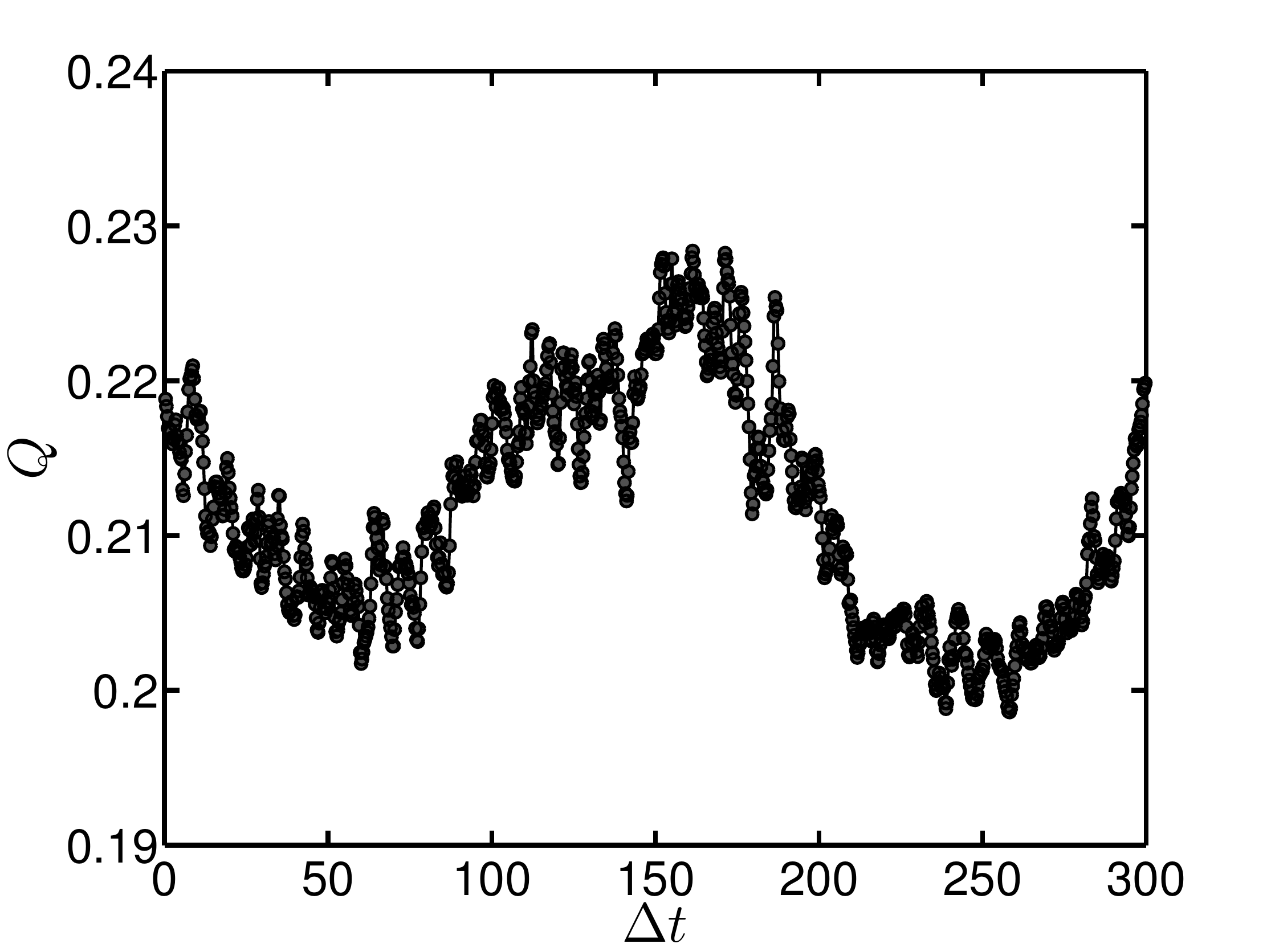}}
  \caption{The evolution for the measure of chaos $Q$, as defined in equation (\ref{eq_entropy2}). For all the cases, a departure from the initial state is shown. Here we choose $\tau = 500$ as the starting state.}
  \label{fig_entr_lowRo}
\end{figure}

\begin{figure}
\begin{minipage}[b]{0.45\linewidth}
 \centering
 \centerline{\includegraphics[width=3.125in]{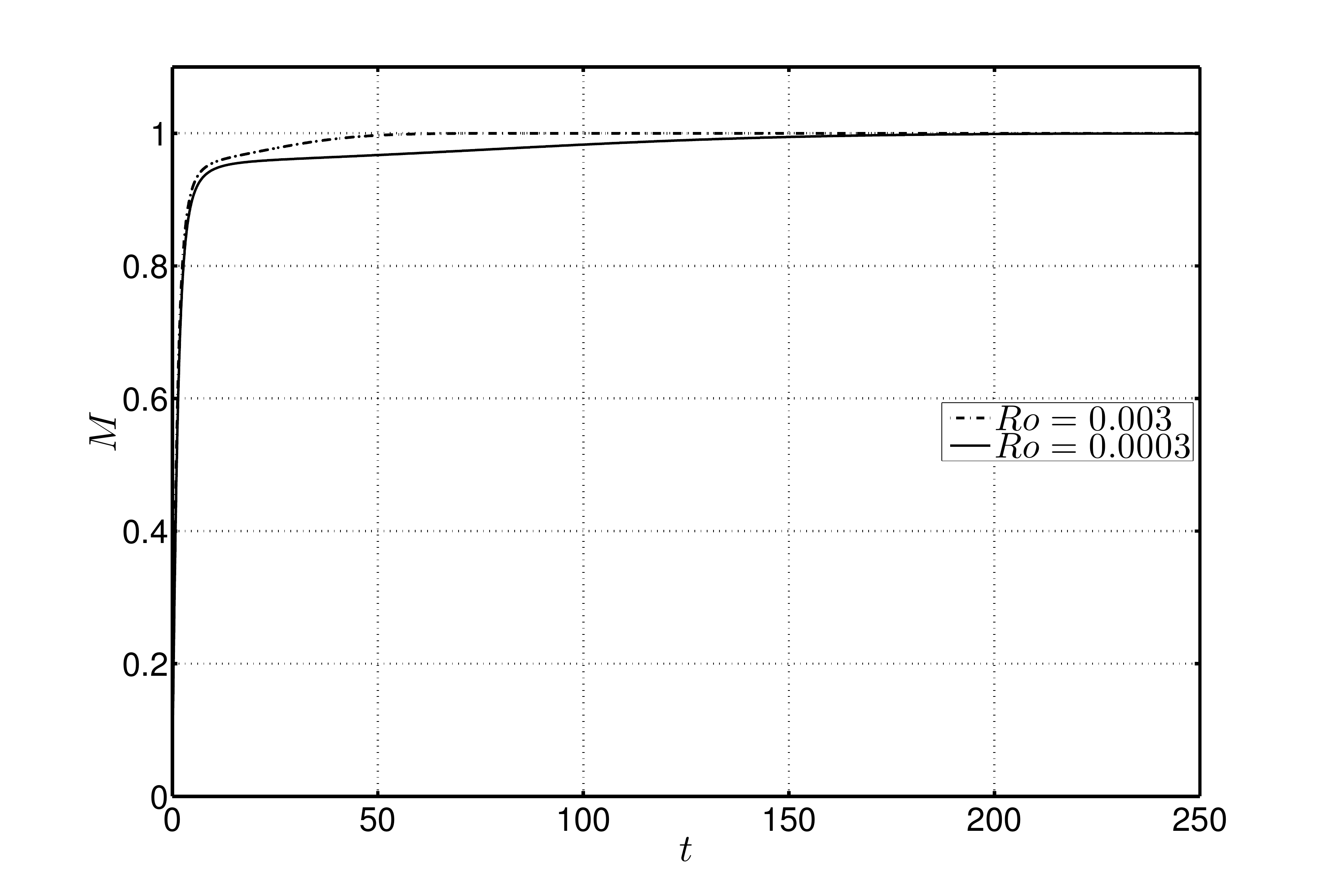}}
 \caption{The evolution of the projection measure $M$ defined in equation (\ref{eq:proj}) for low rotation cases. $M$ becomes equal to 1 at around $t \approx 50$ for $Ro = 0.003$, and at around $t \approx 215$ for $Ro = 3 \times 10^{-4}$. Both cases are within the supercritical regime.}
 \label{fig_M_vs_t_lowRo}
\end{minipage}
\hspace{0.025in}
\begin{minipage}[b]{0.45\linewidth}
 \centering
 \centerline{\includegraphics[width=3.125in]{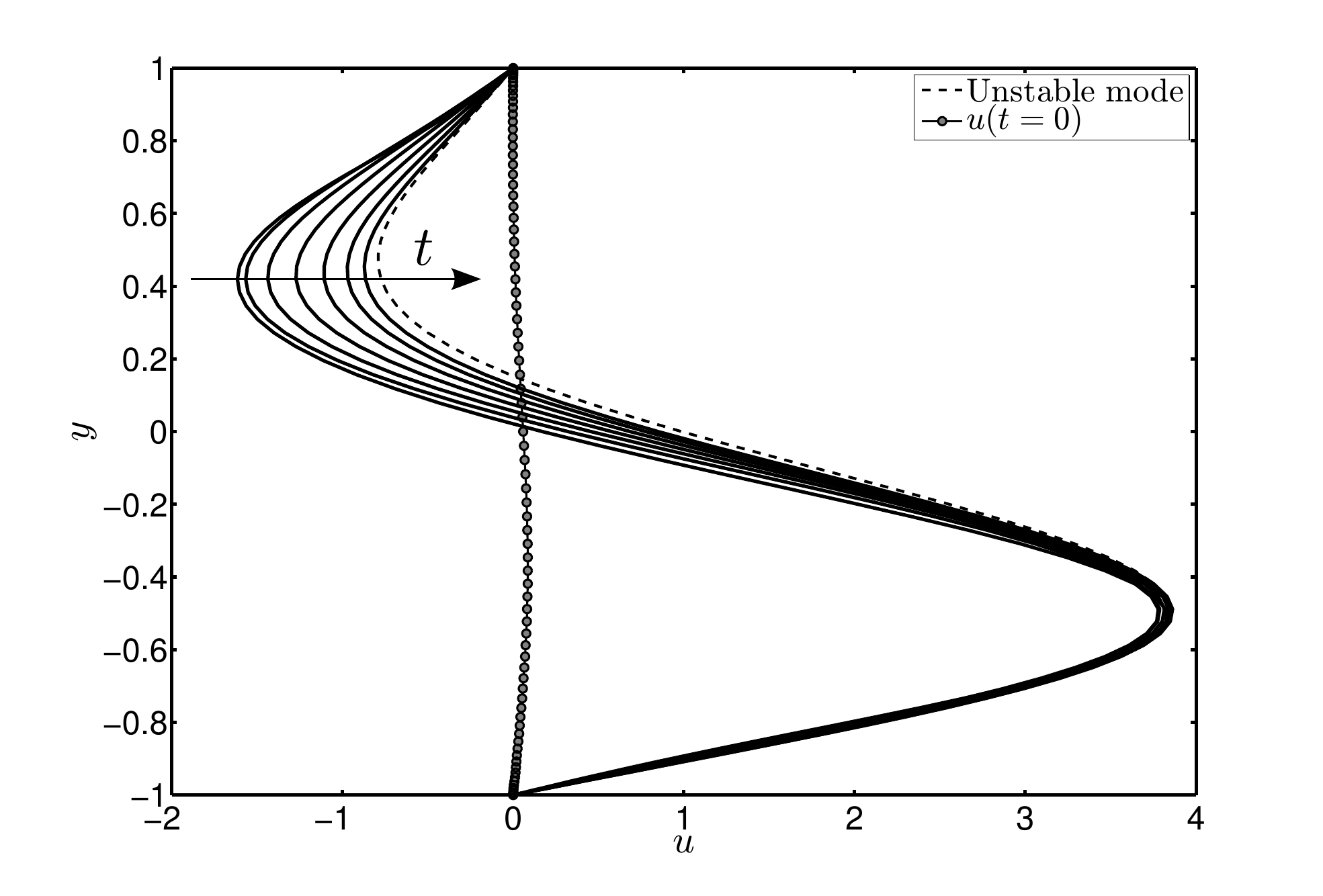}}
 \caption{The normalised streamwise disturbance velocity component $u$ at various times (about 30 time units apart) when flow evolves linearly for $Ro = 3 \times 10^{-4}$. At later times, it is seen that the disturbance start to resemble the unstable mode.}
 \label{fig_u_norm_linevol}
\end{minipage}
\end{figure}

Thus far there has been no indication of whether the unstable mode plays a significant role during the transition process. So now we examine the projection measure $M$ defined by equation \ref{eq:proj} for the two rotation rates that fall in the linearly unstable regime. We see in figure \ref{fig_M_vs_t_lowRo} at later times $M = 1$, indicating that the disturbance has come to comprise the unstable mode alone. At earlier times, the disturbance is a combination of different eigenfunctions of the linearised operator $\mathsfbi{L}$ (defined by equation \ref{eq_pert_eqs}). When the rotation rate is increased, i.e.\, when we go further into the linearly unstable regime, the time period before the unstable mode takes over the dynamics becomes shorter. 

\begin{figure}
\begin{minipage}[b]{0.45\linewidth}
 \centering
 \centerline{\includegraphics[width=3.125in]{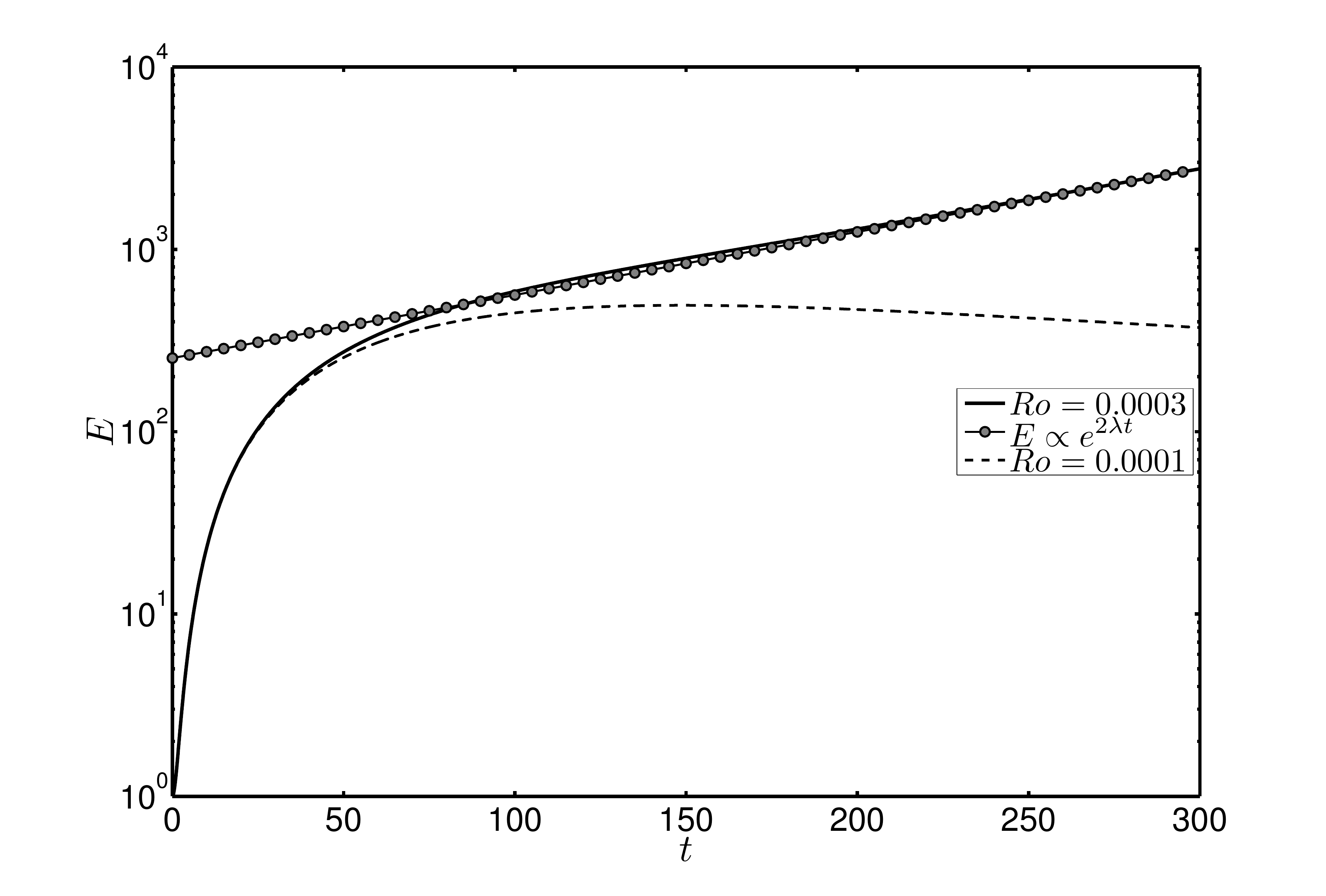}}
 \caption{The linear evolution of the disturbance perturbation energy when $Ro = 3 \times 10^{-4}$, with the unstable mode growth rate $\lambda = 0.00398467$. The (modally stable) $Ro = 1 \times 10^{-4}$ case is also given for reference.}
 \label{fig_en_lin_ev_3em4}
\end{minipage}
\hspace{0.025in}
\begin{minipage}[b]{0.45\linewidth}
 \centering
 \centerline{\includegraphics[width=3.125in]{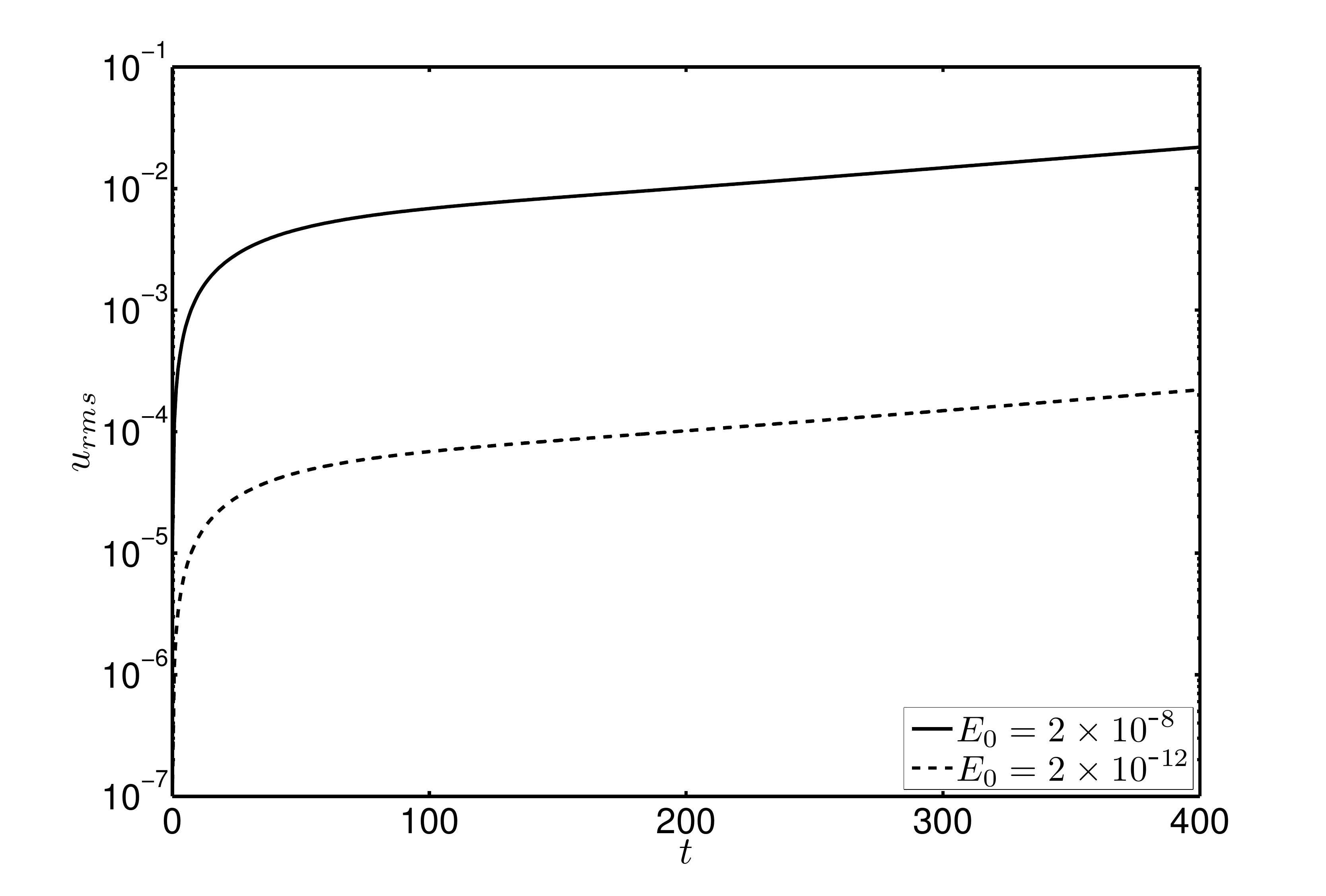}}
 \caption{The unstable mode is eventually excited in the full nonlinear setting when initial perturbation energy is low ($Ro = 3 \times 10 ^{-4}$). Algebraic growth is seen at early times before the unstable mode becomes dominant.}
 \label{fig_rms_lowRo_lindns}
\end{minipage}
\end{figure}

As the operator $\mathsfbi{L}$ is non-normal, the disturbance initially grows algebraically despite the presence of the unstable mode. During the early stages of the evolution, the secondary flow evolves in a manner to form streaks in the flow as is in the non-rotating channel flow. In figure \ref{fig_u_norm_linevol}, we plot the streamwise component of the disturbance velocity field at various times. What was initially a weak secondary flow at $t = 0$ in the streamwise coordinate has become much more pronounced to yield streaks. At later times, after initial transients, we see that it starts to settle towards the unstable mode. This behaviour is captured quantitatively when we look examine the disturbance energy as a function of time (figure \ref{fig_en_lin_ev_3em4}). Exponential growth of the disturbance is seen only after a period when other modes (all decaying) have fallen off. As a reference, the evolution of the disturbance when $Ro = 1 \times 10^{-4}$ (linearly stable) is also given to highlight 
departure due to the presence of the unstable mode. Although the rotation rate is very low, we can already see how an asymmetry develops and increases with time.

The nonlinearity in the governing equations act to limit the linear growth once the secondary flow has become sufficiently strong. Then the peak of the rms values in figure \ref{fig_rms_lowRo} can be considered a marker for when the flow has become fully nonlinear; this is seen to occur at $t \approx 50$ for all the cases. At the same stage in the linear evolution, the unstable mode has not yet come to dominate the dynamics (see figure \ref{fig_M_vs_t_lowRo}). The algebraic evolution of the disturbance has been so strong such that the flow becomes fully nonlinear without having excited the linear unstable mode. 

In the case of the non-rotating channel flow, the optimal perturbation eventually decays if the nonlinear terms are not triggered. This happens when the energy content of the perturbation is initially very low, and the algebraic linear amplification is not strong enough to render the nonlinear terms important. If we were to introduce the perturbation at lower energy levels, the secondary flow can evolve linearly for long enough to coincide with the unstable mode. This is clearly depicted in figure \ref{fig_rms_lowRo_lindns} when $Ro = 3 \times 10^{-4}$. As we increase rotation such that we are further away from the stability boundary, the time taken for the unstable mode to emerge becomes shorter. Upon exciting the unstable mode, the flow would evolve as dictated by the exponential growth rate from linear theory till the nonlinear terms become important.

From figure \ref{fig_rms_lowRo_lindns}, we also see that the initial rise in the rms values happen at a much faster rate than the growth rate specified by the unstable mode. Eventually the unstable mode alone is seen to dominate. At the stage where the unstable mode alone survives, the secondary flow is seen to have already attained a kinetic energy that is larger than what would have achieved had we started with an unstable mode alone. We can then say that given a class of initial conditions with the same kinetic energy, algebraic disturbances can enhance the energy content of the unstable mode and make the flow nonlinear at earlier times.

So in the cases considered above at low rotation rates, algebraically growing disturbances have been shown to be capable of triggering transition in two ways. Firstly, the algebraic growth can be strong enough such that the nonlinear terms come into play. The transition is triggered by the vortex stretching and tilting mechanisms that lead to the formation of streaky structures. Alternatively, if we impose the initial energy content of the disturbance to be very low, we end up with the situation where the unstable mode is eventually excited and transition occurs by the secondary instabilities of the saturated flow. This is akin to a noisy environment from which the unstable mode eventually emerges. With regard to at point of the evolution transition has occured, the nonmodal mechanisms are more dominant at short times. However if the flow has not undergone transition by subcritical mechanisms, the unstable mode will trigger the nonlinearity at later stages.

\subsection{Nonlinear results -- intermediate and higher rotation rates}
\begin{figure}
\centering
  \subfloat[]{
    \includegraphics[width=3.125in]{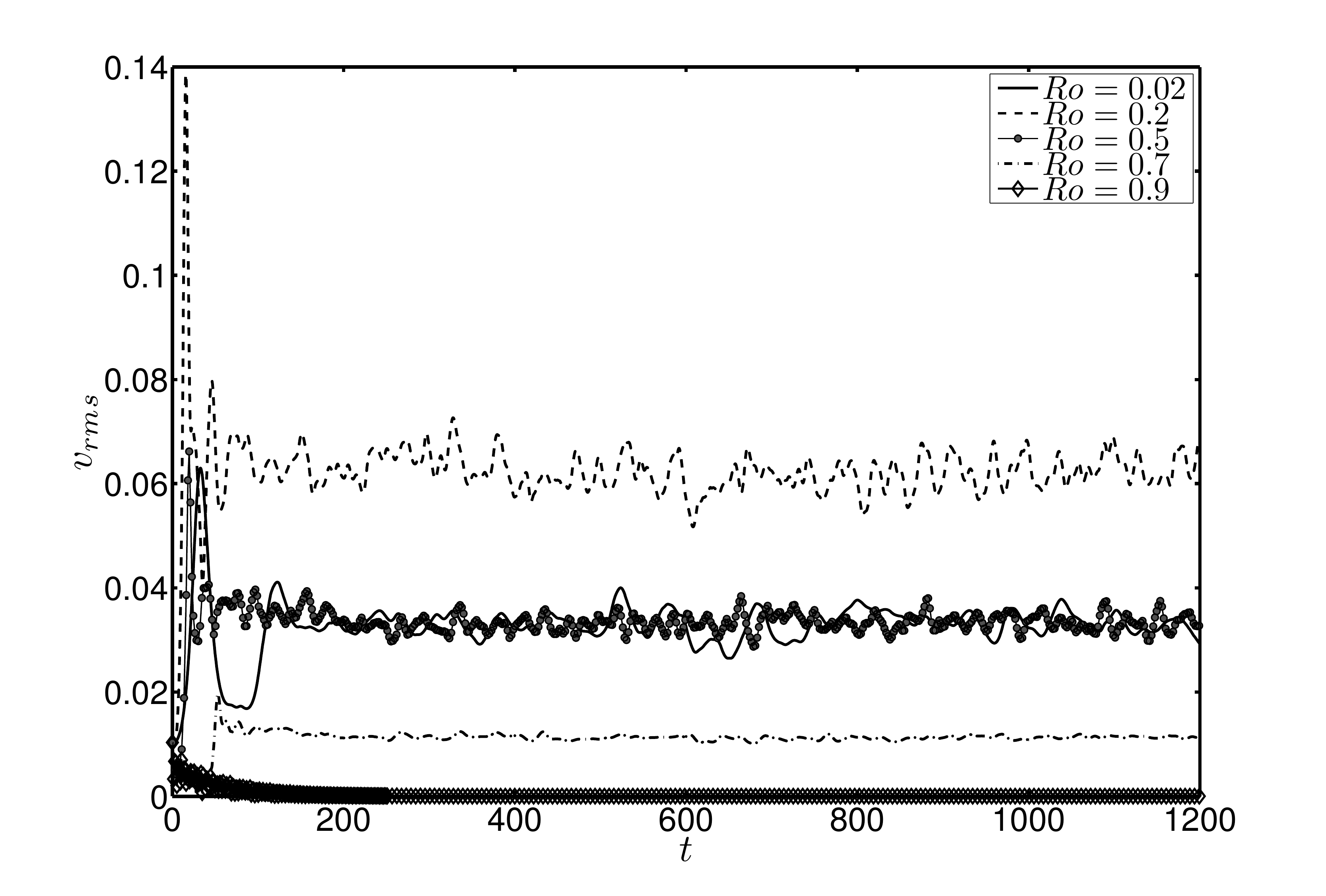}}
    \hspace{0.015in}
  \subfloat[]{
    \includegraphics[width=3.125in]{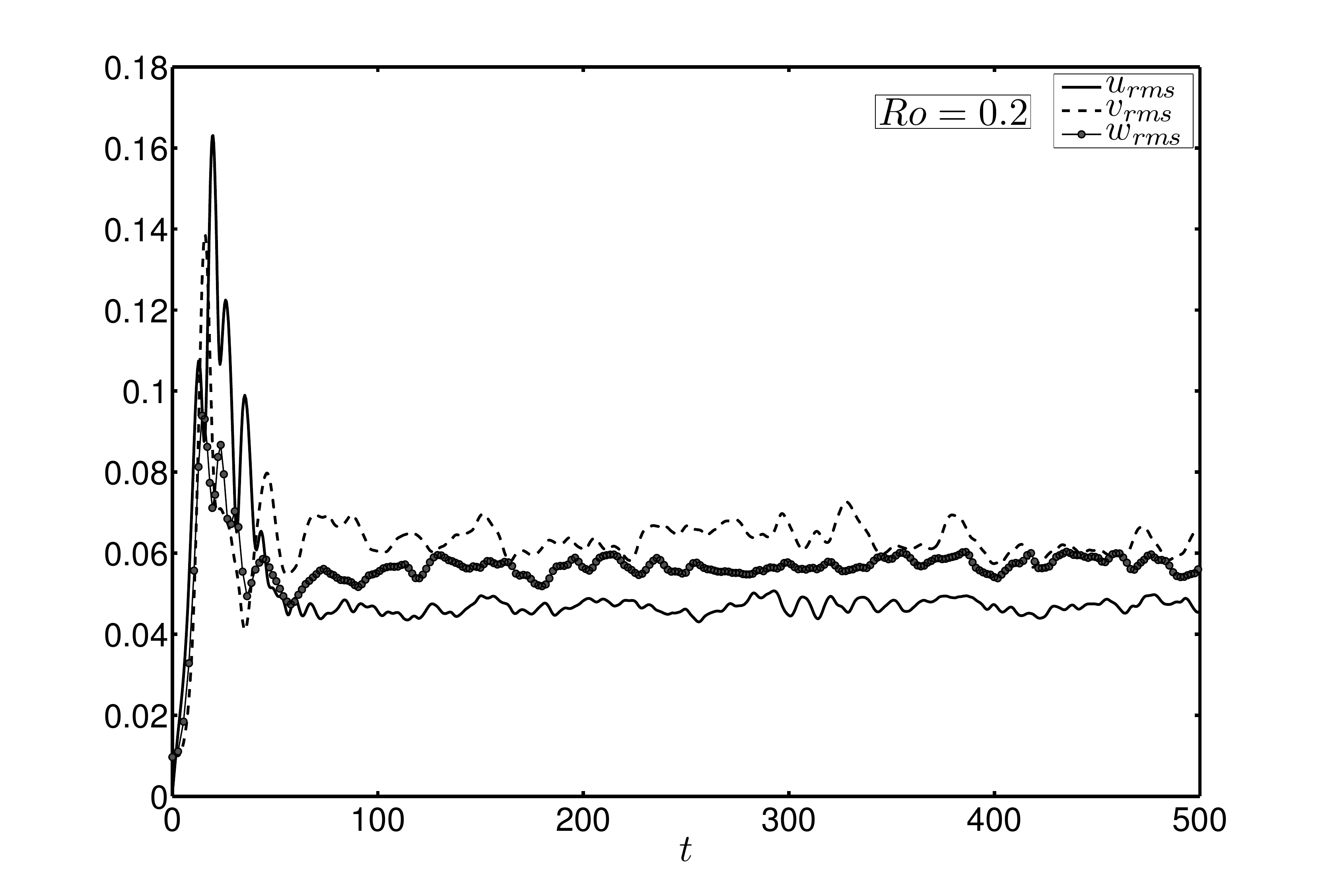}}
  \caption{(a) For the high rotation cases, the evolution of the root mean square (rms) values of the normal and spanwise velocities are not monotonic. Shown here is the rms values of the normal velocity component. (b) The rms values of the different velocity components when $Ro = 0.2$. The secondary flow becomes more isotropic when the rotation rate is increased accompanied by a suppression of the secondary flow directed in the streamwise direction.}
\label{fig_rms_hiRo}
\end{figure}

As the rotation is increased, we are now very much within the modally unstable regime. In the previous subsection, we already did see signs of the unstable mode emerging in the dynamics provided the flow has not become fully nonlinear until a certain time. The range of rotation rates considered here are more in line with earlier studies where strong instabilities and transition to turbulence has been observed. The question then to be posed here would be to see if the non-normal nature of the governing equations and algebraic disturbances have any significant effect on the dynamics of the flow.

As before, we first examine the rms values of the resulting flow at various rotation rates. The rms values of the streamwise velocity get suppressed to a greater extent (not shown). For the other velocity components, the rms behaviour is not monotonic as we increase the rotation rate (see figure \ref{fig_rms_hiRo}(a)). What was initially a more streamwise dominant flow in the case of low rotation rate cases now has comparable rms values of velocity in all directions; the $Ro = 0.2$ case displays this the best (see figure \ref{fig_rms_hiRo}(b)). Inside the linearly unstable region, the rms values of the spanwise and normal velocity components increase in a range of $Ro$. These values once again get suppressed as the rotation is further increased, and the region of linear stability is approached. Also to be noted is that on increasing the rotation rates, the rms values display smaller deviations from their long time average (for instance compare the cases with $Ro = 0.2, 0.7$ in figure \ref{fig_rms_hiRo}(a)). 
The flow has undergone a transition to an unsteady state at all rotation rates except $Ro = 0.9$. For this case the hig[scale=0.275]h rotation rates effectively kills all the fluctuations very quickly in keeping with Taylor--Proudman arguments. The resulting flow then quickly reverts back to the parabolic flow. One must keep in mind that when $Re > 5772$ transition due to the breakdown of the two-dimensional TS waves, which is unaffected by rotation, is still possible \cite{Wallin_etal_13JFM}.

\begin{figure}
\centering
  \subfloat[$Ro = 0.02$]{
    \includegraphics[width=3.15in]{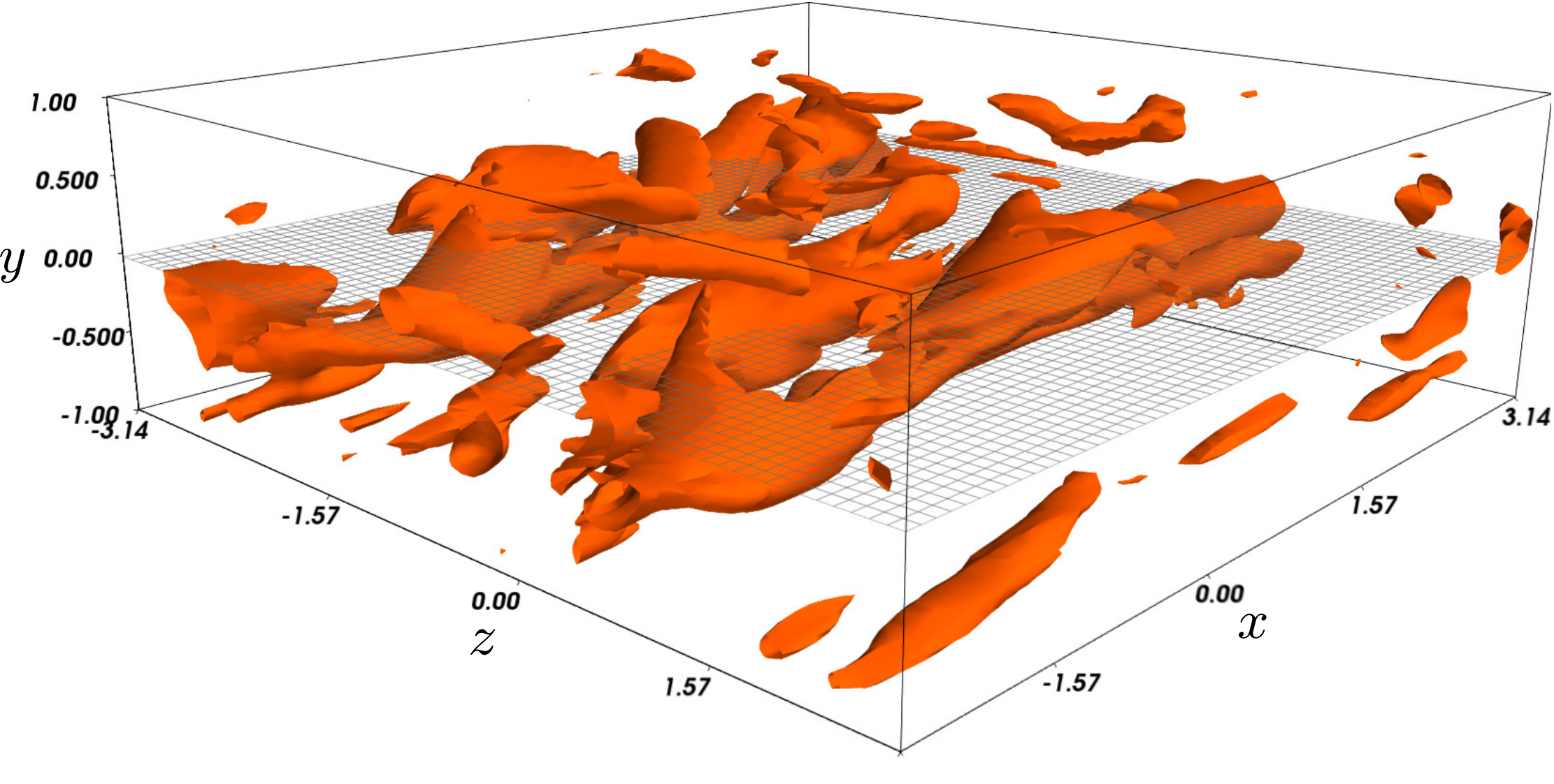}}
    \hspace{0.01in}
  \subfloat[$Ro = 0.2$]{
    \includegraphics[width=3.15in]{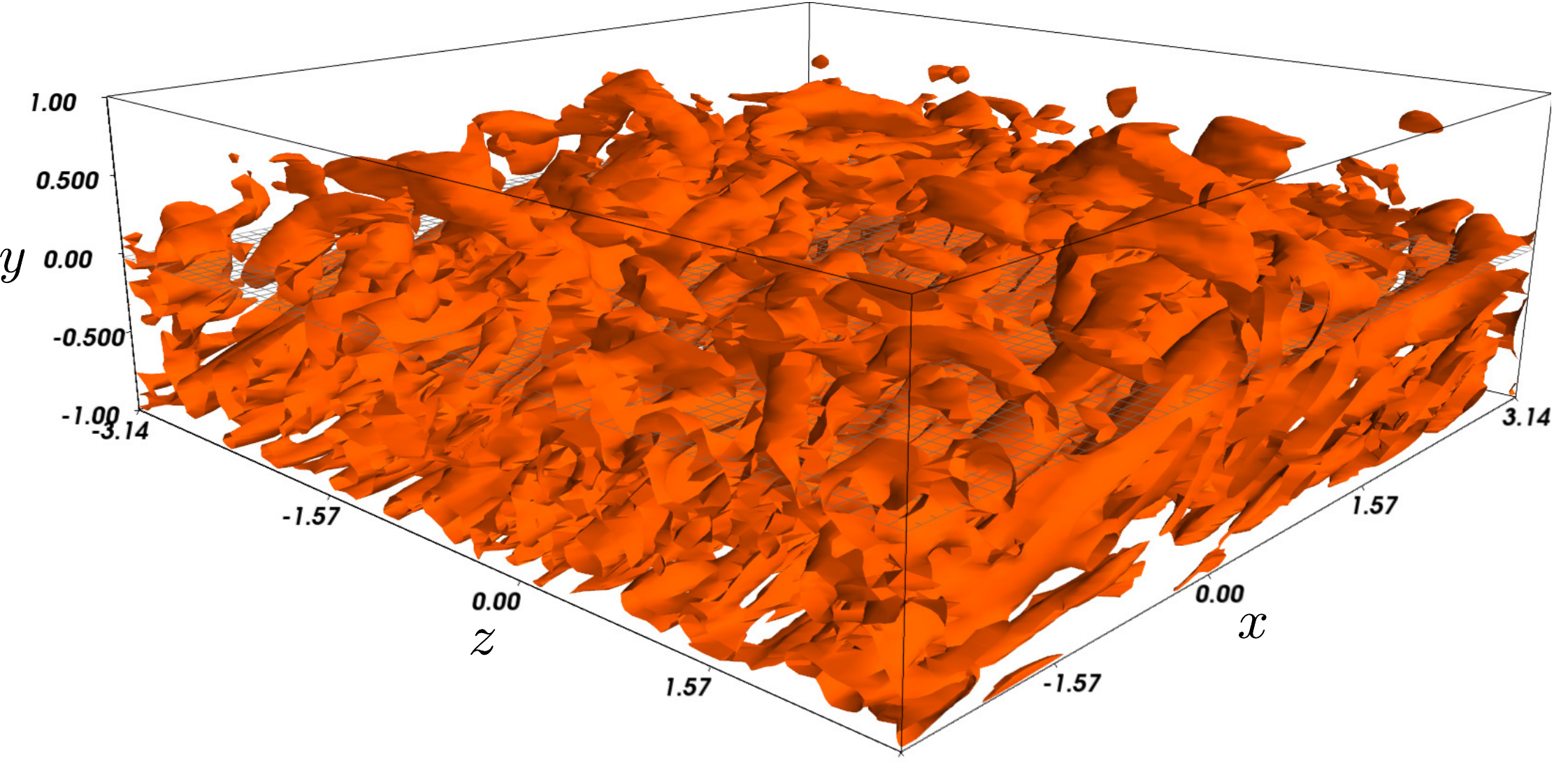}}
    \vspace{0.01in}
  \subfloat[$Ro = 0.5$]{
    \includegraphics[width=3.15in]{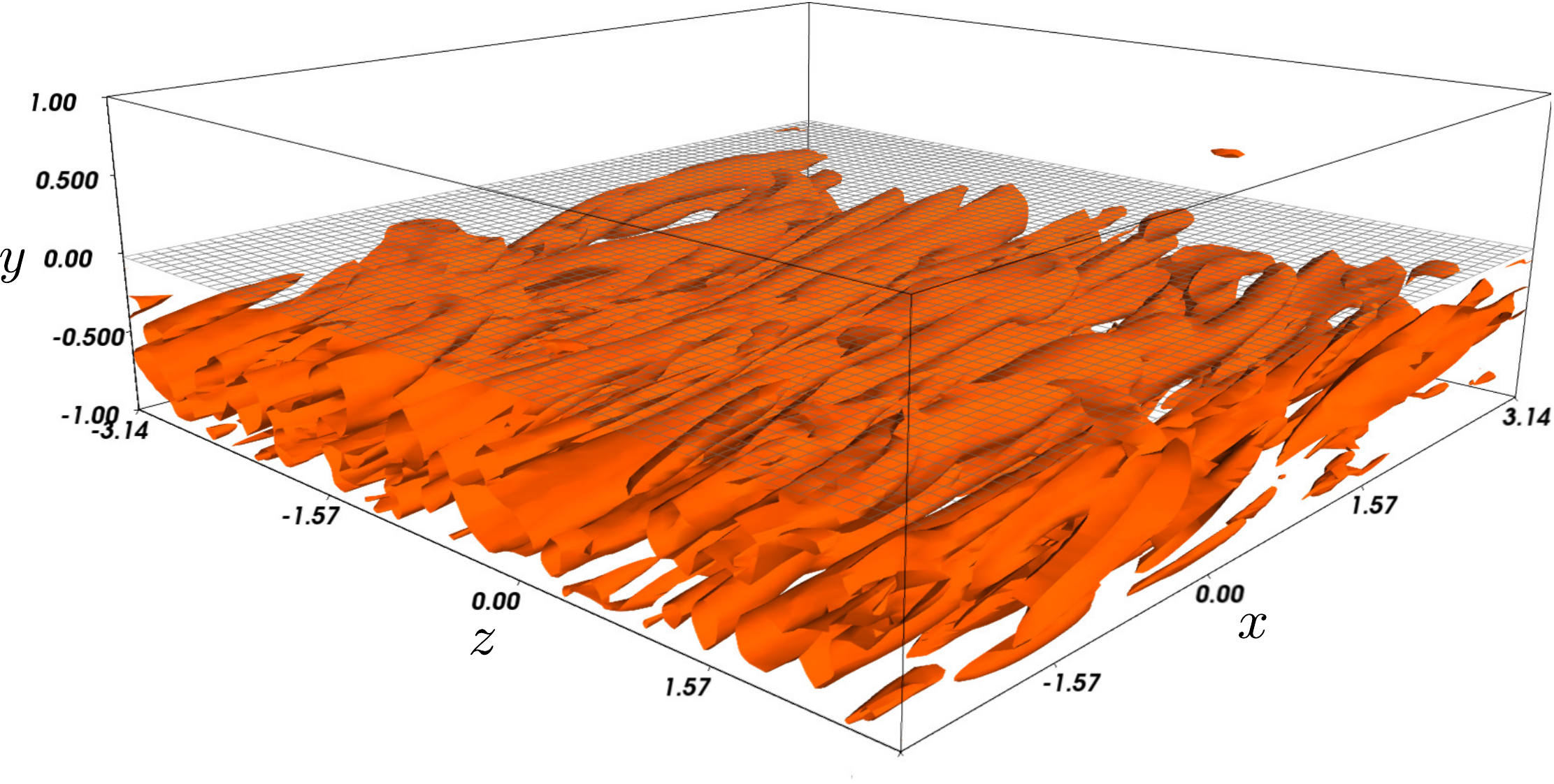}}
    \vspace{0.01in}
  \subfloat[$Ro = 0.7$]{
    \includegraphics[width=3.15in]{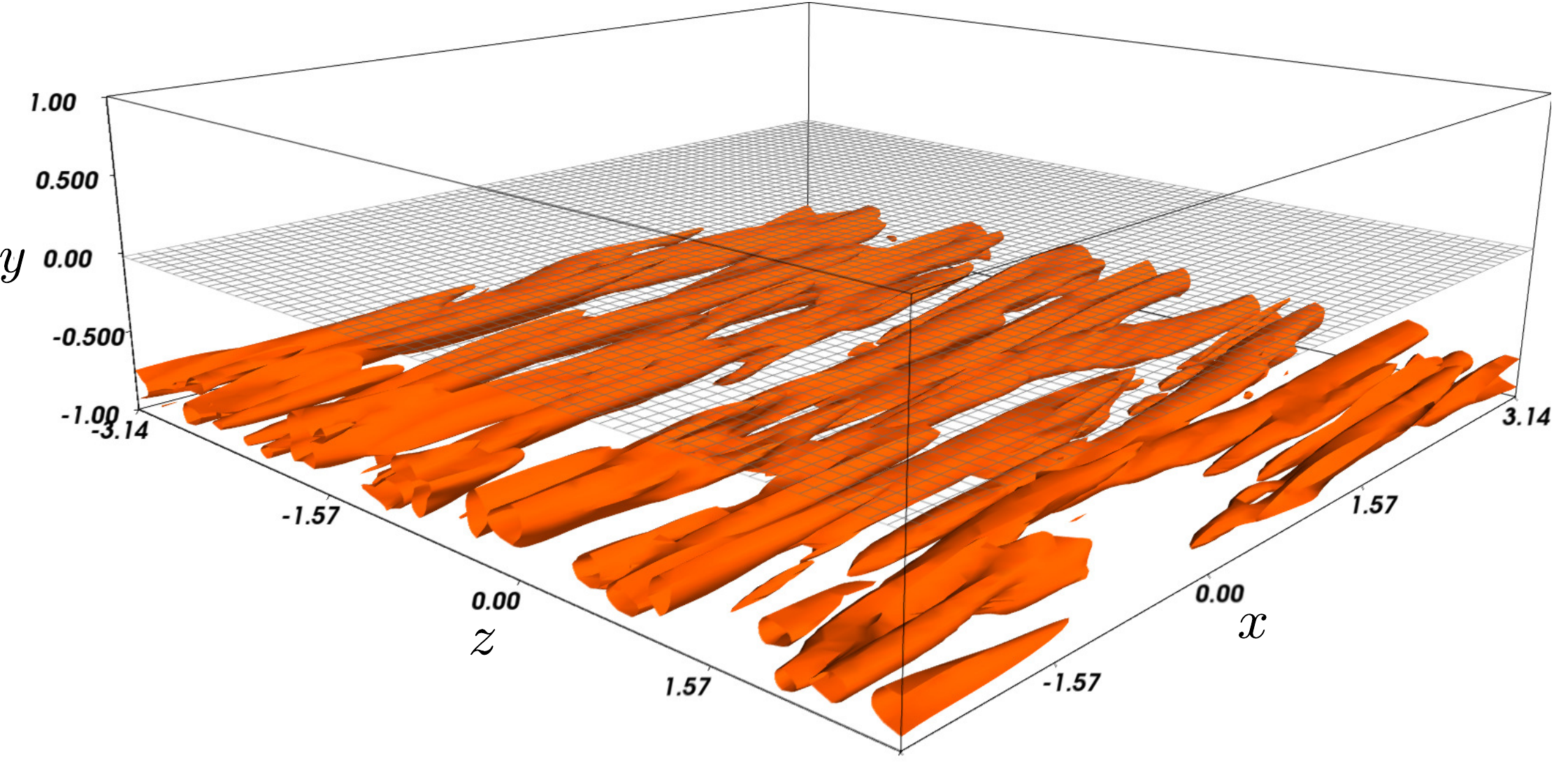}}
  \caption{The isocontours of constant $\lambda_{2} = -0.05$ at various rotation rates. The vortex cores are identified in each of the cases at $t = 500$, a characteristic time after the initial transient behaviour has died out. At low rotation rates, we see that there are no significant changes in the distribution of the identified vortex cores. It is seen that as we increase $Ro$, the vortex structures are getting restricted to the high pressure side of the channel.}
  \label{fig_lambda2}
\end{figure}

When we consider the $\lambda_2$ structures in figure \ref{fig_lambda2}, they are found to be increasingly concentrated at the lower, high pressure side of the channel. The structures appear to be far more disordered when the rotation rate is far from either linear stability boundary ($Ro = 0.2$ for example). The vortex structures formed also appear to be more ordered along the streamwise coordinate as the rotation is further increased. The conspicuous absence of vortical structures on the low pressure side of the channel suggests that the flow remains largely ordered and laminar in that region. Thus the Coriolis force acts to effectively laminarise at least one side of the channel flow. Such behaviour was seen by \cite{Lezius_Johnston_76JFM} who found a reduction in the turbulent intensities near the low-pressure wall. As we increase the rotation rates further, the Coriolis force acts to the confine the secondary motion to smaller regions in the channel. Thus the secondary flow set up is much weaker, as can 
be seen for the $Ro=0.7$ case in figure \ref{fig_lambda2}. 

\begin{figure}
\centering
  \subfloat[$Ro = 0.02$]{
    \includegraphics[width=3.125in]{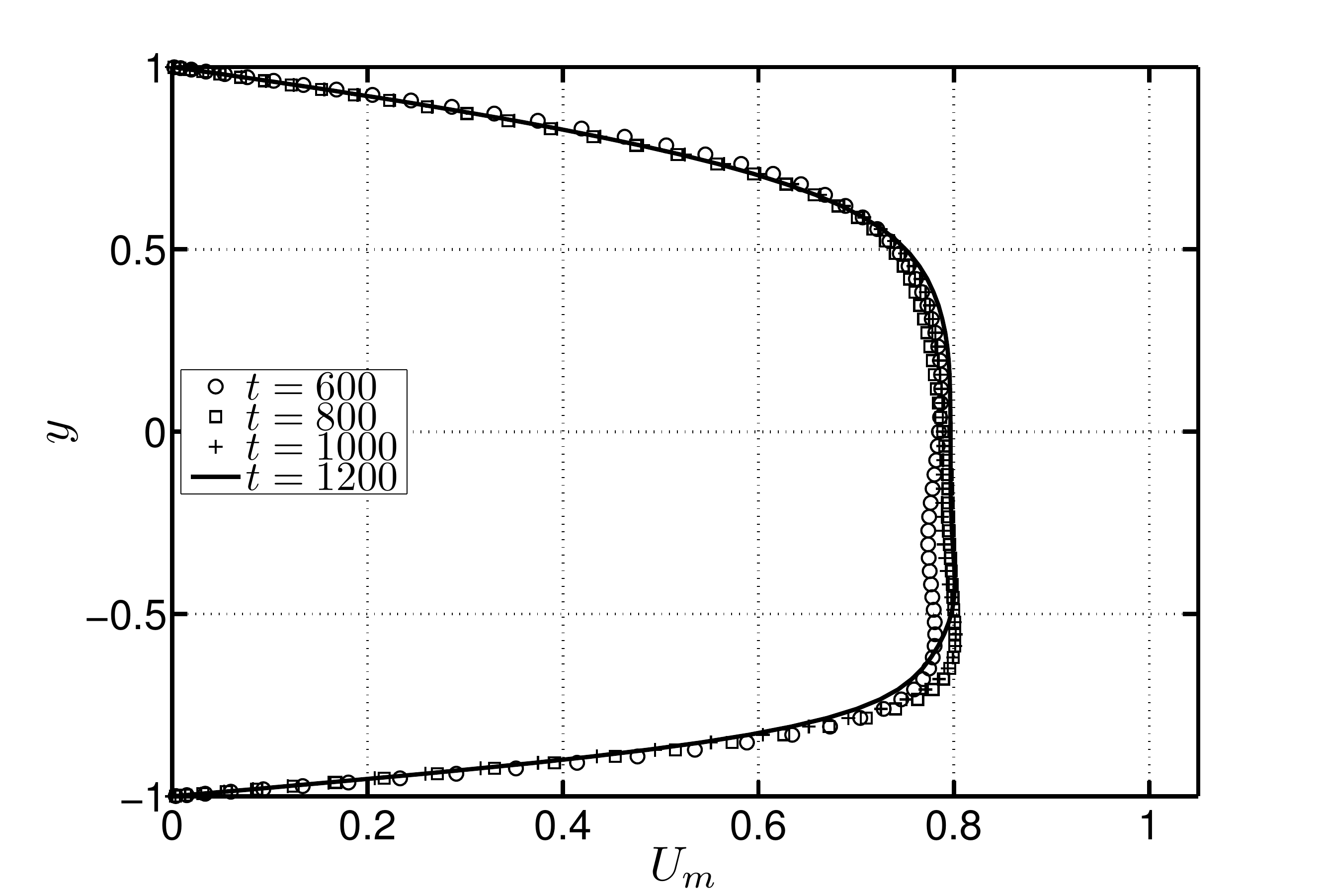}}
  \hspace{0.01in}\noindent
  \subfloat[$Ro = 0.2$]{
    \includegraphics[width=3.125in]{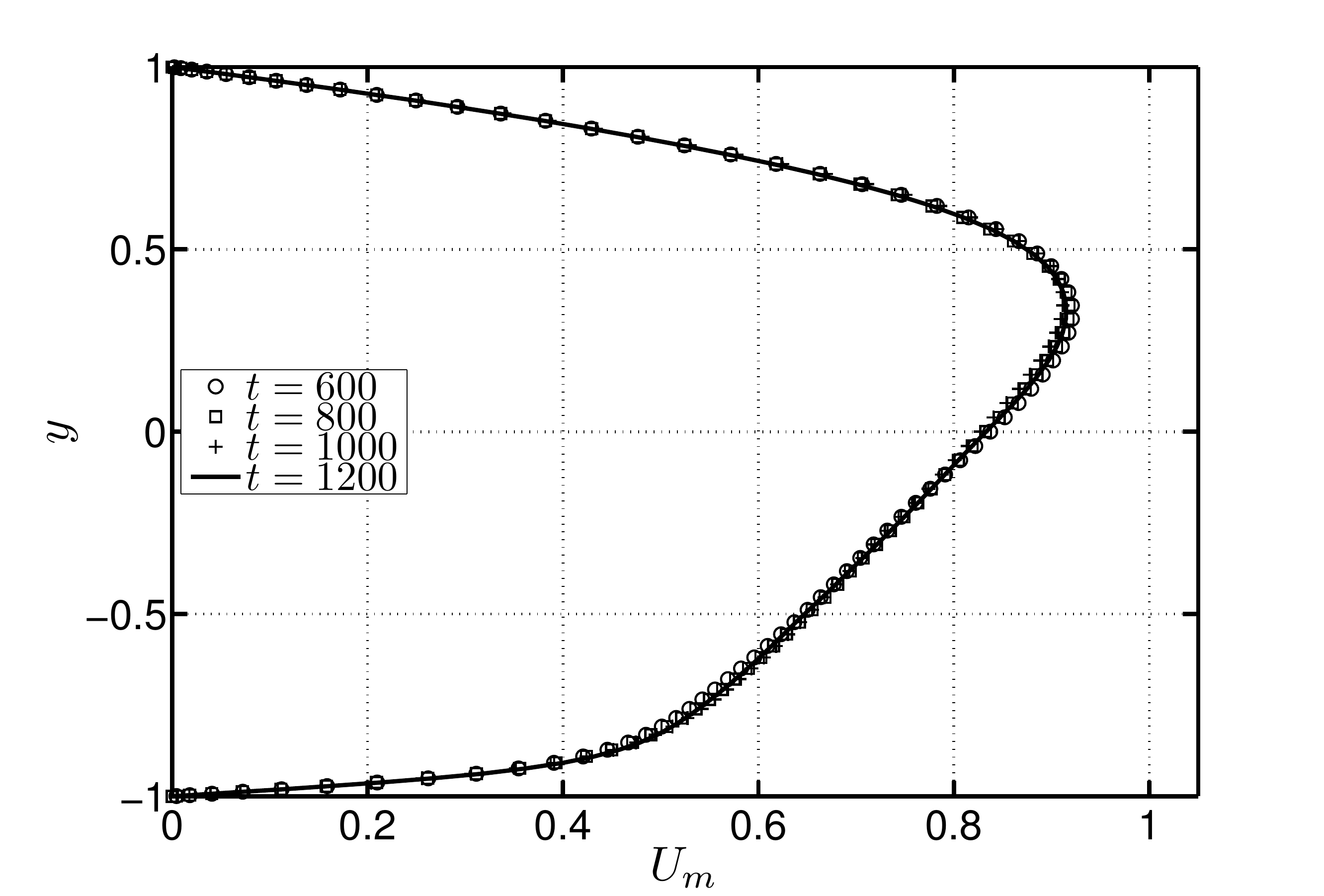}}
  \vspace{0.015in}
  \subfloat[$Ro = 0.5$]{
    \includegraphics[width=3.125in]{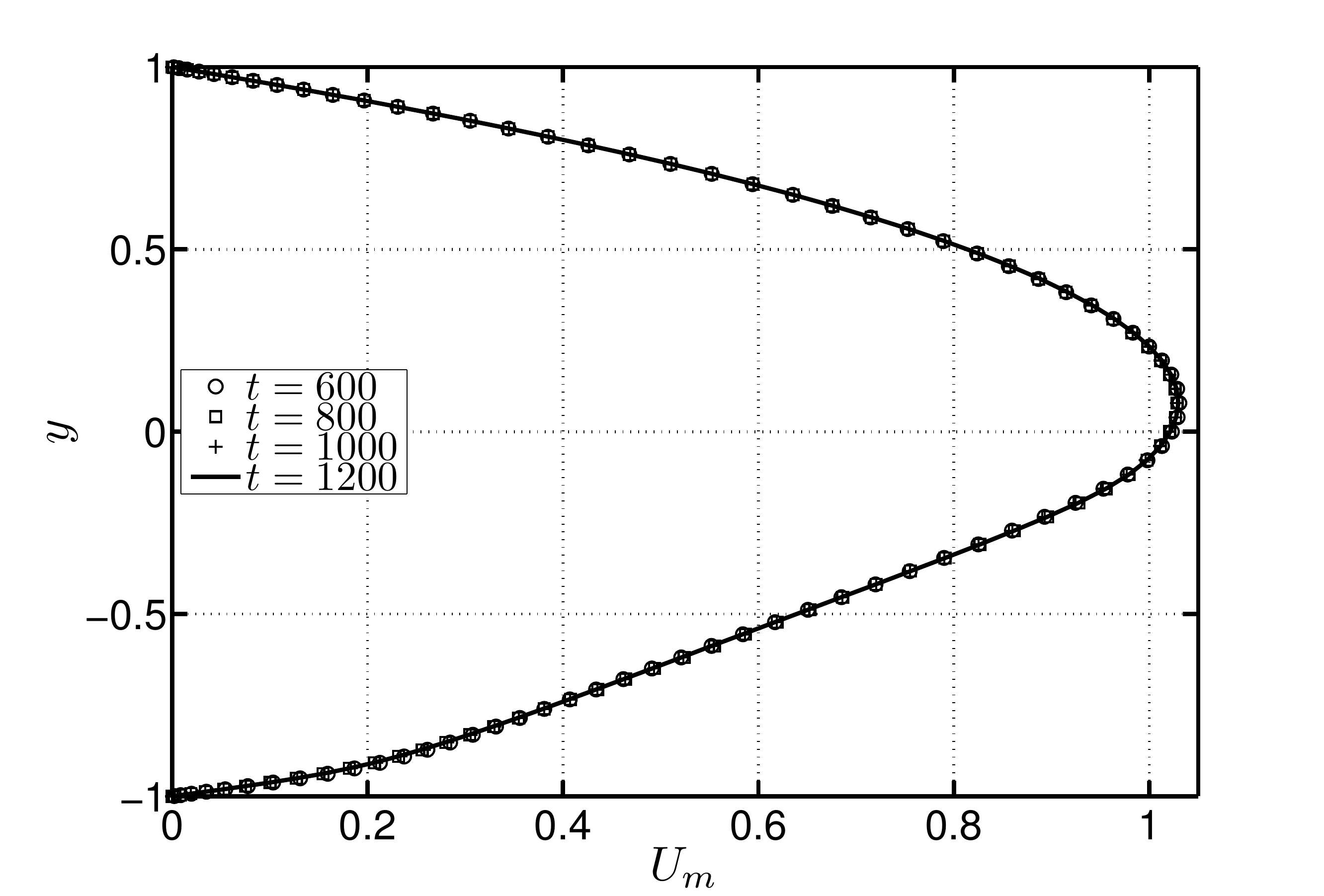}}
  \hspace{0.01in}
  \subfloat[$Ro = 0.7$]{
    \includegraphics[width=3.125in]{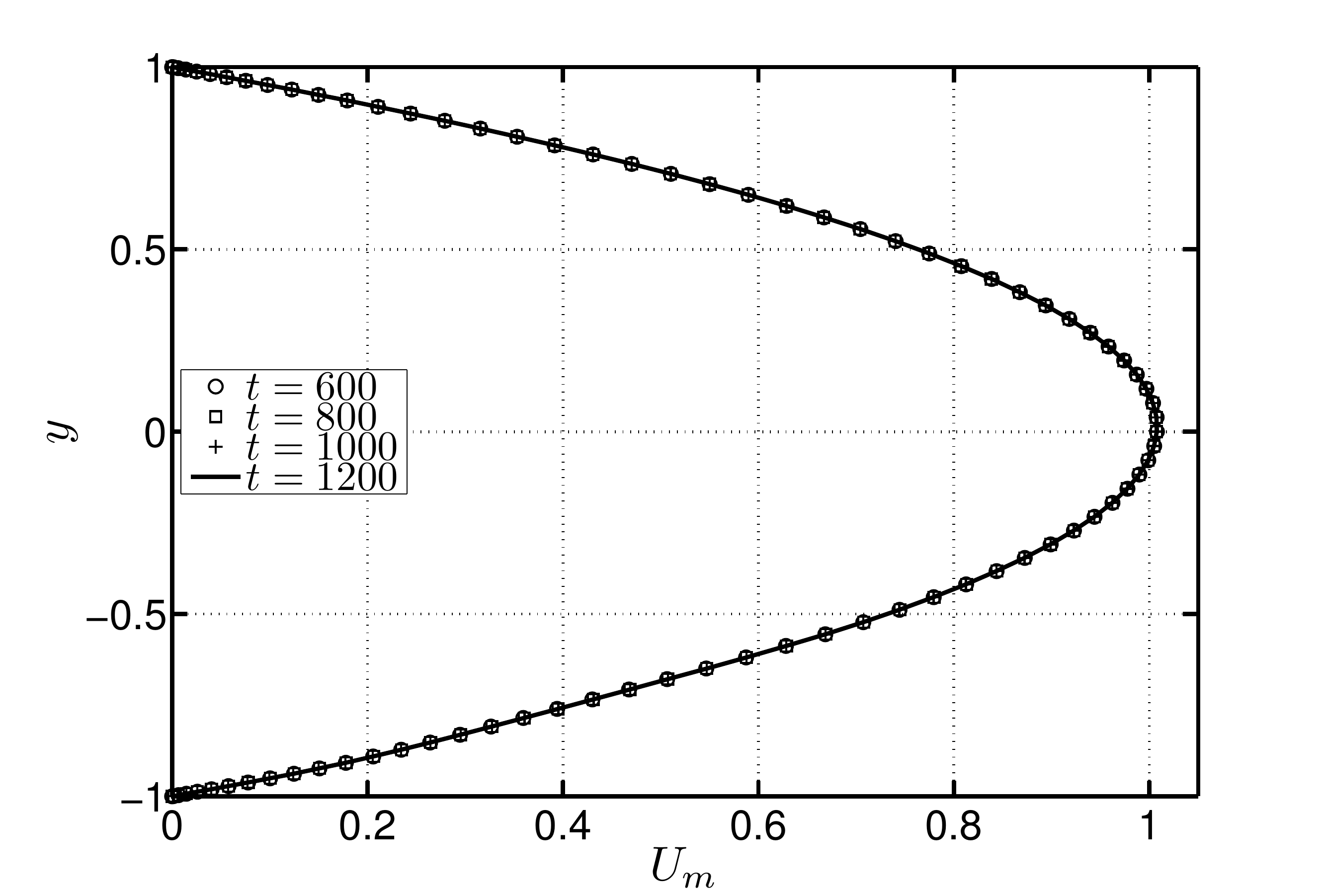}}
  \caption{Spatially averaged streamwise velocity profiles for different rotation rates, shown at different times. The familiar profile of a turbulent channel flow is obtained for low rotation rates. At moderate rotation rates the velocity profiles are markedly asymmetric. The mean velocity is seen to be linear in regions where the secondary vortex structures exist (see figure \ref{fig_lambda2}).}
  \label{fig_U_mean}
\end{figure}

\begin{figure}
 \centering
 \includegraphics[width=4in]{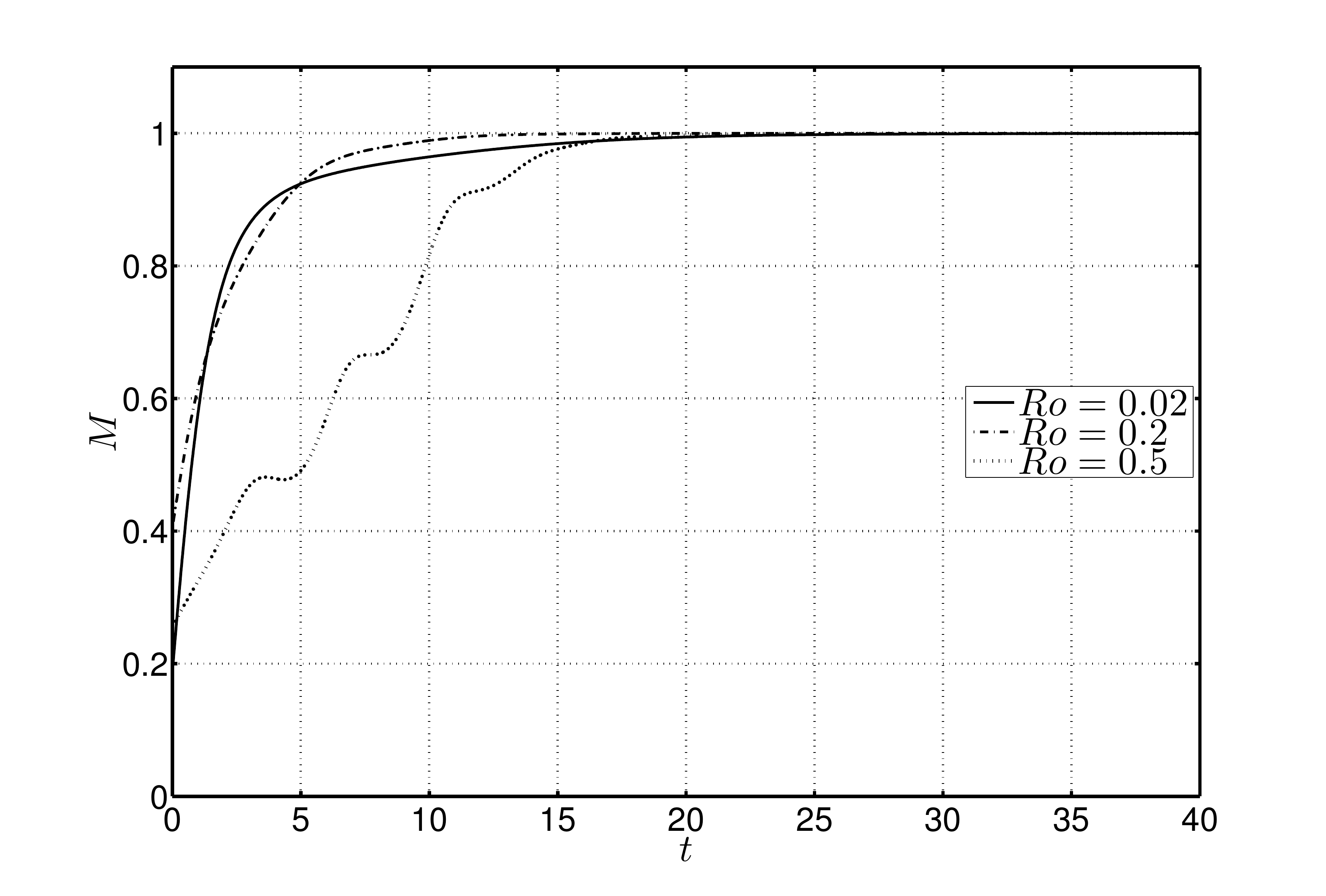}
 \caption{The evolution of the projection measure $M$ defined in equation (\ref{eq:proj}) for different rotation rates. Comparisons of the disturbance and the unstable eigenmode when $M = 1$ are found to give the same structure.}
 \label{fig_M_vs_t}
\end{figure}

The mean flow obtained for these rotation rates are given in figure \ref{fig_U_mean}. At larger rotation rates, smaller variations in time are seen, consistent with the fact that rms fluctuations are low, so shorter averaging is sufficient to obtain the correct mean flow. The case of $Ro=0.2$ displays clear departure from symmetry about the centreline. It is noticed that the velocity profile is linear over a significant portion of the channel width. This portion is the region where the strong vortical structures seen in figure \ref{fig_lambda2} exist. A similar correspondence of linear velocity profiles and strong structures have been reported earlier for turbulent rotating shear flows \cite{Bech_Andersson_97JFM,Lamballais_etal_96IJHFF,Lamballais_etal_98TCFD,Tanaka_etal_2000PF}. We comment here that the resulting mean flow is not different from the case if we induce transition using the unstable mode. The unstable mode is excited relatively quickly as can be surmised from figure \ref{fig_M_vs_t} where the 
projection measure $M$ (equation \ref{eq:proj}) is plotted. Therefore, at higher rotation rates, the initial condition serves as a background out of which the unstable mode emerges. 

\begin{figure}
\centering
  \subfloat[Low $Ro$]{
    \includegraphics[width=3.125in]{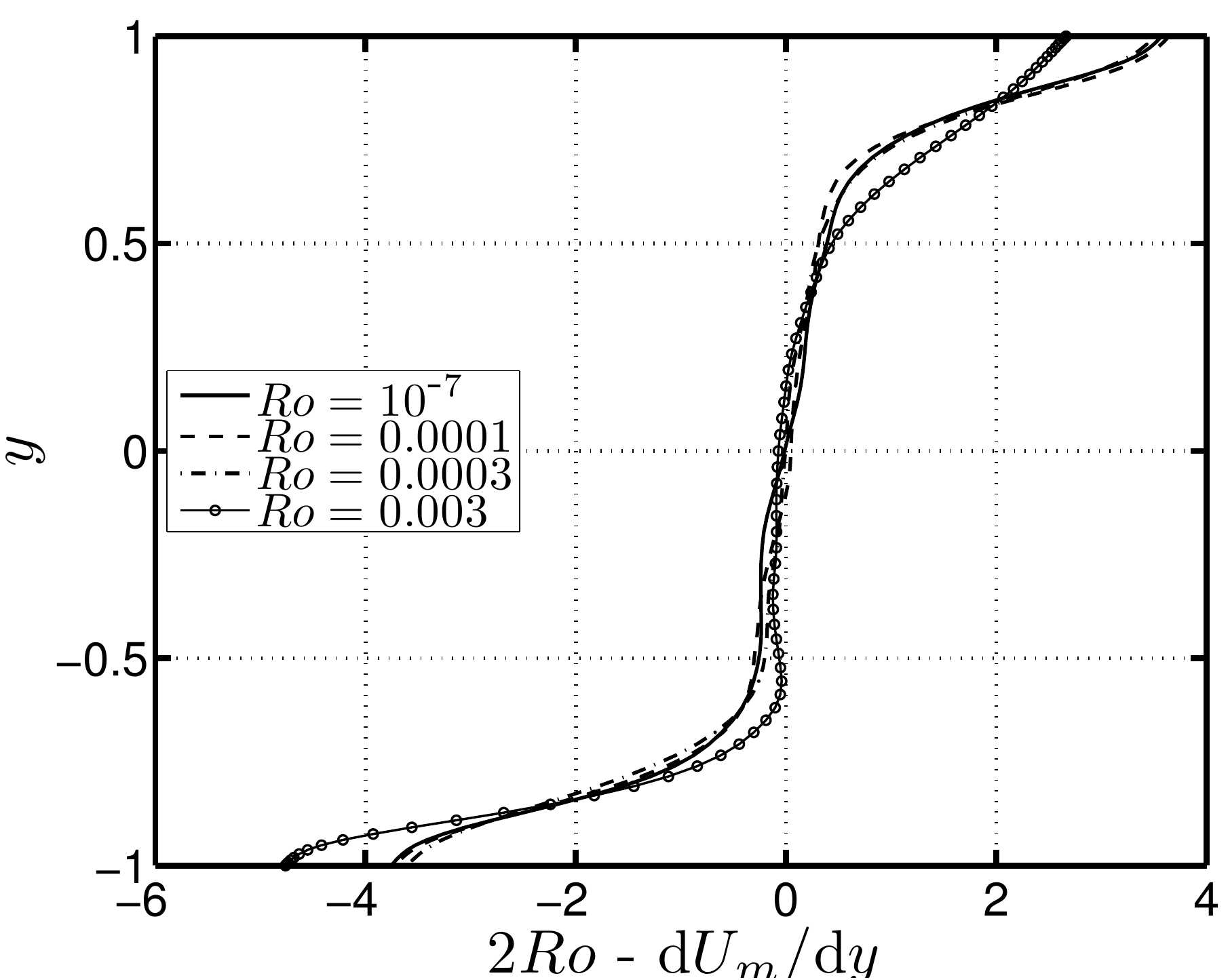}}
  \hspace{0.025in}
  \subfloat[High $Ro$]{
    \includegraphics[width=3.125in]{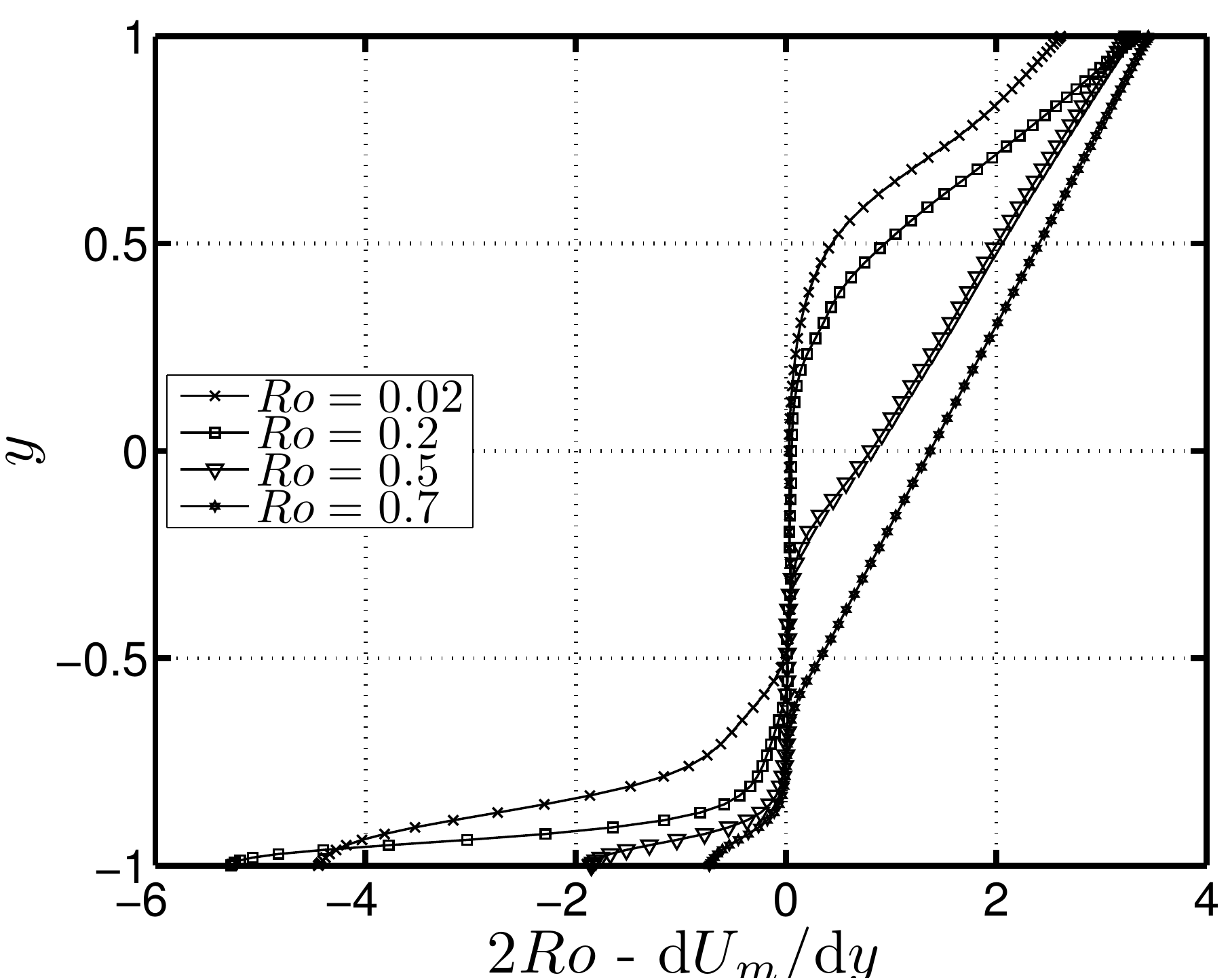}}
  \caption{The mean absolute vorticity of the flow at $t= 500$ for different rotation rates. For low values of $Ro$, there is no significant change with $Ro$. Once we go to a regime of high $Ro$, we see drastic changes in the corresponding field. }
  \label{fig_abs_vort_mean}
\end{figure}

A linear velocity profile implies a constant mean shear, whose value is such that the absolute local vorticity is nearly zero \cite{Suryadi_etal_14PRE,Kawata_Alfredsson_16JFM}. This is evident in figure \ref{fig_abs_vort_mean}, where we plot the mean absolute vorticity of the flow at a characteristic time. When the rotation rates are low, we see that there is no noticeable effect of $Ro$, and there is no region where the velocity profile is linear. As we increase $Ro$, the effect of rotation is pronounced. At moderate $Ro$ we obtain large regions where the absolute vorticity is zero, which are skewed towards the high pressure wall. Further increase in $Ro$ restricts the display of zero absolute vorticity to a small region close to this wall. Consistent with the earlier results of various authors, the fluctuating vortex structures are set up precisely near this region (see figure \ref{fig_lambda2}).

\begin{figure}
\centering
  \subfloat[$Ro = 0.02$]{
    \includegraphics[width=3.125in]{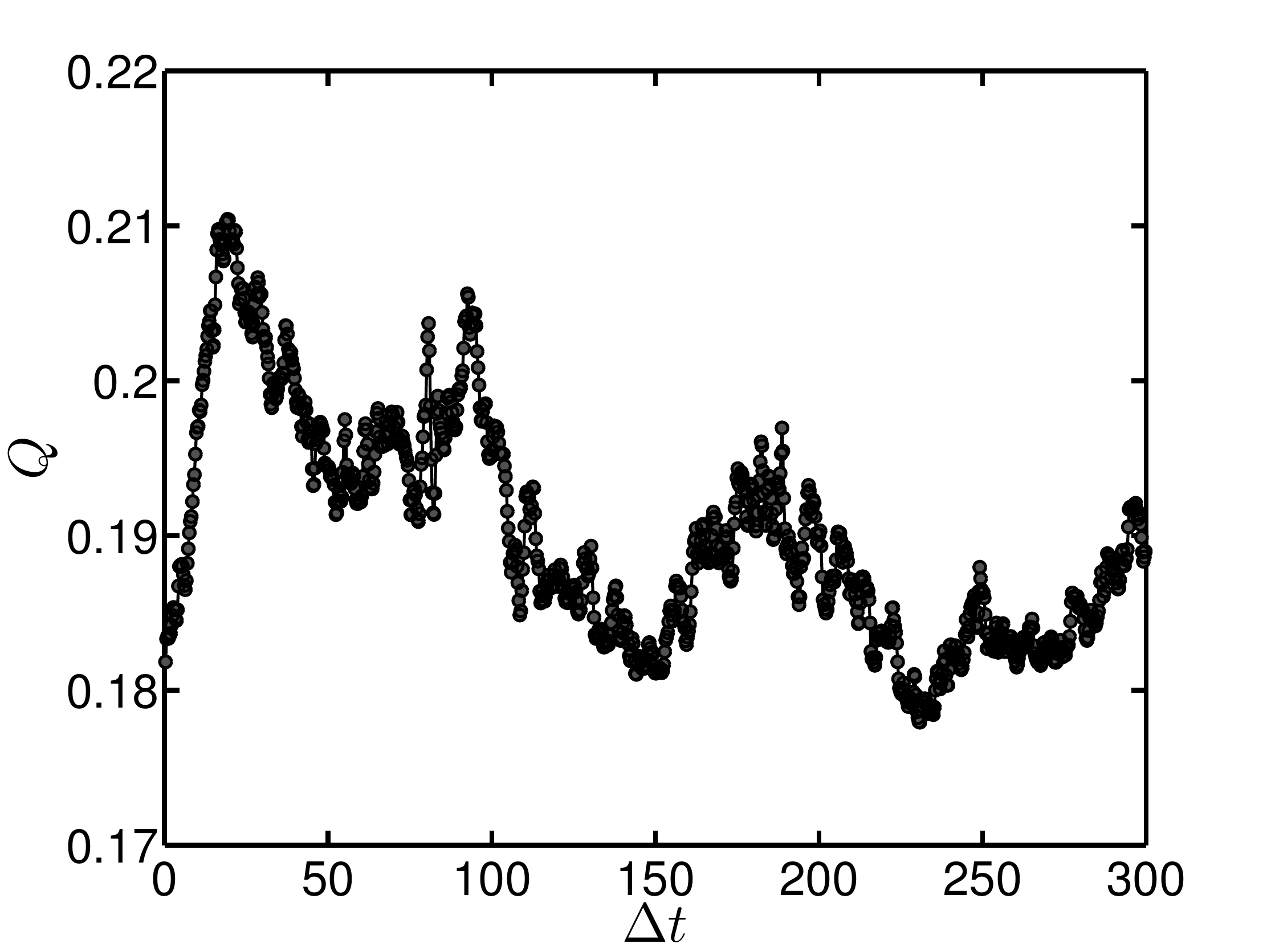}}
  \hspace{0.15in}
  \subfloat[$Ro = 0.2$]{
    \includegraphics[width=3.125in]{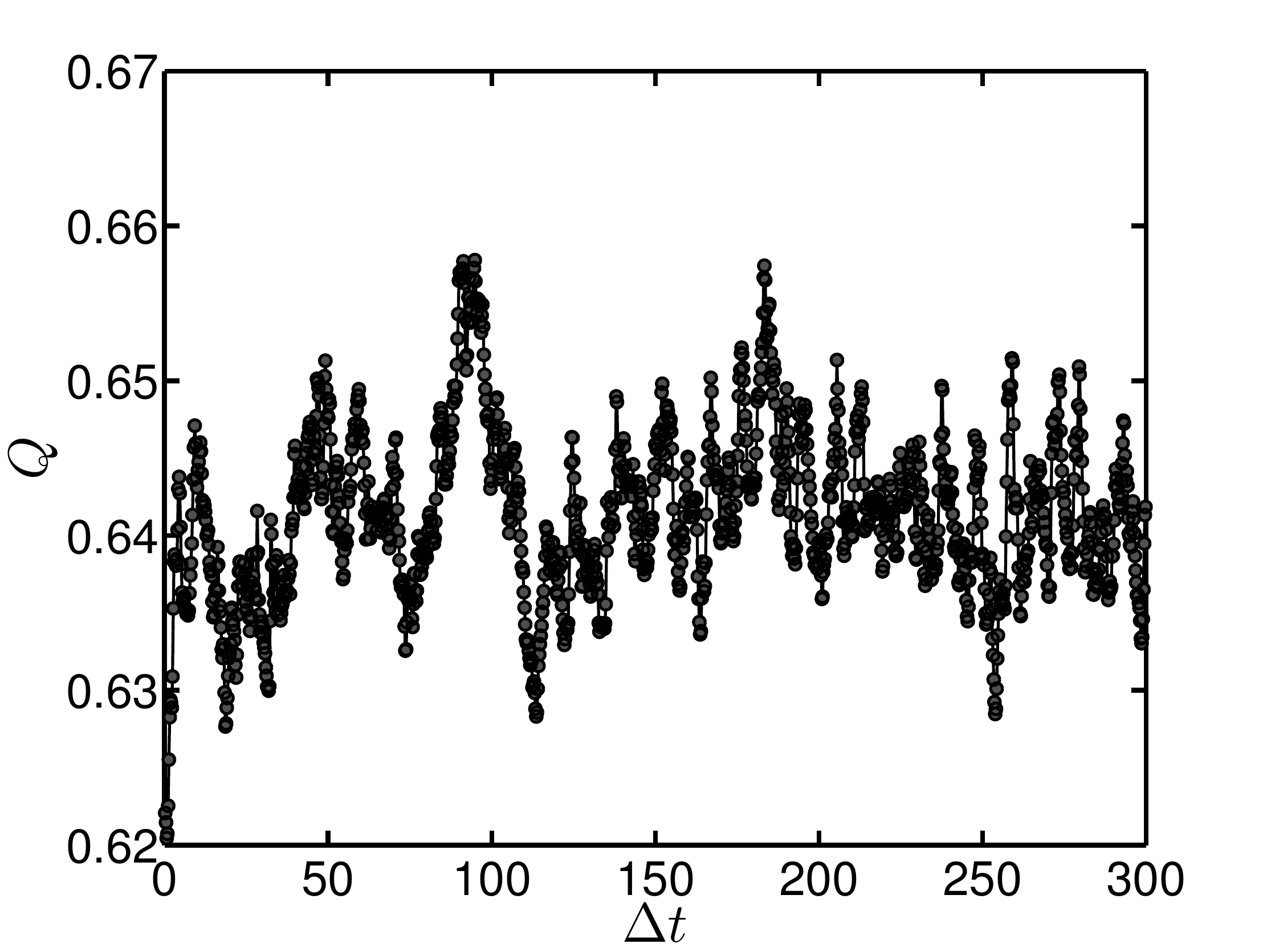}}
  \vspace{0.01in}
  \subfloat[$Ro = 0.5$]{
    \includegraphics[width=3.125in]{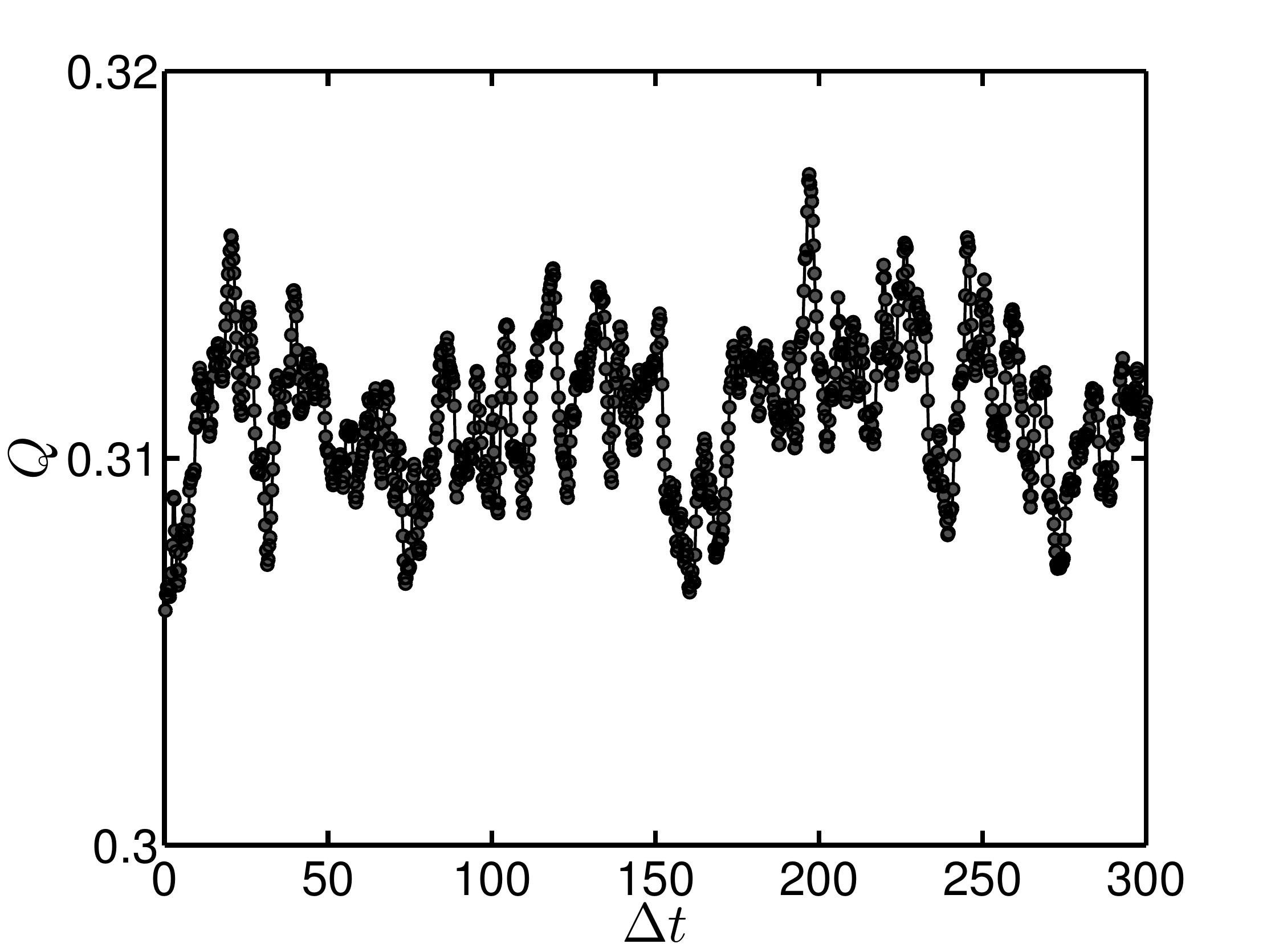}}
  \hspace{0.15in}
  \subfloat[$Ro = 0.7$]{
    \includegraphics[width=3.125in]{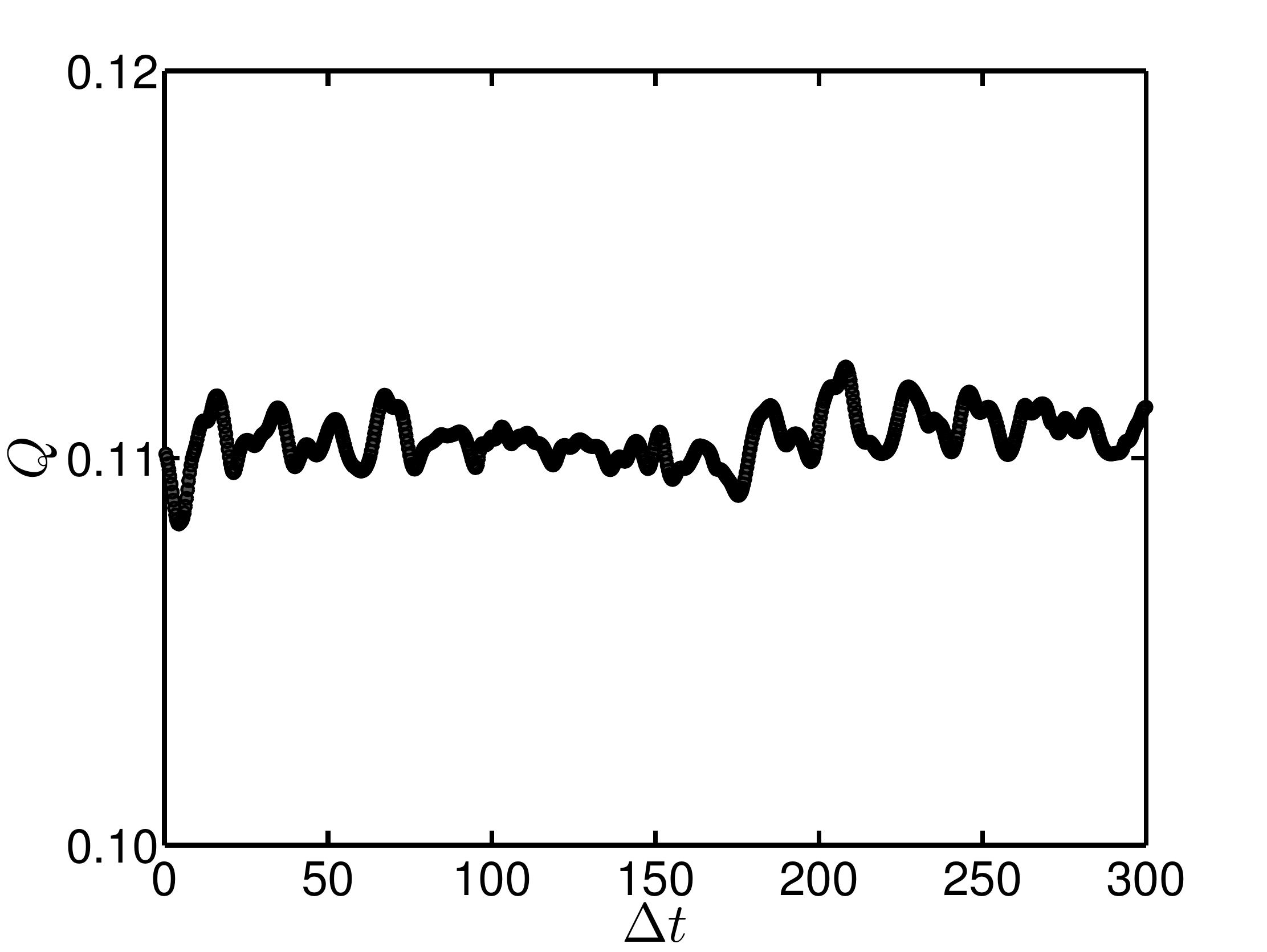}}
  \caption{The evolution for the measure of chaos $Q$, as defined in equation (\ref{eq_entropy2}). For all the cases, a departure from the initial state is shown. Here we choose $\tau = 500$ as the starting state.}
  \label{fig_entr}
\end{figure}

To get a picture of how chaotic the resulting flows are after transition, we plot entropy $Q$ in figure \ref{fig_entr}. It is seen that for none of the cases does $Q$ go to zero, and hence the resulting flows are chaotic. Additionally it must be noted that $Q$ defined in equation \ref{eq_entropy2} is an integral measure over the computational box. At these rotations, we are seeing one side of the channel getting laminarized with the vortex structures concentrated on the other side. This means that the contribution to $Q$ in these cases are not very significant in the laminarized side of the channel. Despite such a situation, the values of $Q$ are fairly high when compared to the low rotation cases. This suggests that the regions where the secondary flow does persist offer extremely chaotic dynamics. 

At this point, let us briefly summarise what has been observed here in this subsection. An initial perturbation not composed of the eigenmodes of the linearised operator has evolved in such a manner that the unstable mode has been excited rapidly. The excited mode is then responsible for driving the flow towards transition. This was seen more readily in cases with high rotation rates. The largest growing algebraic disturbance and the most unstable mode are seen to be streamwise independent. The algebraic disturbance can be considered to be a noisy environment from which the unstable eigenmode is picked up at some point after its introduction. In contrast, for low rotation rate cases within the linearly unstable regime (such as $Ro = 0.0003$, $0.003$), the period over which this process occurs is much longer. In such cases, the transition has already occured in the flow due to sub-critical mechanisms seen in the non-rotating channel flow. 

\section{Conclusion}
In this article, we focus on the role of algebraic disturbances in the transition scenario of the rotating channel in various rotation regimes. We show that the critical modal Reynolds number for the rotating channel flow does not coincide the energy critical Reynolds number for all rotation regimes. As a consequence, transient amplification of disturbances is observed in modally stable regions in the $Re$-$Ro$ phase plane. Interestingly the energy critical Reynolds number is only feebly sensitive to the rotation rate, in contrast to the modal behaviour. On a given side of the modally stable region (i.e., at low and high $Ro$) at a given Reynolds number, the maximum transient growth does not vary much with the rotation rate. It is only in the vicinity of the neutral boundary in $Re$-$Ro$ parameter space that discernible changes in the optimal characteristics are observed.

At low rotation rates, the transient growth of disturbance kinetic energy is due to the lift-up effect and is subsequently shown to be important while considering transition. The optimal transient growth amplitudes, and the corresponding optimal wavenumbers obtained are close to those for the non-rotating channel. However, even at extremely low rotation rates, the optimal structure breaks centreline symmetry due to the Coriolis force, with larger asymmetry closer to the neutral boundary. At extremely large rotation rates, consistent with the Taylor--Proudman theorem, all variation along the axis of rotation is inhibited. The streamwise independent disturbances, which yield the largest transient growth at low $Ro$ are therefore now suppressed. Thus we see only weakly growing perturbations that evolve transiently due to the Orr mechanism.

To study the effect of algebraic disturbances within the modally unstable region, we solve the incompressible Navier--Stokes equations using direct numerical simulations. When the rotation rate is low, we have shown that sub-critical transition similar to the non-rotating case occurs. The transient amplification of the disturbance triggers nonlinearity, and transition ensues. Vortical structures fill the entire domain, but at the higher $Ro$ end of this regime, an asymmetry about the centreline is evident both in the distribution of structures and in the mean flow. This happens over a wide range of initial disturbance amplitudes, except at extremely low initial energy of the disturbance where the unstable mode emerges after the transients die out and the secondary flow grows exponentially till the nonlinear terms become important. This shows that we cannot define a critical rotation number for the switch-over from transient-growth dominated transition to eigenmode-dominated. The switch-over is dependent on 
background perturbations.

On increasing the rotation rate to moderate levels ($Ro \sim 0.2$), the Coriolis force expectedly manifests itself in a more pronounced manner. Initial disturbances rapidly evolve into the most unstable eigenmode, and the resulting transitioned flow is strongly vortical with no characteristic structure or organisation. This is in sharp contrast to the elongated structures seen at both lower and higher rotation rates. With further increase in rotation rate, the secondary (chaotic) flow is increasingly localised towards one wall, becoming smaller until it finally disappears. At high rotation rates, the base flow is extremely stable to non-modal disturbances, as expected.

In summary, for the rotating channel flow, we have shown distinct behaviour patterns at low, intermediate and higher rotation rates, and the switch-over between these is gradual with change in rotation rate. 

This work was carried out under a collaborative Indo-French research project. Funding from Indo-French Centre for the Promotion of Advanced Research-CEFIPRA is gratefully acknowledged. SJ would like thank Prof. Luca Brandt at KTH Mechanics, Stockholm for his tremendous help with regard to understanding the SIMSON code.

\bibliography{ref_rot_chan}
\bibliographystyle{unsrt}

\end{document}